\RequirePackage{luatex85}

\documentclass[a4paper,11pt]{article}
\pdfoutput=1 

\usepackage{jheppub} 
\usepackage[T1]{fontenc} 
\usepackage{enumerate}
\usepackage{amsmath,amsfonts,amssymb}
\usepackage{mathtools}
\usepackage{enumitem}
\usepackage{slashed}
\usepackage{tikz}
\usepackage[mode=buildnew]{standalone}
\usepackage{tikz}

\usetikzlibrary{arrows,calc,graphs,patterns,positioning}
\usetikzlibrary{decorations,decorations.markings,decorations.pathmorphing,decorations.pathreplacing}
\usetikzlibrary{shapes, shapes.geometric}

\tikzset{/pgf/decoration/.cd,
    number of sines/.initial=5,
    angle step/.initial=20,
    snake it/.style={decorate, decoration=snake}
}
\newdimen\tmpdimen

\pgfdeclaredecoration{complete sines}{initial}
{
    \state{initial}[
        width=+0pt,
        next state=move,
        persistent precomputation={
            \pgfmathparse{\pgfkeysvalueof{/pgf/decoration/angle step}}%
            \let\anglestep=\pgfmathresult%
            \let\currentangle=\pgfmathresult%
            \pgfmathsetlengthmacro{\pointsperanglestep}%
                {(\pgfdecoratedremainingdistance/\pgfkeysvalueof{/pgf/decoration/number of sines})/360*\anglestep}%
        }] {}
    \state{move}[width=+\pointsperanglestep, next state=draw]{
        \pgfpathmoveto{\pgfpointorigin}
    }
    \state{draw}[width=+\pointsperanglestep, switch if less than=1.25*\pointsperanglestep to final, 
        persistent postcomputation={
        \pgfmathparse{mod(\currentangle+\anglestep, 360)}%
        \let\currentangle=\pgfmathresult%
    }]{%
        \pgfmathsin{+\currentangle}%
        \tmpdimen=\pgfdecorationsegmentamplitude%
        \tmpdimen=\pgfmathresult\tmpdimen%
        \divide\tmpdimen by2\relax%
        \pgfpathlineto{\pgfqpoint{0pt}{\tmpdimen}}%
    }
    \state{final}{
        \ifdim\pgfdecoratedremainingdistance>0pt\relax
            \pgfpathlineto{\pgfpointdecoratedpathlast}
        \fi
   }
}

\tikzset{
    scalar/.style={draw=black, thick, dashed},
    cscalar/.style={draw=black, postaction={decorate}, decoration={markings, mark=at position .55 with {\arrow[]{latex}}}, thick, dashed},
    fermion/.style={draw=black, postaction={decorate}, decoration={markings, mark=at position .55 with {\arrow[]{latex}}}, thick},
    antifermion/.style={draw=black, postaction={decorate}, decoration={markings, mark=at position .55 with {\arrow[>=latex]{<}}}, thick},
    majorana/.style={draw=black, thick},
    semiloop/.style={draw, thick, dashed, shape=semicircle, minimum size=1cm, outer sep=9pt, postaction={decorate}, decoration={markings, mark=between positions 0.32 and 1 step 1/2. with {\arrow[]{latex}}}},
    photon/.style={draw=none, outer ysep=3pt, postaction={draw=black, thick, decorate}, decoration={complete sines, amplitude=7pt}},
    photonloop/.style={draw=none, outer ysep=3pt, postaction={draw=black, thick, decorate}, decoration={complete sines, number of sines=15, amplitude=5pt}},
    gluon/.style={draw=none, outer ysep=3pt, postaction={draw=black, thick, decorate}, decoration={coil, amplitude=3pt, segment length=5pt}}, 
    composite/.style={draw=gray!70!, line width=5pt},
    vtx/.style={inner sep=0pt},  
    mass/.style={cross out, draw, minimum size=5pt, inner sep=0pt},
    ins/.style={draw, cross, shape=circle, minimum size=5pt, inner sep=0pt}, 
    cross/.style={path picture={\draw[black] (path picture bounding box.south east) -- (path picture bounding box.north west) (path picture bounding box.south west) -- (path picture bounding box.north east);}},
    vertex/.style={draw, shape=circle, fill=black, minimum size=3pt, inner sep=0pt},
    blob/.style={draw, shape=circle, preaction={fill,white}, pattern = north west lines, minimum size=15pt, inner sep=0pt},        
    loop/.style={shape=circle, minimum size=1.7cm, outer sep=9pt},
    site/.style={draw, shape=circle, minimum size=1.5cm, inner sep=0pt},
    brane/.style={trapezium, draw, trapezium stretches=true, minimum height=4cm, minimum width=3.5cm, inner sep=0, trapezium left angle=35, trapezium right angle=145, shape border uses incircle, shape border rotate=17.5},
    momentum/.style={->, dist/.store in=\segDistance, pos/.store in=\segPos, len/.store in=\segLength,
      to path={
      ($(\tikztostart)!\segPos!(\tikztotarget)!\segLength/2!(\tikztostart)!\segDistance!90:(\tikztotarget)$) -- 
      ($(\tikztostart)!\segPos!(\tikztotarget)!\segLength/2!(\tikztotarget)!\segDistance!-90:(\tikztostart)$)  \tikztonodes
      }, 
      pos=.5,
      len=7mm,
      dist=2mm
    },    
}

\usepackage{scalerel}
\usepackage{hyperref}
\usepackage[section]{placeins}
\title{\boldmath Radiative neutrino mass model from a mass dimension-11 $\Delta L =2 $ effective operator}
\usepackage{xcolor}
\usepackage{array}
\usepackage{float}
\usepackage{tabu}
\usepackage{braket}
\usepackage{subcaption}
\usepackage{graphicx}
\usepackage[utf8]{inputenc} 
\usepackage{dsfont}
\newcommand{\RomanNumeralCaps}[1]
    {\MakeUppercase{\romannumeral #1}}

\graphicspath{{figs/}}  

\allowdisplaybreaks

\author{John Gargalionis,}
\author{Iulia Popa-Mateiu,\footnote{Corresponding author.}}
\author{Raymond R. Volkas}

\affiliation{ARC Centre of Excellence for Particle Physics at the Terascale \\ School of Physics \\ The University of Melbourne,\\ Victoria 3010, Australia}

\emailAdd{garj@student.unimelb.edu.au}
\emailAdd{ipopa@student.unimelb.edu.au}
\emailAdd{raymondv@unimelb.edu.au}

\abstract{We present the first detailed phenomenological analysis of a radiative Majorana neutrino mass model constructed from opening up a $\Delta L = 2$ mass-dimension-11 effective operator constructed out of standard model fields. While three such operators are generated, only one dominates neutrino mass generation, namely $O_{47} = \overline{L^C} L \overline{Q^C} Q \overline{Q} Q^C H H$, where $L$ denotes lepton doublet, $Q$ quark doublet and $H$ Higgs doublet. The underlying renormalisable theory contains the scalars $S_1 \sim (\bar{3},1,1/3)$ coupling as a diquark, $S_3 \sim (\bar{3},3,1/3)$ coupling as a leptoquark, and $\Phi_3 \sim (3,3,2/3)$, which has no Yukawa couplings but does couple to $S_1$ and $S_3$ in addition to the gauge fields. Neutrino masses and mixings are generated at two-loop order. A feature of this model that is different from many other radiative models is the lack of proportionality to any quark and charged-lepton masses of the neutrino mass matrix. One consequence is that the scale of new physics can be as high as $10^7$ TeV, despite the operator having a high mass dimension. This raises the prospect that $\Delta L = 2$ effective operators at even higher mass dimensions may, when opened up, produce phenomenologically-viable radiative neutrino mass models. The parameter space of the model is explored through benchmark slices that are subject to experimental constraints from charged lepton flavour-violating decays, rare meson decays and neutral-meson mixing. The acceptable parameter space can accommodate the anomalies in $R_{K^{(*)}}$ and the anomalous magnetic moment of the muon.}

\begin{document} 
\maketitle
\flushbottom

\section{Introduction}
\label{sec:intro}

The minimal Standard Model (SM) features massless neutrinos. However, the experimental observation of neutrino oscillations has established that at least two of the three known neutrinos are massive~\cite{1958JETP....6..429P, Pontecorvo:1957qd, 0004-637X-496-1-505, Poon:2001ee, PhysRevLett.81.1562, Michael:2006rx, Abe:2008aa, Abe:2011sj, Abe:2011fz, An:2012eh, Ahn:2012nd, Abe:2014ugx}. These experiments have measured the squared-mass differences $\Delta m^2_{21} \equiv m_2^2 - m_1^2$ and $| \Delta m^2_{32}| \equiv |m_3^2 - m_2^2|$, but are unable to probe the absolute neutrino mass scale. However, cosmological constraints derived from large-scale structure and cosmic microwave background measurements provide a strong upper bound on the sum of the neutrino masses of about $0.2~\text{eV}$~\cite{Ade:2015xua}. Independently of cosmology, $\beta$-decay endpoint measurements constrain the absolute mass scale to be at most about $1~\text{eV}$~\cite{Aker:2019uuj,Kraus:2004zw,PhysRevD.84.112003}. With or without the cosmological constraint, it is clear that the neutrino mass eigenvalues are at least six orders of magnitude smaller than that of the lightest charged fermion, the electron. The neutrino mass problem is the determination of the dynamical mechanism by which neutrino masses are generated and why those masses are so small. All mechanisms require the introduction of as-yet undiscovered fields, and thus constitute physics beyond the Standard Model (BSM). (We will use BSM and ``new physics'' (NP) interchangeably.) \\\\
A pivotal question for neutrino mass models is whether or not neutrinos are their own antiparticles. Being electrically neutral, neutrinos are the only Majorana fermion candidates in the SM. Thus, neutrino mass models fall into two categories: Dirac and Majorana. Dirac mass can be generated by introducing right-handed neutrino fields into the low-energy spectrum of the SM. Neutrino mass would then be generated through the same mechanism responsible for all SM fermion masses; however, the smallness of the neutrino masses would simply  be due to unusually small Yukawa couplings -- an unsatisfying resolution.  \\\\
Majorana neutrino mass models can provide a more natural explanation.\footnote{As can more complicated Dirac mass models.} The argument is as follows. All Majorana neutrino mass terms must take the form $\overline{\nu_L^c}m_{\nu}\nu_L+ \overline{\nu_L}m_\nu ^*(\nu_L)^c$, where $\nu_L$ is a SM left-handed neutrino field and  $\nu_L^c$ is the CP conjugate which is equivalent to a right-handed antineutrino field. Since both $\nu_L$ and $\overline{\nu_L^c}$ carry a lepton number of $+1$, these mass terms  violate total lepton number by two units ($\Delta L = 2$), as necessary when neutrinos and antineutrinos are identical. As we review below, this feature helps us explore neutrino mass models in a systematic way. Now, recall that the quantum numbers for the left-handed lepton doublet, $L \sim (\mathbf{1}, \mathbf{2}, -1/2)$, which contains the left-handed neutrinos, are such that a Majorana mass term breaks $SU(2)_L \times U(1)_Y$ symmetry. This issue can be resolved by introducing exotic fields that exist at an energy scale above the electroweak scale. These heavy exotic fields couple to SM particles in a gauge invariant and renormalisable way, and generate self-energy Feynman diagrams for the left-handed neutrinos at tree-level or loop-level.  At energy scales below the electroweak scale, neutrino mass manifests (can be understood) through $\Delta L = 2$ effective operators, obtained by integrating out the exotic heavy fields. The mass terms are then suppressed by the scale of new physics leading to a natural explanation for the smallness of neutrino masses. The three seesaw models~\cite{Minkowski:1977sc,Konetschny:1977bn,GellMann:1980vs,PhysRevLett.44.912,Glashow:1979nm,PhysRevD.24.1232,PhysRevD.22.2227, PhysRevD.22.2860, Foot:1988aq}, for example, are all UV-completions of the same mass-dimension 5 effective operator, called the Weinberg operator. The seesaw models are wonderfully minimal, however the high BSM scale typically invoked makes them challenging to experimentally test \cite{Fabbrichesi:2014qca, Dev:2013ff, Guo:2016dzl}. We should, therefore, consider other possibilities, for this reason, as well as for the sake of completeness. \\\\
Classifying Majorana neutrino mass models using $\Delta L = 2$ effective operators, each of which can be ``opened up'' (UV-completed at tree level) to produce neutrino self-energy diagrams, is a systematic way to approach the neutrino mass problem. Babu and Leung  \cite{Babu:2001ex}  have published a near-complete list of $\Delta L = 2$ effective operators which may be opened up using exotic fields such as massive scalars, vector-like fermions and massive Majorana fermions~\cite{PhysRevD.87.073007}. The resulting models generate Majorana neutrino mass either at tree-level or loop-level with most of the operators leading to models that produce the latter. An alternative and complementary approach to neutrino-mass model classification can be structured around loop-level completions of the $\Delta L = 2$ Weinberg-like operators $\overline{L^C}LHH(H^\dagger H)^n$~\cite{Bonnet:2012kz,
  Sierra:2014rxa, Cepedello:2018rfh, Cepedello:2017eqf}. The mass-dimension of an operator is necessarily odd when $(\Delta B - \Delta L)/2$ is odd \cite{Kobach:2016ami}, where $\Delta B$ is the change in baryon number. All effective operators classified by Babu and Leung conserve baryon number and break lepton number by two, thus they all have odd mass-dimension.\\\\
Radiative neutrino mass models, which have mass generated at loop level, introduce additional suppression factors alongside the suppression that     comes from the masses of the heavy exotic particles. Radiative neutrino mass models are attractive because they naturally produce small neutrino masses for three reasons:
\begin{enumerate}[label=\roman*]
\item a suppression of $\frac{1}{(16\pi^{2})^{l}}$, where $l$ is the number of loops in the neutrino self-energy diagram, from the numerical factor which automatically comes with each loop integration,
\item a product of couplings which are potentially all smaller than 1 representing the interaction strengths of the exotic particles, and
\item  a suppression by  $\left(\frac{v}{\Lambda}\right)^p$, where $v$ is the vacuum expectation value (vev) of the Higgs field, $\Lambda$ is the mass scale of the exotica coming from the exotic propagators introduced during the UV-completion, and the exponent $p>0$ is model-dependent.
\end{enumerate} 
The trend is for the higher dimensional effective operators listed in \cite{Babu:2001ex} to include more suppression in the form of i.\ and ii., thus accordingly decreasing the scale of new physics (NP) needed to produce small neutrino masses. A combination of points i.\ and iii.\ is the reason Babu and Leung do not include effective operators of dimension 13 and greater in their list. It was believed that any exotic particles used to complete these models would have to be detectable at an energy scale that has already been probed \cite{Babu:2001ex} and therefore, dimension-11 operators that produce neutrino masses in agreement with current data at two-loop level or more would lie in a sweet spot --- bringing the scale of BSM physics to a few TeV, an energy scale that is being directly probed at the Large Hadron Collider (LHC) and indirectly at precision- or luminosity-frontier experiments, and would be fully accessible at a future $100$ TeV collider. However, in this paper, we present a radiative Majorana neutrino mass model derived from a mass-dimension 11 effective operator with a scale of NP $\Lambda$ that can be as high as about $10^7$ TeV. Our findings suggest that dimension-13, and possibly even dimension-15 effective operators should not be overlooked in the search for viable Majorana neutrino mass models. 
For a comprehensive review of radiative neutrino mass models and the effective operator method see~\cite{Cai:2017jrq}. \\\\
In this paper, we present the first detailed radiative Majorana neutrino mass model derived from a mass-dimension 11 effective operator. In Section~\ref{Sec: The Model}, we define our Model and explain how neutrino masses are generated. Then, in Section~\ref{Sec: Constraints}, we investigate the constraints imposed by experimental results from rare processes involving charged leptons and flavour physics and discuss the results in Section~\ref{sec:results}. In Section~\ref{Sec: Conclusion}, we offer our conclusions. 

\section{The Model}
\label{Sec: The Model}

\begin{figure}[t]
\centering
\begin{subfigure}[c]{0.3\textwidth}
\centering
\includegraphics[width=0.7\linewidth]{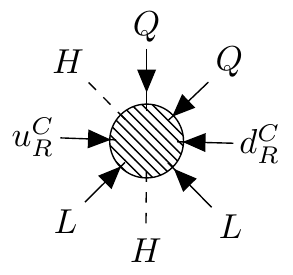}
\caption{}
\end{subfigure}
\hspace{0.1cm}
\begin{subfigure}[c]{0.3\textwidth}
\centering
\includegraphics[width=0.7\linewidth]{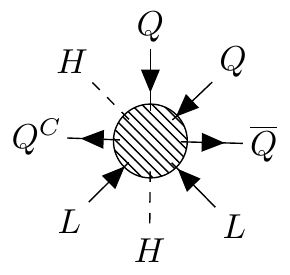}
\caption{}
   \end{subfigure}
\hspace{0.1cm}
\begin{subfigure}[c]{0.3\textwidth}
\centering
\includegraphics[width=0.7\linewidth]{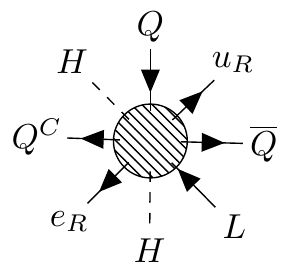}
\caption{}
\end{subfigure}
\caption{Diagrammatic representation of the effective operators $O_{25}$, $O_{47}$, $O_{55}$ (from left to right). The arrows on the fermions indicate the chirality with the convention being that the arrow points in for left-handed fermions.}
\label{fig: effective operators}
\end{figure}

We introduce three exotic colour-triplet scalar fields to the particle content of the SM: an $SU(2)$ singlet, $S_{1}$, and two $SU(2)$ triplets, $S_{3}$ and $\phi_{3}$, with quantum numbers given by
\begin{align*}
 S_{1} \sim (\bar{\textbf{3}}, \textbf{1}, 1/3),~~~~
 S_3 \equiv 
 	\begin{pmatrix}
	\begin{array}{l}
 		S_{3}^{4/3}\\  S_{3}^{1/3}\\ S_{3}^{-2/3}
	\end{array}
 	\end{pmatrix} 
 \sim (\bar{\textbf{3}}, \textbf{3}, 1/3),~~~~
 \boldsymbol{\phi}_3 \equiv
  	\begin{pmatrix}
	\begin{array}{l}
 		\phi_{3}^{5/3}\\  \phi_{3}^{2/3}\\  \phi_{3}^{-1/3}
		\end{array}
	\end{pmatrix}
 \sim (\textbf{3}, \textbf{3}, 2/3),
	\end{align*}
where the subscripts indicate the transformation property of the scalars under $SU(2)_{L}$ and the superscripts indicate the electric charge of each component of the exotic scalars. The first entry in the triples specifies the colour multiplet, the second the weak-isospin allocation, and the third the hypercharge, $Y$, normalised such that electric charge $Q = I_3 + Y$.\\\\
The three exotic scalars listed above generate three separate $\Delta L = 2$, dimension-$11$ effective operators at tree level, and give rise to radiative Majorana neutrino masses at two-loops. It is important to note that these three scalars do not give rise to any lower dimension $\Delta L = 2$ effective operators at tree-level. Thus, the neutrino self-energy diagrams generated in the UV-completion of these dimension-$11$ operators with our three exotic scalars will be the leading order contribution to the neutrino mass~\cite{explodingOperators}. In the notation used by Babu and Leung in \cite{Babu:2001ex}, the operators, depicted in Figure \ref{fig: effective operators}, are 
\begin{align*}
O_{25} &= \overline{L^C} L \overline{Q^C} d_R^C \overline{Q^C} u_R^C HH ,\\
O_{47} &= \overline{L^C} L \overline{Q^C} Q \overline{Q}Q^C HH ,\\
O_{55} &=\overline{L^C}Q\overline{Q}Q^C\overline{e_R^C}u_R HH.
\end{align*}
The scalars $S_1$ and $S_3$ can Yukawa-couple as leptoquarks, diquarks, as one of each, or as both.  As leptoquarks, they appear together in models which tackle the flavour anomalies in the $R_{K^{(*)}}$ and $R_{D^{(*)}}$ observables~\cite{Crivellin:2017zlb, Buttazzo:2017ixm, Marzocca:2018wcf, Bigaran:2019bqv, Crivellin:2019dwb}. $S_1$ coupling as a leptoquark is able to explain $R_{D^{(*)}}$ (see, for example \cite{Sakaki:2013bfa,Kim:2018oih}) while $S_3$ coupling as a leptoquark is able explain  $R_{K^{(*)}}$  \cite{Hiller:2017bzc,Hiller:2014yaa,PhysRevD.98.055003,Kumar:2018era,Dorsner:2017ufx,Angelescu:2018tyl,Gripaios:2014tna}. However, in order to generate neutrino mass in this model, the fermion content of the chosen effective operators forces us to have one of either $S_1$ or $S_3$ coupling as a leptoquark, and the other as a diquark. As we discuss later, the choice that leads to neutrino mass generation at an acceptable loop order has $S_3$ coupling as a leptoquark, and $S_1$ coupling as a diquark, both with flavour dependent couplings. Consequently, our model can only adequately explain the anomalies resolved by the leptoquark $S_{3}$ -- specifically those in $R_{K^{(*)}}$. The scalar $\phi_3$ only couples to other scalars and gauge bosons.

\subsection{The Lagrangian}
\label{sec: the lagrangian}
The general, gauge invariant, renormalisable Lagrangian produced when introducing the three scalars mentioned above can be found in Appendix \ref{App: full lagrangian}. The full Lagrangian has both leptoquark and diquark couplings for $S_1$ and $S_3$, thus explicitly violating baryon number conservation. This is, of course, phenomenologically unacceptable unless the couplings that lead to proton decay are extremely small. In our analysis, we simply impose exact $U(1)_B$ symmetry so that baryon number conservation is exact.\\\\
Two neutrino mass models emerge once baryon number conservation is imposed on the general, gauge invariant, renormalisable Lagrangian, written in full in Appendix \ref{App: full lagrangian}. Two models are produced by the SM particles together with the three exotic scalar fields with baryon number assignments
\begin{table}[H]
\centering 
\begin{tabu}to 1\linewidth{X[l,0.2] X[l] X[l,0.1]}
&\textbf{Model 1}: $B(S_{3})=-1/3$, $B(S_{1})=2/3$, and $B(\phi_{3})=1/3$,~or &\\
&\textbf{Model 2}: $B(S_{3}) = 2/3$, $B(S_{1})= -1/3$, and $B(\phi_{3})= 1/3$.&
\end{tabu}
\end{table}
\noindent
Model 2, in which $S_3$ is a diquark and $S_1$ is a leptoquark, leads to unacceptably small neutrino masses, as will be detailed in Section \ref{Subsec: Model 2 and vanishing neutrino self energies}. Model 1, in which $S_3$ is a leptoquark and $S_1$ is a diquark produces non-vanishing neutrino mass associated with the UV-completion of the dimension-11 operators $O_{25}$, $O_{47}$, and $O_{55}$. Although our model generates all three operators, we now show that, once dressed to become self-energy diagrams, the graph associated with operator $O_{47}$ dominates by several orders of magnitude compared to those associated with $O_{25}$ and $O_{55}$. \\\\
\begin{figure}[t]
\begin{subfigure}[c]{0.5\textwidth}
\centering
\begin{tikzpicture}[thick,scale=1]
	\coordinate[] (a1) at (-1,1.5) {};
		\node[vtx, label=0:$L$] at (-0.2,1) {};
	\coordinate[] (a2) at (-0.5,3) {};
		\node[vtx, label=0:$Q$] at (-0.2,2) {};
	\coordinate[] (b) at (0.5,1.5) {};
	
	\coordinate[] (mid) at (2,1.5) {};
	\coordinate[] (mid2) at (2,3) {};
	\coordinate[] (d) at (3.5,1.5) {};
		\node[vtx, label=0:$u^{c}_{R}$] at (0.7,3.1) {};
		\node[vtx, label=0:$d^{c}_{R}$] at (2.5,3.1) {};
	\coordinate[] (e1) at (5,1.5) {};
		\node[vtx, label=180:$L$] at (4.5,1) {};
	\coordinate[] (e2) at (4.5,3) {};
		\node[vtx, label=180:$Q$] at (4.2,2) {};
	\coordinate[vtx, label=180:$\braket{H}$] (H) at (2.45,0) {};
	\coordinate[vtx, label=0:$\braket{H}$] (H2) at (3.05,0) {};
	\coordinate[](quarter) at (2.75,1.5){};
	
	\node[vtx, label=0:$\textcolor{blue}{{\tiny f}}$] at (0.4,1.75) {};
	\node[vtx, label=0:$\textcolor{blue}{{\tiny g}}$] at (1.9,2.8) {};
	\node[vtx, label=0:$\textcolor{blue}{{\tiny h}}$] at (3,1.75) {};
	\node[vtx, label=0:$\textcolor{blue}{{\tiny \lambda}}$] at (2.75,1.25) {};
	\node[vtx, label=0:$\textcolor{blue}{{\tiny \mu}}$] at (1.7,1.2) {};
	
	\graph[use existing nodes]{
		a1--[fermion]b;
		b--[scalar]mid;
		mid--[scalar]mid2;
		mid--[scalar]d;
		H--[scalar]quarter;
		H2--[scalar]quarter;
		d--[antifermion]e1;
	};
	
   \draw[postaction={decorate},thick,decoration = {markings,
    mark=at position 0.5 with {
        \draw (-2pt,-2pt) -- (2pt,2pt);
        \draw (2pt,-2pt) -- (-2pt,2pt);},
     mark=at position 0.25 with {\arrowreversed {latex}},
     mark=at position 0.75 with {\arrow {latex}}
      }] (3.5,1.5) arc (0:90:1.5);    
   \draw[postaction={decorate},thick,decoration = {markings,
    mark=at position 0.5 with {
        \draw (-2pt,-2pt) -- (2pt,2pt);
        \draw (2pt,-2pt) -- (-2pt,2pt);},
     mark=at position 0.25 with {\arrowreversed {latex}},
     mark=at position 0.75 with {\arrow {latex}}
      }] (0.5,1.5) arc (180:90:1.5);
\end{tikzpicture}
\\
\caption{}
\label{Fig: generic diagram a}
\end{subfigure}
\begin{subfigure}[c]{0.5\textwidth}
\centering
\begin{tikzpicture}[thick,scale=1]
	\coordinate[] (a1) at (-1,1.5) {};
		\node[vtx, label=0:$L$] at (-0.5,1.2) {};
	\coordinate[] (a2) at (-0.5,3) {};
		\node[vtx, label=0:$$] at (0,2.5) {};
		\node[vtx, label=0:$Q$] at (-0.2,2) {};
	\coordinate[] (b) at (0.5,1.5) {};
		
	\coordinate[] (mid) at (2,1.5) {};
	\coordinate[] (mid2) at (2,3) {};
	\coordinate[] (d) at (3.5,1.5) {};
		
	\coordinate[] (e1) at (5,1.5) {};
		\node[vtx, label=180:$L$] at (4.7,1.2) {};
	\coordinate[] (e2) at (4.5,3) {};
		\node[vtx, label=180:$Q$] at (4.2,2) {};
	\coordinate[vtx, label=180:$\braket{H}$] (H) at (2.45,0) {};
	\coordinate[vtx, label=0:$\braket{H}$] (H2) at (3.05,0) {};
	\coordinate[](quarter) at (2.75,1.5){};
	\graph[use existing nodes]{
		a1--[fermion]b;
		b--[scalar]mid;
		mid--[scalar]mid2;
		mid--[scalar]d;
		H--[scalar]quarter;
		H2--[scalar]quarter;
		d--[antifermion]e1;
	};
	
    \draw[postaction={decorate},thick,decoration = {markings,
    mark=at position 0.5 with {\arrow {latex}}}] (2,3) arc (90:0:1.5);    
   \draw[postaction={decorate},thick,decoration = {markings,
    mark=at position 0.5 with {\arrow {latex}}}] (2,3) arc (90:180:1.5);
\end{tikzpicture}
\\
\caption{}
\label{Fig: generic diagram b}
\end{subfigure}\\
\hspace{0.5cm}
\begin{subfigure}[c]{0.99\textwidth}
\centering
\begin{tikzpicture}[thick,scale=1]
	\coordinate[] (a1) at (-1,1.5) {};
		\node[vtx, label=0:$L$] at (-0.4,1.2) {};
	\coordinate[] (a2) at (-0.5,3) {};
		\node[vtx, label=0:$Q$] at (0,2.5) {};
	\coordinate[] (b) at (0.5,1.5) {};
	
	\coordinate[] (mid) at (2,1.5) {};
	\coordinate[] (mid2) at (2,3) {};
	\coordinate[] (d) at (3.5,1.5) {};
		\node[vtx, label=180:$u_{R}$] at (4,2.2) {};
	\coordinate[] (e1) at (5,1.5) {};
		\node[vtx, label=180:$\nu_{L}$] at (5,1.2) {};
	\coordinate[] (e2) at (4.5,3) {};
		\node[vtx, label=180:$e_{R}$] at (4,1.2) {};
	\coordinate[vtx, label=180:$\braket{H}$] (H) at (2.45,0) {};
	\coordinate[vtx, label=0:$\braket{H}$] (H2) at (3.05,0) {};
	\coordinate[](quarter) at (2.75,1.5){};
	\graph[use existing nodes]{
		a1--[fermion]b;
		b--[scalar]mid;
		mid--[scalar]mid2;
		mid--[scalar]d;
		H--[scalar]quarter;
		H2--[scalar]quarter;
		d--[antifermion]e1;
	};
	
   \draw[postaction={decorate},thick,decoration = {markings,
    mark=at position 0.75 with {\arrowreversed {latex}}
    ,
    mark=at position 0.25 with {\arrow {latex}},
     mark=at position 0.5 with {
        \draw (-2pt,-2pt) -- (2pt,2pt);
        \draw (2pt,-2pt) -- (-2pt,2pt);}
    }] (3.5,1.5) arc (0:90:1.5);    
   \draw[postaction={decorate},thick,decoration = {markings,
    mark=at position 0.35 with {\arrowreversed {latex}}}] (0.5,1.5) arc (180:90:1.5);
	\draw[black, thick, decorate, decoration={complete sines, number of sines=7,amplitude=6pt}](4.5,1.5) arc (0:105 : 1.3);
	\draw[postaction={decorate},thick,decoration = {markings, mark= at position 0.35 with {
        \draw (-2pt,-2pt) -- (2pt,2pt);
        \draw (2pt,-2pt) -- (-2pt,2pt);},
	mark=at position 0.25 with {\arrow {latex}}
    }](d)--(e1);
    \end{tikzpicture}
\\
\caption{}
\label{Fig: generic diagram c}
\end{subfigure}
\caption{Generic neutrino mass diagrams from the completions of operators (a) $O_{25}$, (b) $O_{47}$ and (c) $O_{55}$.}
\label{Fig: generic diagram}
\end{figure}
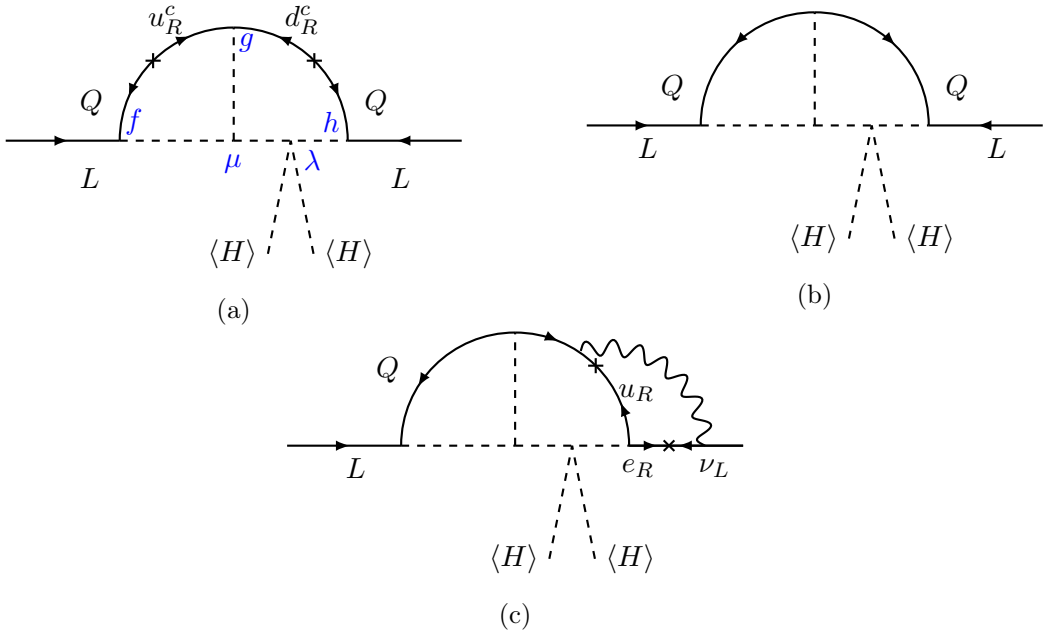
\noindent\textbf{Scale of new physics from effective operators}---
The generic type of neutrino mass diagrams generated from the UV-completion of each operator can be found in Figure \ref{Fig: generic diagram}. Let us start by analysing $O_{25}$, whose neutrino mass diagram is depicted in Figure \ref{Fig: generic diagram a}.\footnote{When the scale of new physics is less than or equal to $2$ TeV, neutrino masses generated through the UV-completion of operator $O_{25}$ will also include an extra contribution from a three-loop diagram obtained by closing the neutral Higgs bosons into a loop.} Due to the chirality structure of $O_{25}$, its UV-completions include two mass insertions. Consequently, contributions to the neutrino mass originating from this operator will depend on the mass of the up and down-type quarks in the loop. The neutrino mass will be 
{\begin{equation} \begin{split}
m^{O_{25}}_{\nu} \propto \frac{fgh \lambda}{(16\pi^{2})^{2}}\frac{Y^{u}Y^{d} v^{2}}{\Lambda}\Big(\ \frac{1}{16\pi^{2}} + \frac{v^{2}}{\Lambda^{2}} \Big),
\label{eqn: O_{25} neutrino mass generic}
\end{split} \end{equation} }%
where $f,~g,~h,$ and $\lambda$ are coupling constants, $v$ is the Higgs vev, $Y^{u}$ and $Y^{d}$ are the Higgs Yukawa couplings for the up-type and down-type quarks, and $\Lambda$ is the scale of new physics. The cubic scalar coupling, $\mu$, we assume to be of the scale of new physics and thus it cancels with a factor of $\Lambda$ in the denominator.\\\\
Operator $O_{47}$ is an interesting operator in that it produces neutrino mass contributions that are not constrained by the masses of SM particles. The neutrino mass generated from tree level completions of operator $O_{47}$ will look like 
{\begin{equation} \begin{split}
m^{O_{47}}_{\nu} \propto \frac{fgh\lambda}{(16\pi^{2})^{2}}\frac{v^{2}}{\Lambda}.
\label{eqn: O_{47} neutrino mass generic}
\end{split} \end{equation} }%
Finally, operator $O_{55}$ produces neutrino mass  
\begin{equation} \begin{split}
m^{O_{55}}_{\nu}\propto \frac{fgh \lambda}{(16\pi^{2})^{3}} \frac{ Y^{u}Y^{e}g_{W}^{2}v^{2}}{\Lambda},
\label{eqn: O_{55} neutrino mass generic}
\end{split} \end{equation} 
where $g_{W}$ is the weak coupling, and $Y^{e}$ is the SM charged lepton Yukawa coupling. As $O_{55}$ only contains one lepton field, a $W$ boson is required to obtain a second neutrino, generating the self-energy diagram in Figure \ref{Fig: generic diagram c}. It is worth noting here that the diquark coupling, $g$ in the expansion of operators $O_{47}$ and $O_{55}$ refers to the left-handed coupling, $z_1^{LL}$, whereas the coupling $g$ in the expansion of operator $O_{25}$ is the right-handed diquark coupling, $z_1^{RR}$. There is no a priori reason for $z_1^{RR}$ to be larger than $z_1^{LL}$, thus we assume them to be the same in this order-of-magnitude analysis. In fact if we set $z_1^{RR}$ to zero, operator $O_{25}$ generates vanishing neutrino mass. \\\\
Comparing the orders of magnitude of neutrino masses generated through the completions of the three operators, given in Equations \ref{eqn: O_{25} neutrino mass generic}, \ref{eqn: O_{47} neutrino mass generic}, and \ref{eqn: O_{55} neutrino mass generic}, we find that $O_{25}$ and $O_{55}$ are very suppressed compared to $O_{47}$:
\begin{equation} \begin{split}
m^{O_{25}}_{\nu} \leq Y^{b}\Big(\frac{1}{16\pi^{2}} + \frac{v^{2}}{\Lambda^{2}}\Big)m^{O_{47}}_{\nu}~~~\mathrm{and}~~~m^{O_{55}}_{\nu} \leq \frac{Y^{\tau}g_{W}^{2}}{16\pi^{2}}m^{O_{47}}_{\nu}.
\label{eqn: comparison of operators}
\end{split} \end{equation} 
That is, the contribution to neutrino masses coming from the insertions of $O_{25}$ is suppressed by at least $10^{-3}$, and those from $O_{55}$ by at least\footnote{In Equation \ref{eqn: comparison of operators}, $Y^{b}$ and $Y^{\tau}$ denote the Yukawa couplings for the bottom quark and the tau lepton respectively. In using the largest Yukawa coupling constants when comparing $m^{O_{47}}_{\nu}$ to $m^{O_{25}}_{\nu}$ and $m^{O_{55}}_{\nu}$, we ensure that we have  a lower bound on the suppression.} $10^{-5}$ compared to $O_{47}$. With this knowledge, we can sensibly approximate the contribution to neutrino mass generated by the UV--completions of $O_{47}$ to be dominant, ignoring the contributions associated with the other two operators. \\\\
From Equation \ref{eqn: O_{47} neutrino mass generic} we expect that for indicative couplings $f=g=h=\lambda=1$, the scale of NP is $\Lambda \sim 10^{7}$ TeV, for couplings $f=g=h=\lambda=0.1$, the scale of NP is $\Lambda \sim 10^{3}$ TeV, and for couplings $f=g=h=\lambda=0.01$, the scale of NP is $\Lambda \sim 10^{-1}$ TeV. Thus, to an order of magnitude precision, we can expect that our neutrino mass model is viable with reasonably-valued exotic couplings. \\\\
This analysis leads us to conclude that, to a good approximation we can, and do, choose to consider only the neutrino mass diagrams associated with the direct closure of operator $O_{47}$ into neutrino self-energies. These are two-loop diagrams, with only exotic scalars and left-handed SM fermions running through the loops, as depicted in Figure \ref{Fig: generic diagram b}. \\\\
Even after imposing $B$-conservation with the Model 1 assignments, the Lagrangian retains a large number of parameters. In order to make exploring that parameter space tractable, we also make the simplifying assumption that all couplings that play no role in neutrino mass generation are zero. \\\\
The general Lagrangian of equation \ref{eqn: full lagrangian} then simplifies to the following:
{\begin{equation} \begin{split}
\mathcal{L} &= \mathcal{L}_{\text{SM}} + \mathcal{L}_{\text{NP}} \\
 &=\mathcal{L}_{\text{SM}} + \mathcal{L}_{\text{gauge}-S} + \mathcal{L}_{S_3-F} +\mathcal{L}_{S_1-F}+\mathcal{L}_{\rm 4SB} +\mathcal{L}_{\rm 3SB}+ \mathcal{L}_{\rm 2SB},
 \label{eqn: Model 1 simplified lagrangian}
\end{split} \end{equation} }%
where the coupling between the gauge bosons and exotic scalars, $\mathcal{L}_{\text{gauge}-S}$ , is defined in equation \ref{eqn: gauge lagrangian for exotics}, S-F represents couplings between the exotic scalar and SM fermions and 4SB, 3SB and 2SB represent scalar only interactions between four, three and two scalar bosons, respectively. In the fermion sector, we define
\begin{subequations}
 \begin{equation}
\begin{split}
\mathcal{L}_{S_3-F} = &+y^{LL}_{3ij}\overline{Q^{C}}^{~i,a}\epsilon_{ab}(\tau^k S^k_3)^{b}_c L^{j,c} + \mathrm{h.c.}\\
 =&-\frac{(y_{3}^{LL})_{ij}}{\sqrt{2}}\overline{d_{L}^{C}}^{~i}S_{3}^{1/3}\nu^{j} - y_{3 ij}^{LL}\overline{d_{L}^{C}}^{~i}S_{3}^{4/3}e_{L}^{j} \\
 &+ (V^{*}_{\text{CKM}}y_{3}^{LL})_{ij}\overline{u_{L}^{C}}^{~i}S_{3}^{-2/3}\nu^{j} - \frac{(V^{*}_{\text{CKM}}y_{3}^{LL})_{ij}}{\sqrt{2}}\overline{u_{L}^{C}}^{~i}S_{3}^{1/3}e_{L}^{j}+\mathrm{h.c.}
 \end{split} 
 \end{equation}\\
\begin{equation}
 \begin{split}
\mathcal{L}_{S_1-F} 
= &\phantom{-}z^{LL}_{1 ij} \overline{Q^{C}}^{~i,a}S^*_1 \epsilon_{ab}Q^{j,b} +\mathrm{h.c.}\\
=& - (2z_{1}^{LL}V_{\text{CKM}}^{\dagger})_{ij}\overline{d_{L}^{C}}^{~i}S_{1}^{*}u_{L}^{j} + \mathrm{h.c.},
\label{eqn: fermion sector S1 our model}
\end{split} 
\end{equation}
\end{subequations}
and in the scalar sector, we have
{\begin{equation} \begin{split}
\mathcal{L}_{4SB} &= \lambda_{\phi_3 S_3 H}[\phi_3^\dagger  S_3^\dagger]_3 [H H]_3 + \mathrm{h.c.},
\label{eqn: 4SB term} \\
\mathcal{L}_{3SB} &= m_{S_3 S_1 \phi_3}[S_3^\dagger  \phi_3 ]_{1}S_1^\dagger+\mathrm{h.c.},~\mathrm{and}\\
\mathcal{L}_{2SB} &=- \mu^2_{S_3}S_3^\dagger S_3 -\mu^2_{S_1}S_1^\dagger S_1 -\mu^2_{\phi_3}\phi_3^\dagger \phi_3.
\end{split} \end{equation}}%
There are no Yukawa interactions allowed by the SM gauge symmetries between the scalar $\phi_3$ and SM fermions.
The $\tau^k$, $k=1,2,3$, are the Pauli matrices; $i,j = 1, 2, 3$ are generation indices; $a,b = 1,2$ are $SU(2)$ flavour indices; $\epsilon^{ab} = (i\tau^2)^{ab}$; and $S_3^k$ are components of $S_3$ in $SU(2)$ space. Colour indices are not explicitly shown. The diquark coupling to $S_{1}$, $z_{1}^{LL}$ is symmetric due to a combination of the antisymmetry of $SU(2)$ structure and the colour structure of the fermion bilinear. We take the leptoquark coupling, $y_3^{LL}$, to be real. In the expansion of the $SU(2)$ structure of the Lagrangian, we have also rotated into the mass eigenbasis of the quarks, using the convention that $u_{L}^{i} \rightarrow (V_{\text{CKM}}^{\dagger})_{ij} u_{L}^{j}$ and $d_{L}^{i} \rightarrow d_{L}^{i}$. Ultimately, this simply amounts to a definition of the relevant coupling matrices: $y_3^{LL}$ and $z_1^{LL}$. In the scalar sector, the notation $[\ldots]_{i}$ indicates that the scalars enclosed couple to form an $SU(2)$ singlet for $i=1$ or triplet for $i=3$.
 
\subsection{Scalar Boson mixing}
After electroweak symmetry breaking, Equation \ref{eqn: 4SB term} produces mass mixing between like-charge components of $S_3$ and $\phi_3$, generating the mass matrices
\begin{equation} \begin{split}
&\begin{pmatrix}
S_{3}^{1/3*} & \phi_{3}^{-1/3}
\end{pmatrix}
\begin{pmatrix}
\mu_{S_{3}}^{2} & -\lambda_{\phi_{3}S_{3}H} \frac{v^{2}}{\sqrt{2}}\\
 -\lambda^{*}_{\phi_{3}S_{3}H} \frac{v^{2}}{\sqrt{2}} & \mu_{\phi_{3}}^{2} 
\end{pmatrix}
\begin{pmatrix}
S_{3}^{1/3}\\
\phi_{3}^{-1/3*}
\end{pmatrix}~\mathrm{and} \\
&\begin{pmatrix}
S_{3}^{-2/3*} & \phi_{3}^{2/3}
\end{pmatrix}
\begin{pmatrix}
 \mu_{S_{3}}^{2} & \lambda_{\phi_{3}S_{3}H} \frac{v^{2}}{\sqrt{2}}\\
  \lambda^{*}_{\phi_{3}S_{3}H} \frac{v^{2}}{\sqrt{2}} & \mu_{\phi_{3}}^{2} 
\end{pmatrix}
\begin{pmatrix}
 S_{3}^{-2/3}\\
\phi_{3}^{2/3*} 
\end{pmatrix}.
\label{eqn: mass matrix for scalars}
\end{split} \end{equation}
We define the mass eigenstate fields $r_{1,2,3,4}$ through
{\begin{equation} \begin{split}
\begin{pmatrix}
S_{3}^{1/3}\\
\phi_{3}^{-1/3*}
\end{pmatrix}
=
\begin{pmatrix}
\mathrm{cos~\theta_{12}}& \mathrm{sin~\theta_{12}}\\
\mathrm{-sin~\theta_{12}}&\mathrm{cos~\theta_{12}}
\end{pmatrix}
\begin{pmatrix}
r_{1}\\
r_{2}
\end{pmatrix}~~~~\mathrm{and}~~~~\begin{pmatrix}
S_{3}^{-2/3}\\
\phi_{3}^{2/3*}
\end{pmatrix}
=
\begin{pmatrix}
\mathrm{cos~\theta_{34}}& \mathrm{sin~\theta_{34}}\\
\mathrm{-sin~\theta_{34}}&\mathrm{cos~\theta_{34}}
\end{pmatrix}
\begin{pmatrix}
r_{3}\\
r_{4}
\end{pmatrix}.
\end{split} \end{equation} }%
The mixing angles $\mathrm{\theta_{12}}$ and $\mathrm{\theta_{34}}$ are related to the squared mass parameters $\mu^{2}_{S_{3}}$, $\mu^{2}_{\phi_{3}}$ and off-diagonal parameters in the mass matrix through
{\begin{equation} \begin{split}
\mathrm{tan~2\theta_{12}} = \frac{\sqrt{2}\lambda_{\phi_{3}S_{3}H}v^{2}}{\mu^{2}_{S_{3}}-\mu^{2}_{\phi_{3}}} = - \mathrm{tan~2\theta_{34}},
\label{eqn: tan relations to parameters}
\end{split} \end{equation} }%
so that $\mathrm{\theta_{12}} = -\mathrm{\theta_{34}}$. \\\\
There are seven BSM physical scalar states in our theory (one diquark, five leptoquarks and one coloured, electrically-charged scalar that does not couple to SM fermions) with squared masses given by 
{\begin{equation} \begin{split}
&m^{2}_{S_{1}}= \mu^{2}_{S_{1}},\quad m^{2}_{S_{3}^{4/3}} = \mu^{2}_{S_{3}}, \quad m^{2}_{\phi_{3}^{5/3}} = \mu^{2}_{\phi_{3}},\\
&m^{2}_{r_{1},r_{2}} = m^{2}_{r_{3},r_{4}} = \frac{1}{\sqrt{2}}\left[\mu^{2}_{\phi_{3}} + \mu^{2}_{S_{3}} \pm (\mu^{2}_{S_{3}}-\mu^{2}_{\phi_{3}})\sqrt{1 + 2\left(\frac{\lambda_{\phi_{3}S_{3}H}v^{2}}{\mu^{2}_{S_{3}} - \mu^{2}_{\phi_{3}}}\right)^{2}} \, \right].
\label{eqn: masses of exotic scalars}
\end{split} \end{equation} }%
The squared masses of all physical particles are required to be positive, placing a bound on $\lambda_{\phi_{3}S_{3}H}$ such that 
{\begin{equation} \begin{split}
|\lambda_{\phi_{3}S_{3}H}|\leq \frac{\sqrt{2}\mu_{S_{3}}\mu_{\phi_{3}}}{v^{2}}. 
\end{split} \end{equation} }%
Thus, the parameters $\mu_{S_{3}}$, $\mu_{\phi_{3}}$ and $\lambda_{S_{3}\phi_{3}H}$ determine the masses of the five leptoquarks and one charged scalar, and the mixing angles. We can also derive the relationship between the mixing angle and the squared mass difference of the leptoquarks involved in mixing from Equations \ref{eqn: tan relations to parameters} and \ref{eqn: masses of exotic scalars}, 
\begin{equation} \begin{split}
\mathrm{sin}~2\theta = \frac{\sqrt{2}\lambda_{\phi_{3}S_{3}H}v^{2}}{m^{2}_{r_{1,3}}-m^{2}_{r_{2,4}}} 
\end{split} \end{equation} 
where $ \mathrm{\theta} \equiv \mathrm{\theta}_{12} = -\mathrm{\theta}_{34}$.

\subsection{Neutrino mass generation}
\label{sec: neutrino mass generation}
\begin{figure}[t]
\begin{subfigure}[t]{0.45\textwidth}
\centering
\begin{tikzpicture}[thick,scale=1]
	\coordinate[] (a1) at (-1,1.5) {};
	
	\coordinate[] (a2) at (-2,0.5) {};
		\node[vtx, label=180:$L$] at (-2,0.5) {};
	\coordinate[] (a3) at (-2,2.5) {};
		\node[vtx, label=180:$Q$] at (-2,2.5) {};
	\coordinate[] (mid1) at (0.5,1.5) {};
	
	\coordinate[] (mid2) at (0.5,2.5) {};
	\coordinate[] (c1) at (-0.5,3.5) {};
	\node[vtx, label=180:$\overline{Q}$] at (-0.5,3.5) {};
	\coordinate[] (c2) at (1.5,3.5) {};
	\node[vtx, label=0:$Q^{C}$] at (1.5,3.5) {};
	
	\coordinate[] (b1) at (2,1.5) {};
	\coordinate[] (b2) at (3,0.5) {};
		\node[vtx, label=0:$L$] at (3,0.5) {};
	\coordinate[] (b3) at (3,2.5) {};
		\node[vtx, label=0:$Q$] at (3,2.5) {};
	\coordinate[] (midH) at (1.25,1.5) {};
	\coordinate[] (H1) at (0.9,-0.2) {};
		\node[vtx, label=180:$\braket{H}$] at (0.9,-0.2) {};
	\coordinate[] (H2) at (1.6,-0.2) {};
		\node[vtx, label=0:$\braket{H}$] at (1.6,-0.2) {};
		\node[vtx, label=0:$S_{1}$] at (0.5,2) {};
		\node[vtx, label=0:$S_{3}$] at (-0.5,1.2) {};
		\node[vtx, label=0:$S_{3}$] at (1.5,1.2) {};
		\node[vtx, label=0:$\phi_{3}$] at (0.5,1.2) {};
		
		\graph[use existing nodes]{
		a1--[antifermion]a2;
		a1--[antifermion]a3;
		a1--[scalar]mid1;
		mid1--[scalar]mid2;
		mid1--[scalar]b1;
		b1--[antifermion]b2;
		b1--[antifermion]b3;
		mid2--[fermion]c1;
		mid2--[fermion]c2;
		midH--[scalar]H1;
		midH--[scalar]H2;
	};

\end{tikzpicture}
%
%
%
%
\caption{The UV-completion of operator $O_{47}$ with the introduction of three exotic scalars; $S_{1}$, $S_{3}$, $\phi_{3}$.}
\label{fig: neutrino mass diagram for our model a}
\end{subfigure}
\hfill
\begin{subfigure}[t]{0.45\textwidth}
\centering
\begin{tikzpicture}[thick,scale=1.1]
	\coordinate[] (a1) at (-1,1.5) {};
		\node[vtx, label=0:$L$] at (-0.5,1.2) {};
	\coordinate[] (a2) at (-0.5,3) {};
		\node[vtx, label=0:$Q$] at (0,2.5) {};
	\coordinate[] (b) at (0.5,1.5) {};
		
	\coordinate[] (mid) at (2,1.5) {};
	\coordinate[] (mid2) at (2,3) {};
	\coordinate[] (d) at (3.5,1.5) {};
		\node[vtx, label=180:$S_{1}$] at (2.65,2.3) {};
	\coordinate[] (e1) at (5,1.5) {};
		\node[vtx, label=180:$L$] at (4.7,1.2) {};
	\coordinate[] (e2) at (4.5,3) {};
		\node[vtx, label=180:$Q$] at (4,2.5) {};
	\coordinate[vtx, label=180:$\braket{H}$] (H) at (2.45,0) {};
	\coordinate[vtx, label=0:$\braket{H}$] (H2) at (3.05,0) {};
	\coordinate[](quarter) at (2.75,1.5){};
		\node[vtx, label=180:$S_{3}$] at (1.6,1.2) {};
		\node[vtx, label=180:$\phi_{3}$] at (2.7,1.2) {};
		\node[vtx, label=180:$S_{3}$] at (3.5,1.2) {};

	\graph[use existing nodes]{
		a1--[fermion]b;
		b--[scalar]mid;
		mid--[scalar]mid2;
		mid--[scalar]d;
		H--[scalar]quarter;
		H2--[scalar]quarter;
		d--[antifermion]e1;
	};
	
    \draw[postaction={decorate},thick,decoration = {markings,
    mark=at position 0.5 with {\arrow {latex}}}] (2,3) arc (90:0:1.5);    
   \draw[postaction={decorate},thick,decoration = {markings,
    mark=at position 0.5 with {\arrow {latex}}}] (2,3) arc (90:180:1.5);
\end{tikzpicture}
\caption{Closing the loops by joining the quarks leads to a neutrino mass diagram.}
\label{fig: neutrino mass diagram for our model b}
\end{subfigure}\\
\begin{subfigure}[b]{0.99\textwidth}
\centering
\begin{tikzpicture}[thick,scale=1.1]
	\coordinate[] (a1) at (-1,1.5) {};
		\node[vtx, label=0:$\nu_{i}$] at (-0.5,1.2) {};
	\coordinate[] (a2) at (-0.5,3) {};
		\node[vtx, label=0:$u_{L}$] at (0,2.5) {};
	\coordinate[] (b) at (0.5,1.5) {};
		
	\coordinate[] (mid) at (2,1.5) {};
	\coordinate[] (mid2) at (2,3) {};
		\node[vtx, label=180:$S_{1}$] at (2.65,2.3) {};
	\coordinate[] (d) at (3.5,1.5) {};
		\node[vtx, label=180:$r_{3,4}$] at (1.6,1.2) {};
		\node[vtx, label=180:$r_{1,2}$] at (3.2,1.2) {};

	\coordinate[] (e1) at (5,1.5) {};
		\node[vtx, label=180:$\nu_{j}$] at (4.7,1.2) {};
	\coordinate[] (e2) at (4.5,3) {};
		\node[vtx, label=180:$d_{L}$] at (4,2.5) {};
		
	\coordinate[](quarter) at (2.75,1.5){};
	\graph[use existing nodes]{
		a1--[fermion]b;
		b--[scalar]mid;
		mid--[scalar]mid2;
		mid--[scalar]d;
		d--[antifermion]e1;
	};
	
    \draw[postaction={decorate},thick,decoration = {markings,
    mark=at position 0.5 with {\arrow {latex}}}] (2,3) arc (90:0:1.5);    
   \draw[postaction={decorate},thick,decoration = {markings,
    mark=at position 0.5 with {\arrow {latex}}}] (2,3) arc (90:180:1.5);
\end{tikzpicture}
\caption{ The neutrino self-energy, after electroweak symmetry breaking and rotating into the mass basis of the exotic scalars.}
\label{fig: neutrino mass diagram for our model c}
\end{subfigure}
\caption{The UV--completion of operator $O_{47}$ by the exotic scalars $S_{3}$ (leptoquark), $S_{1}$ (diquark) and $\phi_{3}$ and its closure forming the neutrino self-energy Feynman diagrams for our model.}
\label{fig: neutrino mass diagram for our model}
\end{figure}
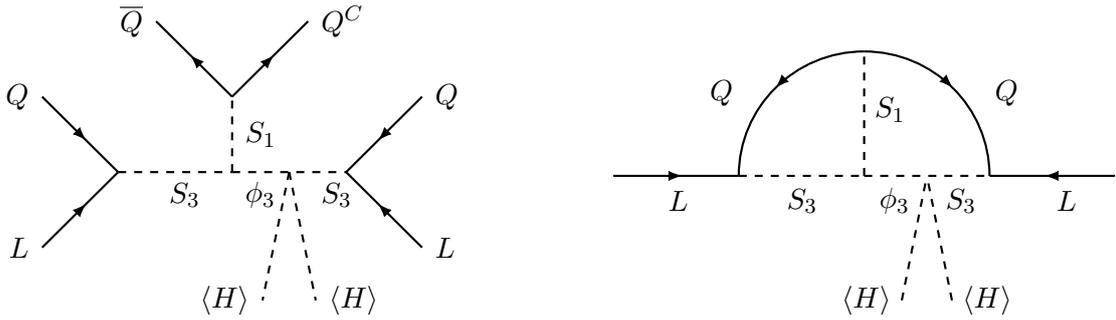
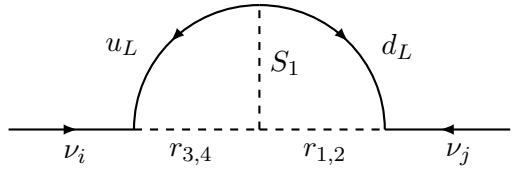
\begin{figure}[t]
    \centering
        \includegraphics[width=\linewidth]{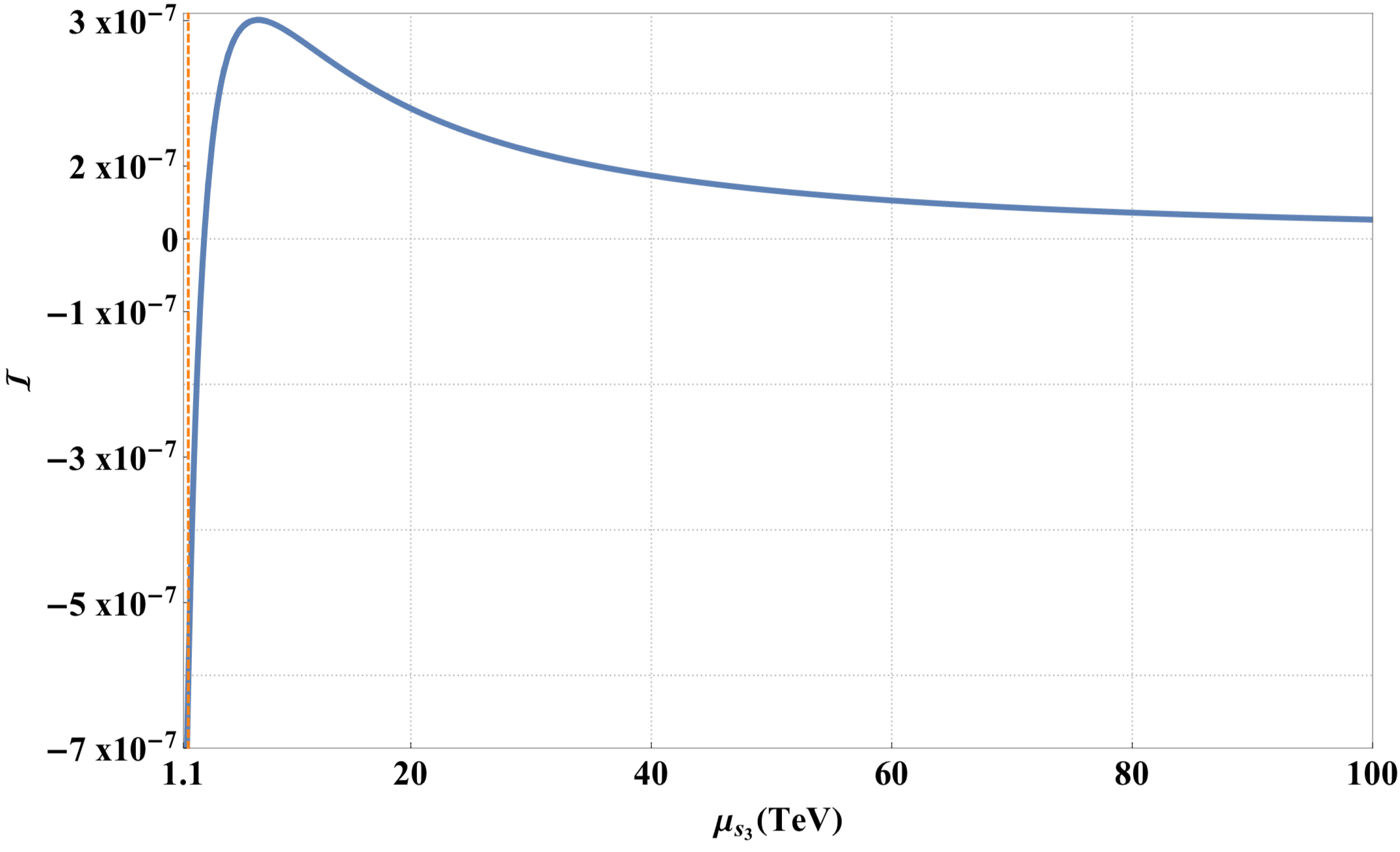}
        \caption{Plot showing the behaviour of the integral, $\mathcal{I}$, (see Equation \ref{eq: integral in limit of zero quark mass for Model 1}), for leptoquark mass parameter $\mu_{S_3}$ ranging from $1.1-100$ TeV. This corresponds to a physical mass range of $m_{r_1}=m_{r_3} = 1.3- 118.9$ TeV, while $m_{r_2}=m_{r_4}$ remains constant for this range. The other exotic mass parameters and the quartic and cubic scalar coupling values have been fixed: $\mu_{S_1} = 7.5$ TeV, $\mu_{\phi_3} = 1.5$ TeV, $\lambda_{S_3\phi_3 H}$ = 1 and $m_{S_1S_3\phi_3} = 1.5$ TeV. The dashed orange line corresponds to a singularity which exists when $\mu_{S_3}=\mu_{S_1}$.}
        \label{Fig: integral vs mS3}
\end{figure}

As discussed in Section \ref{sec: the lagrangian}, the dominant contributions to the neutrino mass for Model 1 are two-loop neutrino self-energy graphs, with exotic scalars and left-handed SM fermions running through the loops, generated by the completion of operator $O_{47}$ \footnote{For Model 1 the $SU(2)$ structure of $O_{47}$ is $O_{47}^j= \overline{L^C}^i L^j \overline{Q^C}^k Q^l \overline{Q}_p Q^C_q H^m H^n \epsilon_{im} \epsilon_{jn}\epsilon_{kl}\epsilon^{pq}$.}.
The UV-completion of $O_{47}$ by the diquark $S_{1}$, the leptoquark $S_{3}$, and the scalar $\phi_{3}$, is depicted in the tree-level diagram of Figure \ref{fig: neutrino mass diagram for our model a}. Joining the quark $Q$ to $\overline{Q}$ lines, and the second $Q$ to $Q^{C},$ gives two-loop self-energy diagrams, which generate the neutrino mass matrix. There is no mass insertion necessary in the quark lines for a chirality flip, thus the neutrino mass matrix generated by this model is not proportional to the mass of any SM fermion; this interesting feature characterises this model. In terms of the physical mass eigenstates, $S_{1}$, $r_{1}$, $r_{2}$, $r_{3}$, and $r_{4}$, there are eight diagrams; half of them are obtained by reversing the flow of charge arrows in both loops of Figure \ref{fig: neutrino mass diagram for our model c}. Individually, each diagram is divergent. However, due to the absence of a bare neutrino mass in the Lagrangian, the divergences are guaranteed to cancel.\\\\
The neutrino mass matrix is obtained from the flavour sum of the self-energy diagrams with the freedom to choose the external momentum to be zero:
{\begin{equation} \begin{split}
i\Sigma_{ij} = 9\sqrt{2}m_{S_{1}S_{3}\phi_{3}}\left[(V_{\text{CKM}}^{*}y_{3}^{LL})^{T}_{ik}(\mathcal{I}z_{1}^{LL}V_{\text{CKM}}^{\dagger})^{\dagger}_{kl}y_{3lj}^{LL} \right] + (i \leftrightarrow j).
\end{split} \end{equation} }%
The PMNS matrix can then be used to obtain the physical masses of the neutrinos, the factor of $9$ is a QCD colour factor, and the loop integral, $\mathcal{I}_{kl}$, is 
 \begin{equation} \begin{split}
\mathcal{I}_{kl} = &-\int \frac{d^{4}k_{1}}{(2\pi)^{4}} \int \frac{d^{4}k_{2}}{(2\pi)^{4}} \frac{2i \slashed k_{1}\slashed k_{2}}{(k_{1}^{2}-m_{u_{k}}^{2})(k_{2}^{2}-m_{d_{l}}^{2})}\frac{\mathrm{cos~\theta}~\mathrm{sin~\theta} }{(k_{1}-k_{2})^{2}-m_{S_{1}}^{2}}\\&\Bigg( \frac{\mathrm{cos^{2}~\theta}}{(k_{1}^{2} - m_{r_{3}}^{2})(k_{2}^{2}-m_{r_{1}^{2}})}- \frac{\mathrm{cos^{2}~\theta}}{(k_{1}^{2} - m_{r_{3}}^{2})(k_{2}^{2}-m_{r_{2}^{2}})} \\
&+\frac{\mathrm{sin^{2}~\theta}}{(k_{1}^{2} - m_{r_{4}}^{2})(k_{2}^{2}-m_{r_{1}^{2}})}-\frac{\mathrm{sin^{2}~\theta}}{(k_{1}^{2} - m_{r_{4}}^{2})(k_{2}^{2}-m_{r_{2}^{2}})}\Bigg).
\label{eq: integral unsimplified}
\end{split} \end{equation} 
The tensor structure of the numerator arises from the chiral projection operators at the vertices, and the lack of proportionality to the SM fermion masses is a good sanity check when cross-referenced with the lack of the relevant mass insertions in the self-energy diagram in Figure \ref{fig: neutrino mass diagram for our model}. Although $m_{r_1} = m_{r_3}$ and $m_{r_2} = m_{r_4}$ at our level of approximation, we denote these individually so that the correspondence between terms in the loop integral and diagrams is manifest.\\\\
When evaluating the integral it is convenient to work in terms of the dimensionless parameters 
\begin{equation} \begin{split}
s_{k} = \frac{m^{2}_{u_{k}}}{m_{S_{1}}^{2}},~~~~s'_{l}= \frac{m^{2}_{d_{l}}}{m^{2}_{S_{1}}},~~ \mathrm{and} ~~t_{\alpha}=\frac{m^{2}_{r_{\alpha}}}{m^{2}_{S_{1}}}.
\end{split} \end{equation}
The numerator of Equation \ref{eq: integral unsimplified} can be rewritten as: $\slashed k_{1}\slashed k_{2} = k_\mu k_\nu \gamma^\mu \gamma^\nu =  k_\mu k_\nu(\frac{1}{2}\{\gamma^\mu, \gamma^\nu \} + \frac{1}{2} [\gamma^\mu, \gamma^\nu])=  k_\mu k_\nu \eta^{\mu \nu} I_4 = k_1\cdot k_2~I_4 $, where the antisymmetric commutator term vanishes due to the fact the integral is $\mu,\nu$ symmetric, and $I_4$ is the $4\times4$ identity matrix in Lorentz-spinor space. Factoring out $m_{S_{1}}^{2}$ and rescaling the momenta to dimensionless quantities allows us to write the integral as
\begin{equation} \begin{split}
\mathcal{I}_{kl} = &-\frac{2i}{(2\pi)^{8}}\int d^{4}k_{1} \int d^{4}k_{2} \frac{ k_{1}\cdot k_{2}}{(k_{1}^{2}-s_{k})(k_{2}^{2}-s'_{l})}\frac{\mathrm{cos\theta}~\mathrm{sin\theta} }{(k_{1}-k_{2})^{2}-1}\\&\Bigg( \frac{\mathrm{cos^{2}\theta}}{(k_{1}^{2} - t_{3})(k_{2}^{2}-t_{1})}- \frac{\mathrm{cos^{2}\theta}}{(k_{1}^{2} - t_{3})(k_{2}^{2}-t_{2})} \\
&+\frac{\mathrm{sin^{2}\theta}}{(k_{1}^{2} - t_{4})(k_{2}^{2}-t_{1})}-\frac{\mathrm{sin^{2}\theta}}{(k_{1}^{2} - t_{4})(k_{2}^{2}-t_{2})}\Bigg).
\label{eqn: two-loop integral for O47}
\end{split} \end{equation}
This is a sum of four integrals, each of which is evaluated in Appendix \ref{sec:Calculation of integral}, both in full generality and in the sensible limit that the quark masses are much smaller than the scale of new physics, $m_{S_{1}}$, i.e.\ in the limit $s_{k}, s_l'\rightarrow 0$. In this limit, the integral is independent of $k,l$, and we thus obtain $\mathcal{I}_{kl} = \mathcal{I}\ \forall k,l$, which we calculate to be
\begin{align}
    \begin{split}
   \mathcal{I} = &\phantom{+} \frac{ 2\pi^4
   \cos{\theta}\sin{\theta}}{(2\pi)^8}\Bigg( \cos^2\theta
      \ln t_3 (\ln t_1 - \ln t_2) + \sin^2\theta\ln t_4(\ln t_1 - \ln t_2) \\
      &+ \frac{ \cos^2\theta\Big(\hat{g}(t_3,t_1)(1 + t_3 + t_1) - \hat{g}(t_3,0)(1 + t_3)\Big)}{t_3\, t_1} \\
      &- \frac{\cos^2\theta \Big(\hat{g}(t_3,t_2)(1 + t_3 + t_2 ) - \hat{g}(t_3,0)(1 + t_3)\Big)}{t_3\,t_2} \\
      &+   \frac{\sin^2\theta\Big(\hat{g}(t_4, t_1)(1 + t_4 + t_1) - \hat{g}(t_4,0) (1 + t_4)\Big)}{t_4\,t_1}\\
      &- \frac{\sin^2 \theta\Big(\hat{g}(t_4,t_2)(1 + t_4 + t_2) -  \hat{g}(t_4,0)(1 + t_4)\Big)}{t_4\, t_2}\Bigg),
    \end{split}
    \label{eq: integral in limit of zero quark mass for Model 1}
\end{align}
where $\hat{g}(t_{\alpha},t_{\beta})$ and $\hat{g}(t_{\alpha},0)$ are defined in Appendix \ref{sec:Calculation of integral}, specifically in Equations \ref{eqn: defn of g}  and \ref{eqn: defn of g0} respectively. The behaviour of this integral for leptoquark mass parameter $\mu_{S_3}$ ranging from $1.1-100$ TeV is shown in Figure \ref{Fig: integral vs mS3}. A combination of the loop suppression factor and suppression coming from the mass of the heavy exotic scalars allow the integral to give neutrino mass a substantial suppression. For this plot, the other exotic mass parameters and the quartic and cubic scalar coupling values have been fixed: $\mu_{S_1} = 7500$ GeV, $\mu_{\phi_3} = 1500$ GeV, $\lambda_{S_3\phi_3 H}=1$ and $m_{S_1S_3\phi_3} = 1500$ GeV. Figure \ref{Fig: sum of neutrino masses vs mS3} shows an example of the calculated sum of neutrino masses for leptoquark mass parameter $\mu_{S_3}$ ranging from $1.1-100$ TeV. The other parameters are set as for Figure \ref{Fig: integral vs mS3}, with leptoquark and diquark Yukawa couplings set to $y_{3ij}^{LL}=0.001$ and $z_{1ij}^{LL}=0.01$ respectively. Note that neutrino mass goes to zero as $\mu_{S_3} \rightarrow \infty$, as expected. It should be understood that Figure \ref{Fig: sum of neutrino masses vs mS3} shows only the typical scale of the neutrino mass, and our model has enough freedom to allow for more precise fitting to the experimental results, including the correct mass differences between the neutrino mass states --- which can be achieved by enforcing a relationship between the leptoquark and diquark coupling matrices, as described below in Section \ref{sec: Casas-Ibarra Parametrisation }. 
\begin{figure}[!ht]
    \centering
        \includegraphics[width=\linewidth]{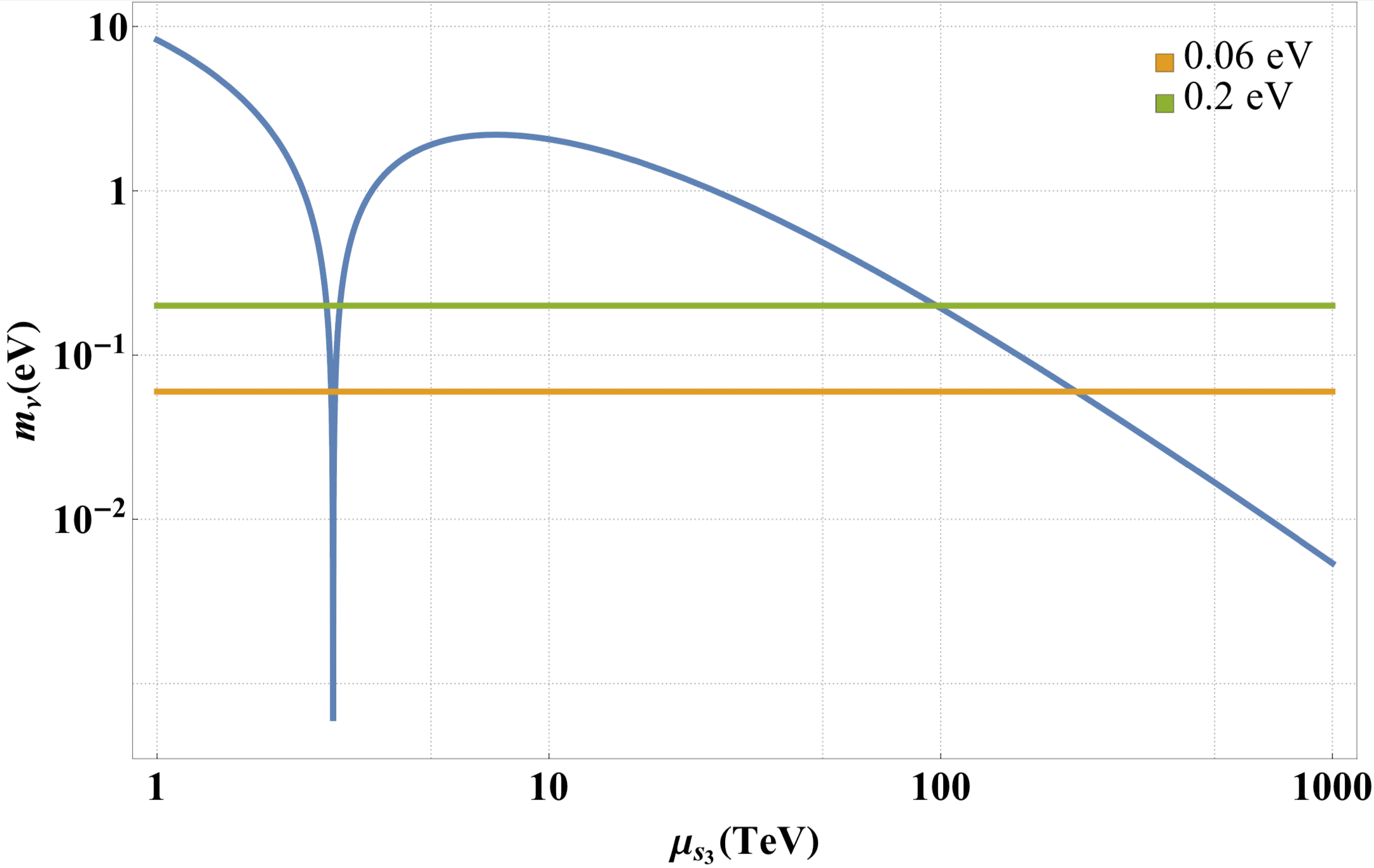}
        \caption{Plot of the sum of neutrino masses, $m_\nu$, in eV, for leptoquark mass parameter $\mu_{S_3}$ ranging from $1-100$ TeV. The other exotic mass parameters and the quartic and cubic scalar coupling values have been fixed: $\mu_{S_1} = 7.5$ TeV, $\mu_{\phi_3} = 1.5$ TeV, $\lambda_{S_3\phi_3 H}$ = 1 and $m_{S_1S_3\phi_3} = 1.5$ TeV. All entries in the leptoquark  and diquark coupling matrices have been set to $y_{3ij}^{LL}=0.001$ and $z_{1ij}^{LL}=0.01$ respectively. The solid horizontal orange line indicates the experimental lower bound of $0.06$ eV, while the green line shows the approximate cosmological upper limit of $0.2$ eV.}
        \label{Fig: sum of neutrino masses vs mS3}
\end{figure}
\subsection{Casas-Ibarra Parametrisation}
\label{sec: Casas-Ibarra Parametrisation }
After these simplifications the neutrino mass, in the flavour basis, is 
\begin{equation} \begin{split}
\boldsymbol{m}_{\nu} = m_{0}(\boldsymbol{y}_{3}^{LL})^{T}\boldsymbol{z}_{1}^{LL} \boldsymbol{y}_{3}^{LL},
\label{eqn: neutrino mass equation}
\end{split} \end{equation} 
where we have absorbed all constants into $m_{0}= 18m_{S_{1}S_{3}\phi_{3}}\mathcal{I}$ and, for convenience, we define the dimensionless matrix $\kappa \equiv \frac{m_{\nu}}{m_{0}}$.
For a given $\kappa$, we thus see that the two coupling matrices, $y_{3}^{LL}$ and $z_{1}^{LL}$, must be related. Their relationship can be obtained using the parametrisation method originally described by Casas and Ibarra \cite{Casas:2001sr}.
Recalling that the diquark couplings $z_{1}^{LL}$ must be symmetric, we can use Takagi's factorisation method to diagonalise $z_{1}^{LL} = S^{T}D_{z}S$, where $S$ is a unitary matrix, and $D_z$ has positive diagonal values.  Using $\kappa = U^*_{\text{PMNS}} D_\kappa U^\dagger_{\text{PMNS}}$, where $D_\kappa$ is diagonal with positive entries, it follows that 
{\begin{equation} \begin{split}
D_{\kappa} &= U_{\text{PMNS}}^{T}(y_{3}^{LL})^{T}S^{T}D_{z}Sy_{3}^{LL}U_{\text{PMNS}}\\
&= (\sqrt{D_{z}}Sy_{3}^{LL}U_{\text{PMNS}}
)^{T}(\sqrt{D_{z}}Sy_{3}^{LL}U_{\text{PMNS}}).
\end{split} \end{equation} }
Multiplying both sides of the equality by $\sqrt{D_{\kappa}^{-1}}$ on the left and the right, we get 
{\begin{equation} \begin{split}
\mathds{1}_{3\times 3}= (\sqrt{D_{z}}Sy_{3}^{LL}U_{\text{PMNS}}
\sqrt{D_{\kappa}^{-1}})^{T}(\sqrt{D_{z}}Sy_{3}^{LL}U_{\text{PMNS}}\sqrt{D_{\kappa}^{-1}}),
\end{split} \end{equation} }%
where $\mathds{1}_{3\times 3}$ is the identity matrix. This implies that 
\begin{equation} \begin{split}
\sqrt{D_{z}}Sy_{3}^{LL}U_{\text{PMNS}}\sqrt{D_{\kappa}^{-1}} = R,
\end{split} \end{equation} 
where $R$ is an orthogonal matrix (in general with complex entries). Thus, to produce the measured light neutrino masses contained in $D_{\kappa}$, with mixing parameters contained in $U_{\text{PMNS}}$, the most general leptoquark coupling is given by
{\begin{equation} \begin{split}
y_{3}^{LL} = S^{\dagger}\sqrt{D_{z}^{-1}}R\sqrt{D_{\kappa}}U_{\text{PMNS}}^{\dagger}.
\label{eq: casas ibarra for y}
\end{split} \end{equation} }%
Based on this equation, we see that $y_{3}^{LL}$ depends on the known low-energy parameters contained in $D_{\kappa}$ and $U_{\text{PMNS}}$, as well as the following free parameters: six real parameters from the symmetric diquark coupling, $z_{1}^{LL}$, and three, generally complex, parameters in $R$. Alternatively, since we initially place constraints on $y_{3}^{LL}$, we can rearrange Equation \ref{eq: casas ibarra for y} to find $z_{1}^{LL}$ as a function of $y_{3}^{LL}$, such that
{\begin{equation} \begin{split}
z_{1}^{LL}= (\sqrt{D_{\kappa}}U_{\text{PMNS}}(y_{3}^{LL})^{-1})^{T}(\sqrt{D_{\kappa}}U_{\text{PMNS}}(y_{3}^{LL})^{-1}).
\label{eqn: casas Ibarra for z}
\end{split} \end{equation} }%
Notice that, due to its symmetric nature, $z_{1}^{LL}$ is independent of the orthogonal matrix $R$. This makes sense since we still have nine free parameters, now all contained in $y_{3}^{LL}$. 
\subsection{Parameters and notation for analysis}
\label{notation}
Our model, given the simplifying assumptions, has 14 free parameters: four coming from the mass-dimension 1 couplings $\mu_{S_{1}}$, $\mu_{S_{3}}$, $\mu_{\phi_{3}}$, and $m_{S_{1}S_{3}\phi_{3}}$, and another 10 coming from the dimensionless coupling constants $z_{1}^{LL}$, $y_{3}^{LL}$, and $\lambda_{S_{3}S_{1}H}$, with $z_{1}^{LL}$ and $y_{3}^{LL}$ being related by Equation \ref{eqn: casas Ibarra for z}. In Section \ref{Sec: Constraints}, we discuss several phenomenological constraints on the leptoquark couplings. From here on in, we will simplify our notation such that leptoquark couplings read $(y_3^{LL})_{ij} \equiv y_{ij}$, with the index $i$ ($j$) representing the generation of the contributing quark (lepton). We will similarly denote diquark couplings by $(z_1^{LL})_{ij} \equiv z_{ij}$. \\\\
Given that 14 parameters is too large a space to sample properly, and the results would be difficult to visually present, we are forced to fix the majority of the parameters at benchmark values. We choose to scan over four leptoquark coupling constants, $y_{11}$, $y_{12}$, $y_{21}$ and $y_{22}$, and one mass parameter, $\mu_{S_3}$. The benchmark values allocated to $\mu_{S_{1}}$, $\mu_{\phi_{3}}$, $m_{S_{1}S_{3}\phi_{3}}$ and $\lambda_{S_{3}S_{1}H}$ can be found in Table \ref{table: parameters}. In order to give a representative idea of our model's robustness as well as investigate a variety of possible conclusions drawn from future particle experiments, Table \ref{table: parameters} also includes three benchmark textures for the leptoquark coupling matrix.
{\renewcommand{\arraystretch}{1.1}
\begin{table}[H]
\begin{center}
\begin{tabular}[h]{c}
\hline
\hline
$\lambda_{S_3 S_1 H}=1$, \ $\mu_{S_1}= 7.5 $ TeV, \ $\mu_{\phi_3} = 1.5$ TeV, \  $m_{S_1 S_3\phi_3}= 1.5$ TeV \\
\end{tabular}
\begin{tabular}[h]{c|c|c}
\hline
Texture A
& Texture B
& Texture C \\	
$y=\begin{pmatrix}
y_{11}&y_{12}&\epsilon_A
\\
y_{21}&y_{22}&5\epsilon_A\\
\epsilon_A&1&10\epsilon_A
\end{pmatrix}$	& $y=\begin{pmatrix}
y_{11}&y_{12}&\epsilon_B
\\
y_{21}&y_{22}&5\epsilon_B\\
\epsilon_B&1&10\epsilon_B
\end{pmatrix} $&$y=\begin{pmatrix}
y_{11}&y_{12}&\epsilon_C^2
\\
y_{21}&y_{22}&\epsilon_C\\
\epsilon_C^2&1&1
\end{pmatrix}$ \\
	\hline
	\hline
		\end{tabular}
		\end{center}
		\caption{The parameters $\lambda_{S_3 S_1 H}$, $\mu_{S_1}$, $\mu_{\phi_3}$ and $m_{S_1 S_3\phi_3}$ have been fixed to the values indicated. Three different leptoquark coupling matrix textures were used to scan the parameter space, each matrix texture having five of the nine leptoquark couplings fixed, as above, while the other four leptoquark couplings, specifically $y_{11}$, $y_{12}$, $y_{21}$ and $y_{22}$, are scanned over. In texture A, $\epsilon_A = 10^{-5}$, in texture B, $\epsilon_B = 10^{-3}$, and in texture C, $\epsilon_C = 10^{-1}$.}
	\label{table: parameters}
	\end{table}
	}

\subsection{Model 2 and vanishing neutrino self-energies}
\label{Subsec: Model 2 and vanishing neutrino self energies}
Before investigating the phenomenology of Model 1, we will use this section to tie up the loose end of the discarded alternative completion of $O_{47}$. Recall that in Model 2 $S_3$ couples as a diquark and $S_1$ couples as a leptoquark. When integrated out, these exotic fields give rise to an operator with $\mathrm{SU}(2)$-structure $O_{47}^k = \overline{L^C}^i L^j \overline{Q^C}^k Q^l \overline{Q}_m Q^C_n H^m H^n \epsilon_{ik} \epsilon_{jl}$.\footnote{The index $k$ here is used to extend the list of Ref.~\cite{de_Gouv_a_2008}.} Working in two-component Weyl spinor notation, the specific Lorentz structure generated is $(LL)(QQ)(Q^\dagger Q^\dagger) HH$, where parentheses indicate contracted spinors. In Appendix~\ref{App: vanishing model} we present the calculation of the two-loop contribution to the neutrino mass in this model. Curiously, we find that the neutrino masses vanish due to the symmetry properties of the integral and the antisymmetry of a set of couplings. (This antisymmetry is enforced by Fermi--Dirac statistics.) This suggests that the neutrino mass for this model arises at some higher-loop order. Below we show that the leading-order contribution to the neutrino masses arising from this particle content vanishes. This leading-order argument contains essentially the same ingredients as those required to see the behaviour in the UV theory, and we point readers to the calculation in the appendix for more detail.\\\\
Writing all indices ($SU(2)$, $SU(3)$, Lorentz, and flavour) explicitly, \textbf{Model~1} generates
\begin{equation} \begin{split}
\underset{r}{L}^{\alpha i }\underset{s}{L}^{\beta j}\underset{t}{Q}^{\gamma k A}\underset{u}{Q}^{\delta l B}\underset{v}{Q}^\dagger_{ \dot{\epsilon}m C} \underset{w}{Q}^\dagger_{ \dot{\zeta}nD} H^mH^n \epsilon_{ik}\epsilon_{jl}\epsilon_{\alpha \beta}\epsilon_{\gamma \delta}\epsilon^{\dot{\epsilon}\dot{\zeta}}\epsilon_{ABE}\epsilon^{CDE}
\end{split} \end{equation} 
at tree level, where Greek letters ($\alpha, \beta, \ldots$) represent spinor indices, Latin letters from the middle of the alphabet ($i,j, \ldots$) represent $SU(2)$ indices, Latin letters from the end of the alphabet ($r,s...$) represent flavour indices and capital letters represent $SU(3)$ indices. The neutrino masses arising from a single insertion of operators of this type will vanish since they depend on integrals with an odd number of loop momenta in the numerator~\cite{de_Gouv_a_2008}. We thus consider neutrino masses arising from insertions of the dimension-13 operator with a derivative acting on each $Q^\dagger$. This operator will also be generated by the particle content of the UV-completion of operator $O_{47}^k$.
However, we now show that this contribution also vanishes.\\\\
We first show that the operator must be anti-symmetric under exchange of the $v$ and $w$ flavour indices:
\begin{equation*}
    \begin{aligned}
&\cdots\underset{v}{\partial Q}^\dagger_{\epsilon m C} \underset{w}{\partial Q}^\dagger_{ \zeta nD} H^m H^n \epsilon^{{\epsilon} {\zeta}}\epsilon^{CDE}\cdots \\
=&\cdots\underset{v}{\partial Q}^\dagger_{ {\zeta}n D} \underset{w}{\partial Q}^\dagger_{ {\epsilon}mC} H^nH^m \epsilon^{{\zeta}{\epsilon}}\epsilon^{DCE}\cdots&\text{relabel ${\epsilon}\leftrightarrow {\zeta}$, $m\leftrightarrow n$ and $C\leftrightarrow D$}\\
=&\cdots\underset{v}{\partial Q}^\dagger_{ {\zeta}n D} \underset{w}{\partial Q}^\dagger_{ {\epsilon}mC} H^nH^m \epsilon^{{\epsilon}{\zeta}}\epsilon^{CDE}\cdots&\text{reorder indices }\\
=&-\cdots\underset{w}{\partial Q}^\dagger_{ {\epsilon}mC}\underset{v}{\partial Q}^\dagger_{ {\zeta}n D}  H^nH^m \epsilon^{{\epsilon}{\zeta}}\epsilon^{CDE}\cdots&\text{reorder fields }\\
\end{aligned}
\end{equation*}
which confirms that the diquark coupling of $S_3$ is anti-symmetric in flavour as stated in \cite{Dorsner:2016wpm}. The neutrino mass generated by this operator is then represented by 
\begin{align*}
 m_\nu &=   C^{rstu[vw]}(\delta_{tv}\delta_{uw}I_{v}I_{w} + \delta_{tw}\delta_{uv}I_{v}I_{w}) +  C^{srtu[vw]}(\delta_{tv}\delta_{uw}I_{v}I_{w} + \delta_{tw}\delta_{uv}I_{v}I_{w})\\
 &= (C^{rstu[tu]} +C^{srtu[ut]})I_{u}I_{t} +  (C^{srtu[tu]}+C^{srtu[ut]})I_{u}I_{t}\\
 &= 0
\end{align*}
where $C^{rstu[vw]}$ is a Wilson coefficient obtained from the evaluated self-energy diagrams. The square brackets indicate the anti-symmetry under interchange of $v$ and $w$ discussed above.\\\\
It should be noted that $O_{47}^k$ is missing from the list of $\Delta L = 2$ effective operators listed in \cite{de_Gouv_a_2008}. This may be due to an implicit assumption that the number of SM fermion generations is not more than one. In this case $O_{47}^k$ itself vanishes since $Q_m Q_n H^m H^n = 0$. \\\\
One might worry about the validity of this claim in light of the ``extended black box'' theorem \cite{Hirsch_2006}, which states that any non-vanishing $\Delta L = 2$ effective operator leads to non-vanishing Majorana neutrino mass. This is remedied by the fact that we are only closing off the effective operator in the simplest way to generate neutrino masses. The theorem tells us that there must be non-zero contributions to neutrino mass coming from $O_{47}^k$ since the operator itself does not vanish when flavour is considered. We therefore surmise that neutrino masses arise at higher loop order, and are probably too small to meet the lower bound of $\sqrt{|\Delta m^2_{32}|} \simeq 0.05$ eV with phenomenologically acceptable exotic particle masses.\\\\
There are several other $\Delta L = 2 $ effective operators which exhibit the same property, including, but not limited to $O_{11b}$, $O_{12a}$ and $O_{48}$. The two-loop contributions coming from completions of all these operators vanish, implying that there must be nonzero higher-loop contributions. Similar remarks about the $0.05$ eV lower bound pertain. This observation could potentially be used to eliminate a sizable number of effective operators from the pool of neutrino-mass-model candidates.  

\section{Constraints from Rare Processes and Flavour Physics}
\label{Sec: Constraints}
In this section, we investigate the phenomenology of our Model 1, and place constraints on the values of the coupling constants responsible for generating neutrino mass. This investigation is conducted in three parts. First, the leptoquark couplings $y_{ij}$ are constrained via the model's BSM contribution to rare processes of charged leptons, including $\mu$ to $e$ conversion in nuclei, the decays $\mu \to e \gamma$ and $\mu \to eee$, and the anomalous magnetic moment of the muon. Second, the leptoquark couplings are constrained via BSM contributions to rare meson decays. Finally, the diquark couplings, $z_{ij}$, are constrained via experimental results from neutral meson anti-meson mixing. 
\subsection{Rare processes of charged leptons}
In the absence of neutrino flavour oscillations, lepton number is conserved in the SM.  While lepton flavour has been shown to be violated by neutrino oscillations, it has as-yet not been observed in the charged lepton sector. The lepton flavour violating (LFV) terms in our Lagrangian are thus constrained by charged LFV processes. The most stringent upper bounds on LFV processes in leptoquark models come from $\mu \rightarrow eee$ and $\mu \rightarrow e \gamma $ decays and $\mu -e$ conversion in nuclei. 

\subsubsection{$\mu \to e$ conversion in nuclei}
\label{sec: mu to eee}
The strongest bounds on the branching ratio $\mathrm{Br}(\mu \rightarrow e)$ come from  $\mu \rightarrow e$ conversion off titanium and gold nuclei. The current constraints, which were set by the SINDRUM Collaboration \cite{Dohmen:1993mp, Bertl:2006up}, are of order $10^{-12}$ (Table \ref{table: mu to e conversion}), with future experimental sensitivities predicted to improve by several orders of magnitude. The most promising are the COMET \cite{doi:10.1093} and Mu2e/COMET \cite{Carey:2008zz} experiments, aiming for sensitivities of order $10^{-16}$, and the PRISM/PRIME proposal \cite{Yoshitaka2011}, boasting a possible sensitivity of $10^{-18}$.\\\\
The most general interaction Lagrangian for this process, in the notation of \cite{Kitano:2002mt}, is 
{\begin{equation} \begin{split}
\mathcal{L}_{\mu e}^{\rm eff} =&-\frac{4G_{F}}{\sqrt{2}}(m_{\mu}A_{R}\overline{\mu}\sigma^{\mu \nu}P_{L}eF_{\mu \nu} + m_{\mu}A_{L}\overline{\mu}\sigma^{\mu \nu}P_{R}eF_{\mu\nu} + h.c)\\
&-\frac{G_{F}}{\sqrt{2}}\sum_{q=u,d,s}\Big[ (g_{LS(q)}\overline{e}P_{R}\mu + g_{RS(q)}\overline{e}P_{L}\mu)\overline{q}q\\
&+(g_{LP(q)}\overline{e}P_{R}\mu + g_{RP(q)}\overline{e}P_{L}\mu)\overline{q}\gamma_{5}q\\
&+ (g_{LV(q)}\overline{e}\gamma^{\mu}P_{L}\mu + g_{RV(q)}\overline{e}\gamma^{\mu}P_{R}\mu)\overline{q}\gamma_{\mu}q\\
&+(g_{LA(q)}\overline{e}\gamma^{\mu}P_{L}\mu + g_{RA(q)}\overline{e}\gamma^{\mu}P_{R}\mu)\overline{q}\gamma_{\mu}\gamma_{5}q\\
&+ \frac{1}{2}(g_{LT(q)}\overline{e}\sigma^{\mu \nu}P_{R}\mu + g_{RT(q)}\overline{e}\sigma^{\mu \nu}P_{L}\mu)\overline{q}\sigma_{\mu \nu}q +\text{h.c.}
 \Big],
 \label{eqn: mu to e conversion eff lagrangian}
\end{split} \end{equation} }%
where $G_{F}$ is the Fermi constant, $m_{\mu}$ is the muon mass, and the $A_{L,R}$ and $g$'s are all dimensionless coupling constants corresponding to the relevant operators.  \\\\
The branching ratio is defined to be 
\begin{equation} \begin{split}
    \mathrm{Br}(\mu \rightarrow e ) = \frac{\omega_{\mathrm{conv}}}{\omega_{\mathrm{capt}}},
\end{split} \end{equation} 
where $\omega_{\mathrm{conv}}$ is the $\mu$ to $e$ conversion rate, and $\omega_{\mathrm{capt}}$ is the total muon capture rate.  
The conversion rate, $\omega_{\mathrm{conv}}$ is calculated from the effective Lagrangian in Equation \ref{eqn: mu to e conversion eff lagrangian} to be 
{\begin{equation} \begin{split}
\omega_{\mathrm{conv}} &= 2G_{F}^{2} \Big| A^{*}_{R}D + \sum_{q}G^{(q,p)}g_{LS(q)}S^{(p)}+\sum_{q}G^{(q,n)}g_{LS(q)}S^{(n)} \\
 &+ (2g_{LV(u)} +g_{LV(d)})V^{(p)}+(g_{LV(u)} +2g_{LV(d)})V^{(n)} \Big|^{2}  + (R \leftrightarrow L),
\end{split} \end{equation} }%
where $S$, $D$ and $V$ are overlap integrals, and $n$ and $p$ superscripts refer to processes interacting with a neutron or proton respectively. The coefficients $G^{q,p}$ and $G^{q,n}$, associated with $S^{(p,n)}$, are calculated in \cite{Kosmas:2001mv}, but do not play a role in our model. This is due to the fact that $\mu \rightarrow e$ conversion does not generate scalar operators in our model, thus $g_{LS(q)} = 0$. Similarly, the coefficient associated with tensor operators, namely $A_R$, vanishes. Accordingly, we only provide values relevant to our model: the $V$ overlap integral values for titanium and gold can be found in Table \ref{table: mu to e conversion}, while other values can be found in \cite{Kitano:2002mt}.\\\\
	\begin{table}[t]
		\begin{center}
	\begin{tabu}to 0.8\linewidth{X[c] | X[0.5,c] X[0.5,c] |X[c] | X[c]}
	\hline
	\hline
	 & $V^{(p)}$ & $V^{(n)}$ &$\mathrm{Br}(\mu \rightarrow e)$  &$\omega_{\mathrm{capt}}(10^{6}s^{-1})$ \\
	\hline
	$\phantom{-}^{197}_{79}\mathrm{Au}$ & 0.0610 & 0.0859 & $<7.0 \times 10^{-13}$ &13.07  \\
	$\phantom{-}^{48}_{22}\mathrm{Ti}$ & 0.0462 & 0.0399 & $<4.3 \times 10^{-12}$ &2.59 \\
							\hline
	\hline
		\end{tabu}
		\end{center}
		\caption{The relevant overlap integrals in units of $m_{\mu}^{5/2}$, the branching ratio at $90\%$ confidence level and the total capture rates for different nuclei.}
	\label{table: mu to e conversion}
	\end{table}
	{\begin{figure}[t]
\centering
\begin{subfigure}[l]{0.45\textwidth}
\centering
\begin{tikzpicture}[thick,scale=0.8]
	\coordinate[] (a) at (-1,1.5) {};
	\coordinate[] (a1) at (-2,2.5) {};
		\node[vtx, label=180:$\overline{u_{L}^{c}}$] at (-2,2.5) {};
	\coordinate[] (a2) at (-2,0.5) {};
		\node[vtx, label=180:$\mu$] at (-2,0.5) {};
	\node[vtx, label=0:$r_{1,2}$] at (-0.3,1.2) {};
	\coordinate[] (b) at (1,1.5) {};
	\coordinate[] (b1) at (2,0.5) {};
		\node[vtx, label=0:$\overline{u_{L}^{c}}$] at (2,0.5) {};
	\coordinate[] (b2) at (2,2.5) {};
		\node[vtx, label=0:$e$] at (2,2.5) {};
		
	\graph[use existing nodes]{
		a1--[fermion]a;
		a2--[fermion]a;
		a--[scalar]b;
		b--[fermion]b1;
		b--[fermion]b2;
	};
	    \end{tikzpicture}
\caption{}
\end{subfigure}
\hspace{0.1cm}
\begin{subfigure}[c]{0.45\textwidth}
\centering
\begin{tikzpicture}[thick,scale=0.8]
	\coordinate[] (a) at (-1,1.5) {};
	\coordinate[] (a1) at (-2,2.5) {};
		\node[vtx, label=180:$\overline{d_{L}^{c}}$] at (-2,2.5) {};
	\coordinate[] (a2) at (-2,0.5) {};
		\node[vtx, label=180:$\mu$] at (-2,0.5) {};
	\node[vtx, label=0:$S_{3}^{4/3}$] at (-0.3,1.1) {};
	\coordinate[] (b) at (1,1.5) {};
	\coordinate[] (b1) at (2,0.5) {};
		\node[vtx, label=0:$\overline{d_{L}^{c}}$] at (2,0.5) {};
	\coordinate[] (b2) at (2,2.5) {};
		\node[vtx, label=0:$e$] at (2,2.5) {};
		
	\graph[use existing nodes]{
		a1--[fermion]a;
		a2--[fermion]a;
		a--[scalar]b;
		b--[fermion]b1;
		b--[fermion]b2;
	};
	    \end{tikzpicture}
\caption{}
\end{subfigure}
\caption{Tree level processes contributing to $\mu$ to $e$ conversion in nuclei. Note the notation in the left diagram represents two diagrams, one mediated by $r_{1}$ and another by $r_{2}$. The arrows here, as in all other diagrams in this paper. represent the chirality of the field. Arrows pointing towards the vertex represent left-handed fields.}
\label{Fig: mu-e conversion.}
\end{figure}
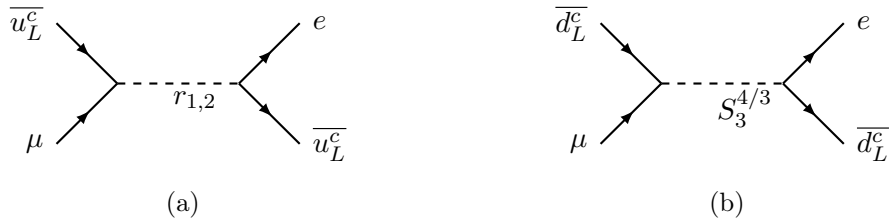}%
In our model, the dominant contributions to $\mu-e$ conversion in nuclei come from diagrams with the leptoquark $S_{3}$ mediating interactions between the charged leptons and the three lightest quarks, as can be seen in Figure \ref{Fig: mu-e conversion.}. The effective Lagrangian, calculated using Feynman rules for fermion number violating interactions found in \cite{Denner:1992vza}, is
{\begin{equation} \begin{split}
\mathcal{L}^{\text{eff}}_{\mu e} = &-\frac{3}{2}(V_{\text{CKM}}^{*}y)_{12}(V^{*}y)^{\dagger}_{11} \Big(\frac{\cos{\theta}}{m_{r_{1}}^{2}} + \frac{\sin{\theta} }{m_{r_{2}}^{2}} \Big)[\overline{e}P_{R}u^{c}][\overline{u^{c}}P_{L}\mu]\\
&- \frac{3y_{12}y^{\dagger}_{11}}{m_{S_{3}^{4/3}}^{2}}[\overline{e}P_{R}d^{c}][\overline{d^{c}}P_{L}\mu].
\end{split} \end{equation} }%
After performing a Fierz transformation and separating out the axial vector components (which vanish) from the vector components, we find
{\begin{equation} \begin{split}
\mathcal{L}^{\text{eff}}_{\mu e} = &-\frac{3}{8}(V_{\text{CKM}}^{*}y)_{12}(V^{*}y)^{\dagger}_{11} \Big(\frac{\cos{\theta}}{m_{r_{1}}^{2}} + \frac{\sin{\theta}}{m_{r_{2}}^{2}} \Big)[\overline{e}\gamma^{\mu}P_{L}\mu][\overline{u}\gamma_{\mu}u]\\*
&- \frac{3y_{12}y_{11}}{4 m_{S_{3}^{4/3}}^{2}}[\overline{e}\gamma^{\mu}P_{L}\mu][\overline{d}\gamma_{\mu}d].
\end{split} \end{equation} }
Comparison with Equation \ref{eqn: mu to e conversion eff lagrangian} shows that the nonzero Wilson coefficients are
{\begin{equation} \begin{split}
g_{LV(u)} &= \frac{3\sqrt{2}}{8G_{F}}(V_{\text{CKM}}^{*}y)_{12}(V^{*}y)^{\dagger}_{11} \Big(\frac{\cos{\theta}}{m_{r_{1}}^{2}} + \frac{\sin{\theta}}{m_{r_{2}}^{2}} \Big).
\end{split} \end{equation} }%
and
{\begin{equation} \begin{split}
g_{LV(d)} &=\frac{3\sqrt{2}y_{12}y_{11}}{4G_{F} m_{S_{3}^{4/3}}^{2}}.
\end{split} \end{equation} }%
The matrix element involving strange quarks vanishes as coherent conversion processes dominate and the vector coupling to sea quarks is zero. This leads to
{\begin{equation} \begin{split}
\frac{\omega_{\mathrm{conv}}}{\omega_{\mathrm{capt}}} &= \frac{2G_{F}^{2}}{\omega_{\mathrm{capt}}} \Big| (2g_{LV(u)} +g_{LV(d)})V^{(p)}+(g_{LV(u)} +2g_{LV(d)})V^{(n)} \Big|^{2}.
\end{split} \end{equation} }%
For fixed leptoquark masses, this process places the most stringent constraints on the product $y_{11}y_{12}$ through $\frac{\omega_{\mathrm{conv}}}{\omega_{\mathrm{capt}}} < 7.0 \times 10^{-13}$ for gold and $<4.3\times 10^{-12}$ for titanium. 

\subsubsection{$\mu \to e\gamma$}
The most stringent constraints on this process are obtained from the non-observation of LFV muonic decays by the MEG experiment \cite{TheMEG:2016wtm}, with a measured branching ratio of $\mathrm{Br}(\mu \rightarrow e \gamma)= 4.2~\times~10^{-13}\, \text{ at }$~$90\%$~CL. Future prospects are looking to improve on this by an order of magnitude. Specifically, the MEG-\RomanNumeralCaps{2} experiment \cite{Baldini:2018nnn,Cattaneo:2017psr} is predicted to start searching for $\mu \rightarrow e \gamma$ decays this year, with a target sensitivity of $4 \times 10^{-14}$. \\ \\
The effective Lagrangian for $\mu \rightarrow e \gamma$ is 
\begin{equation} \begin{split}
\mathcal{L}^{\rm eff}_{l \rightarrow l' \gamma} = \frac{e}{2} \overline{l'}i \sigma^{\mu \nu} F_{\mu \nu}(\sigma_L P_L +\sigma_R P_R)l,
\label{eqn: mu to e gamma effective lagrangian}
\end{split} \end{equation} 
where $l = \mu$ and $l'=e$,  $\sigma^{\mu \nu} = \frac{i}{2}\left[\gamma^\mu,\gamma^\nu\right]$, $F_{\mu \nu}$ and $\sigma_{L(R)}$ are Wilson coefficients. The partial decay width for  $\mu \rightarrow e \gamma$ is 
\begin{equation} \begin{split}
\Gamma(\mu \rightarrow e \gamma) = \frac{(m_\mu^2 + m_e^2)^3(|\sigma_L|^2 + |\sigma_R|^2)}{16\pi m_\mu^3}.
\end{split} \end{equation} 
Our model will have contributions from the leptoquark mass states $r_1$ and $r_2$ with up-type quarks running in the loop and leptoquark $S_3^{4/3}$ with down-type quarks running in the loop, for each of the four diagrams in Figure \ref{Fig: mu to e gamma contributions}. In total there are 36 contributing diagrams leading to Wilson coefficients 
\begin{subequations}
\begin{equation}
\begin{split}
\sigma_L = &\phantom{+} \frac{3ie}{16\pi^2 m^2_{S_3^{4/3}}}\sum_{q=d_i} y_{q1}^*y_{q2}m_e \left[ \frac{4}{3}f_S(x_q) - f_F(x_q)\right] \\
    &+ \frac{3ie}{16\pi^2}\left(\frac{\mathrm{cos^2\theta}}{m^2_{r_1}} +\frac{\mathrm{sin^2\theta}}{m^2_{r_2}}\right)\sum_{q=u_i} y_{q1}^*y_{q2}m_e \left[ \frac{1}{3}f_S(x_q) - f_F(x_q)\right],
    \label{eqn: mu to e gamma wilson coeff left}
    \end{split}
\end{equation} \\
\begin{equation}
\begin{split}
\sigma_R = &\phantom{+} \frac{3ei}{16\pi^2 m^2_{S_3^{4/3}}}\sum_{q=d_i} y_{q1}^*y_{q2}m_\mu \left[ \frac{4}{3}f_S(x_q) - f_F(x_q)\right] \\
    &+ \frac{3ei}{16\pi^2}\left(\frac{\mathrm{cos^2\theta}}{m^2_{r_1}} +\frac{\mathrm{sin^2\theta}}{m^2_{r_2}}\right)\sum_{q=u_i} y_{q1}^*y_{q2}m_\mu \left[ \frac{1}{3}f_S(x_q) - f_F(x_q)\right].
\label{eqn: mu to e gamma wilson coeff right}
\end{split}
\end{equation} 
\end{subequations}
{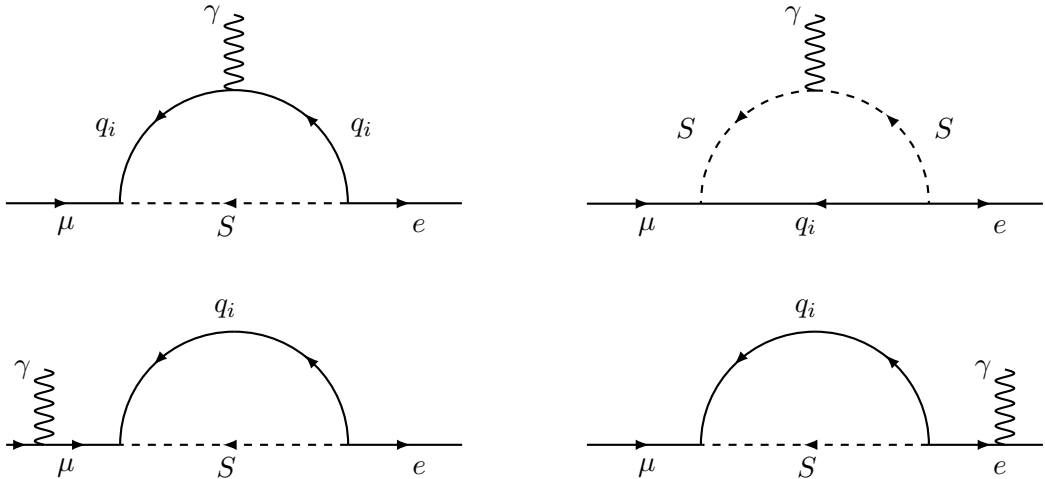
\begin{figure}[t]
\begin{minipage}[c]{0.5\textwidth}
\centering
\begin{tikzpicture}[thick,scale=1]
	\coordinate[] (a1) at (-1,1.5) {};
		\node[vtx, label=0:$\mu$] at (-0.5,1.2) {};
	\coordinate[] (a2) at (-0.5,3) {};
		\node[vtx, label=0:$$] at (0,2.5) {};
		\node[vtx, label=0:$q_{i}$] at (0,2.5) {};
	\coordinate[] (b) at (0.5,1.5) {};
		
	\coordinate[] (mid) at (2,1.5) {};
	\coordinate[] (mid2) at (2,3) {};
	\coordinate[] (d) at (3.5,1.5) {};
		\node[vtx, label=180:$S$] at (2.2,1.2) {};

	\coordinate[] (e1) at (5,1.5) {};
		\node[vtx, label=180:$e$] at (4.7,1.2) {};
	\coordinate[] (e2) at (4.5,3) {};
		\node[vtx, label=180:$q_{i}$] at (4,2.5) {};
	
	\coordinate[] (mid3) at (2,4) {};
	\node[vtx, label=180:$\gamma$] at (2,4) {};
		
	\coordinate[](quarter) at (2.75,1.5){};
	\graph[use existing nodes]{
		a1--[fermion]b;
		d--[cscalar]b;
		mid2--[photon]mid3;
		d--[fermion]e1;
	};
	
    \draw[postaction={decorate},thick,decoration = {markings,
    mark=at position 0.5 with {\arrow {latex[reversed]}}}] (2,3) arc (90:0:1.5);    
   \draw[postaction={decorate},thick,decoration = {markings,
    mark=at position 0.5 with {\arrow {latex}}}] (2,3) arc (90:180:1.5);
\end{tikzpicture}

\end{minipage}
\begin{minipage}[c]{0.5\textwidth}
\centering
\begin{tikzpicture}[thick,scale=1]
	\coordinate[] (a1) at (-1,1.5) {};
		\node[vtx, label=0:$\mu$] at (-0.5,1.2) {};
	\coordinate[] (a2) at (-0.5,3) {};
		\node[vtx, label=0:$$] at (0,2.5) {};
		\node[vtx, label=0:$S$] at (0,2.5) {};
	\coordinate[] (b) at (0.5,1.5) {};
		
	\coordinate[] (mid) at (2,1.5) {};
	\coordinate[] (mid2) at (2,3) {};
	\coordinate[] (d) at (3.5,1.5) {};
		\node[vtx, label=180:$q_i$] at (2.2,1.2) {};

	\coordinate[] (e1) at (5,1.5) {};
		\node[vtx, label=180:$e$] at (4.7,1.2) {};
	\coordinate[] (e2) at (4.5,3) {};
		\node[vtx, label=180:$S$] at (4,2.5) {};
	
	\coordinate[] (mid3) at (2,4) {};
	\node[vtx, label=180:$\gamma$] at (2,4) {};
		
	\coordinate[](quarter) at (2.75,1.5){};
	\graph[use existing nodes]{
		a1--[fermion]b;
		b--[antifermion]d;
		mid2--[photon]mid3;
		d--[fermion]e1;
	};
	
    \draw[postaction={decorate},thick,dashed,decoration = {markings,
    mark=at position 0.5 with {\arrow {latex[reversed]}}}] (2,3) arc (90:0:1.5);    
   \draw[postaction={decorate},thick,dashed,decoration = {markings,
    mark=at position 0.5 with {\arrow {latex}}}] (2,3) arc (90:180:1.5);
\end{tikzpicture}
\end{minipage}\\\\\\
\hspace{0.5 cm}
\begin{minipage}[c]{0.5\textwidth}
\centering
\begin{tikzpicture}[thick,scale=1]
	\coordinate[] (a1) at (-1,1.5) {};
		\node[vtx, label=0:$\mu$] at (-0.5,1.2) {};
	\coordinate[] (a2) at (-0.5,3) {};
		\node[vtx, label=0:$$] at (0,2.5) {};
	\coordinate[] (b) at (0.5,1.5) {};
		
	\coordinate[] (mid) at (2,1.5) {};
	\coordinate[] (mid2) at (2,3) {};
	\coordinate[] (d) at (3.5,1.5) {};
		\node[vtx, label=180:$S$] at (2.2,1.2) {};

	\coordinate[] (e1) at (5,1.5) {};
		\node[vtx, label=180:$e$] at (4.7,1.2) {};
	\coordinate[] (e2) at (4.5,3) {};
		\node[vtx, label=180:$q_{i}$] at (2.2,3.3) {};
	
	\coordinate[] (midb1) at (-0.5,1.5) {};
	\coordinate[] (midb2) at (-0.5,2.5) {};
	\node[vtx, label=180:$\gamma$] at (-0.5,2.5) {};
		
	\coordinate[](quarter) at (2.75,1.5){};
	\graph[use existing nodes]{
		a1--[fermion]midb1;
		midb1--[fermion]b;
		d--[cscalar]b;
		midb1--[photon]midb2;
		d--[fermion]e1;
	};
	
    \draw[postaction={decorate},thick,decoration = {markings,
    mark=at position 0.5 with {\arrow {latex[reversed]}}}] (2,3) arc (90:0:1.5);    
   \draw[postaction={decorate},thick,decoration = {markings,
    mark=at position 0.5 with {\arrow {latex}}}] (2,3) arc (90:180:1.5);
\end{tikzpicture}
\end{minipage}
\begin{minipage}[c]{0.5\textwidth}
\centering
\begin{tikzpicture}[thick,scale=1]
	\coordinate[] (a1) at (-1,1.5) {};
		\node[vtx, label=0:$\mu$] at (-0.5,1.2) {};
	\coordinate[] (a2) at (-0.5,3) {};
		\node[vtx, label=0:$$] at (0,2.5) {};
	\coordinate[] (b) at (0.5,1.5) {};
		
	\coordinate[] (mid) at (2,1.5) {};
	\coordinate[] (mid2) at (2,3) {};
	\coordinate[] (d) at (3.5,1.5) {};
		\node[vtx, label=180:$S$] at (2.2,1.2) {};

	\coordinate[] (e1) at (5,1.5) {};
		\node[vtx, label=180:$e$] at (4.7,1.2) {};
	\coordinate[] (e2) at (4.5,3) {};
		\node[vtx, label=180:$q_{i}$] at (2.2,3.3) {};
	
	\coordinate[] (midb1) at (4.5,1.5) {};
	\coordinate[] (midb2) at (4.5,2.5) {};
	\node[vtx, label=180:$\gamma$] at (4.5,2.5) {};
		
	\coordinate[](quarter) at (2.75,1.5){};
	\graph[use existing nodes]{
		a1--[fermion]b;
		d--[cscalar]b;
		midb1--[photon]midb2;
		d--[fermion]e1;
	};
	
    \draw[postaction={decorate},thick,decoration = {markings,
    mark=at position 0.5 with {\arrow {latex[reversed]}}}] (2,3) arc (90:0:1.5);    
   \draw[postaction={decorate},thick,decoration = {markings,
    mark=at position 0.5 with {\arrow {latex}}}] (2,3) arc (90:180:1.5);
\end{tikzpicture}
\end{minipage}
\caption{There are four types of contributing diagrams to the $\mu \to e\gamma$ process. For our model, each of the four diagrams represents diagrams with either the $S_3^{4/3}$ leptoquark and down-type quarks, or $r_1$ and $r_2$ (electric charge 1/3) and up-type quarks running through the loop.}
\label{Fig: mu to e gamma contributions}
\end{figure}}%
The Wilson coefficients are summed over the virtual up-type ($u_i$) or down-type ($d_i$) quark flavours, the factor of 3 comes from the colour contribution and the factors of $4/3$ and $1/3$ come from the electric charges of the relevant leptoquarks. Equations \ref{eqn: mu to e gamma wilson coeff left} and \ref{eqn: mu to e gamma wilson coeff right} include contributions from both $r_1$ and $r_2$, proportional to $\frac{\cos^2\theta}{m_{r_1}^2}$ and $\frac{\sin^2\theta}{m_{r_2}^2}$ respectively. The relevant loop functions are 
\begin{equation} \begin{split}
    f_S (x) &= \frac{x+1}{4(1-x)^2}+ \frac{x\, \ln x}{2(1-x)^3},\\
    f_F(x) &= \frac{x^2 -5x - 2}{12(x-1)^3} + \frac{x\, \ln x}{2(x-1)^4},
\end{split} \end{equation} 
with $x_q = m_q^2/m_{LQ}^2$.\\\\
We thus obtain the following constraint on the leptoquark coupling constants for first and second generation leptons:
\begin{equation} \begin{split}
\mathrm{Br}(\mu \rightarrow e \gamma) = \frac{\Gamma(\mu \rightarrow e \gamma)}{\Gamma_\mu ^{\mathrm{tot}}} < 4.2 \times 10^{-13}.
\end{split} \end{equation} 
where $\Gamma_\mu^{\mathrm{tot}} = 2.99\times 10^{-19}$ GeV.

\subsubsection{Anomalous magnetic moment of the muon}
The SM predicts the anomalous magnetic moment of the muon to be $a_\mu^{\mathrm{SM}}= 1.16591803(70) \times 10^{-3}$ \cite{Olive_2014,Davier:2010nc}, while the most precise experimental measurement, which comes from the E821 experiment, is $a_\mu^{\mathrm{exp}}=1.16592080(63)\times 10^{-3}$. The difference between the SM prediction and the experimental measurement, $\Delta a_\mu \equiv a_\mu^{\mathrm{exp}} - a_\mu^{\mathrm{SM}} = (2.8 \pm 0.9)\times 10^{-9}$, suggests the possible presence of BSM contributions. In our model, the leptoquark couplings with the muon provide such a contribution, given by
\begin{equation} \begin{split}
    a_l = &\phantom{+}\frac{3m_\mu}{8\pi^2 m^2_{S_3^{4/3}}}\sum_{q=d_i} |y_{q2}|^2 m_
    {\mu}\left[ \frac{4}{3}f_S(x_q) - f_F(x_q)\right] \\
    &+ \frac{3m_\mu}{8\pi^2 }\left(\frac{\mathrm{cos^2\theta}}{m^2_{r_1}} +\frac{\mathrm{sin^2\theta}}{m^2_{r_2}}\right)\sum_{q=u_i} |y_{q2}|^2 m_{\mu} \left[ \frac{1}{3}f_S(x_q) - f_F(x_q)\right].
\end{split} \end{equation} 
The contribution lies inside the bounds of $\Delta a_\mu$, ameliorating the anomaly, without placing a strong constraint on the leptoquark couplings involved. This is consistent with previous results found in literature \cite{Cheung_2001,Queiroz_2014,Chakraverty_2001}. However, when combined with other leptoquark solutions, the leptoquark $S_3$ has been shown to explain the discrepancy between theory and experiment in the anomalous magnetic moment of the muon \cite{Dorsner:2019itg}.

\subsubsection{$\mu \to eee$}
To date, the strongest constraint on $\mathrm{Br}(\mu \rightarrow eee)$  remains the $1.0\times 10^{-12}$ achieved by the SINDRUM collaboration in 1988 \cite{Bellgardt:1987du}. Looking ahead, the Mu3e collaboration \cite{Baldini:2018uhj} is promising to improve the current constraint by four orders of magnitude.\\\\
The interaction Lagrangian for this process involves interactions between the $S_3$ leptoquark and both the gauge sector and the quark sector. At one-loop level, $\mu \rightarrow eee$ decays receive contributions from three types of Feynman diagrams: $\gamma$- penguins, $Z$-penguins and box diagrams, as depicted in Figure \ref{Fig: mu to eee contributions}. Thus, the $\mu \rightarrow eee$ probability amplitude consists of three parts
\begin{equation} \begin{split}
    \mathcal{A}(\mu \rightarrow eee) = \mathcal{A_{\gamma-\mathrm{penguin}}} + \mathcal{A_{Z-\mathrm{penguin}}} + \mathcal{A_{\mathrm{box}}}.
\end{split} \end{equation} 
\textbf{Photon-penguins} --- The $\mu \rightarrow eee$ photon penguin diagrams closely resemble the $\mu \rightarrow e\gamma$ decay diagrams, however this time the photon is internal, and thus not on-shell. The amplitude for the $\mu \rightarrow eee$ photon penguin diagrams is~\cite{Hisano:1995cp, Hisano:1998fj, Arganda:2005ji}
\begin{equation} \begin{split}
  \mathcal{A}_{\gamma -\mathrm{penguin}} &= \overline{\mathbf{u}}_e(p_2)[q^2 \gamma_\mu(A^L_1 P_L + A^R_1 P_R)+ im_\mu \sigma_{\mu \nu}q^\nu(A^L_2 P_L + A^R_2 P^R)]\mathbf{u}_\mu (p_1)\\
  &\times \frac{e^2}{q^2} \overline{\mathbf{u}}_e(p_4)[\gamma_\mu]\mathbf{v}_e (p_3) - (p_2 \leftrightarrow p_4),
\end{split} \end{equation} 
with the Wilson coefficients as follows: 
\begin{equation} \begin{split}
    A^L_1 &= \frac{3}{16 \pi^2}\sum_{i=1}^3 \Bigg(\frac{y_{i1}^*y_{i2}}{36 m_{S_3^{4/3}}}F_1(x_i) + \frac{(V_{\text{CKM}}^*y)^*_{i1}(V_{\text{CKM}}^*y)_{i2}\, \mathrm{cos}^2\theta}{72 m_{r_1}^2}F_2(t_{1i})\\
    &+\frac{(V_{\text{CKM}}^*y)^*_{i1}(V_{\text{CKM}}^*y)_{i2}\, \mathrm{sin}^2\theta}{72 m_{r_2}^2}F_2(t_{2i})\Bigg),\\
    A^R_1 &= 0,\\
    A^L_2 &= \sigma_L,\\
    A^R_2 &= \sigma_R.
\end{split} \end{equation} 
{\begin{figure}[t] 
\begin{subfigure}[c]{0.5\textwidth}
\centering
\begin{tikzpicture}[thick,scale=1]
	\coordinate[] (a1) at (-1,1.5) {};
		\node[vtx, label=0:$\mu$] at (-0.5,1.2) {};
	\coordinate[] (a2) at (-0.5,3) {};
		\node[vtx, label=0:$q_{i}$] at (0,2.5) {};
	\coordinate[] (b) at (0.5,1.5) {};
		
	\coordinate[] (mid) at (2,1.5) {};
	\coordinate[] (mid2) at (2,3) {};
	\coordinate[] (d) at (3.5,1.5) {};
		\node[vtx, label=180:$S$] at (2.2,1.2) {};

	\coordinate[] (e1) at (5,1.5) {};
		\node[vtx, label=180:$e$] at (4.7,1.2) {};
	\coordinate[] (e2) at (4.5,3) {};
		\node[vtx, label=180:$q_{i}$] at (4,2.5) {};
	
	\coordinate[] (mid3) at (2,4.2) {};
	\node[vtx, label=180:$\gamma$ or $Z$] at (2,3.5) {};
	
	\coordinate[] (f1) at (3.5,4.7) {};
		\node[vtx, label=180:$e$] at (3.5,4.4) {};
	\coordinate[] (f2) at (3.5,3.7) {};
		\node[vtx, label=180:$e$] at (3.5,3.5) {};
		
	\coordinate[](quarter) at (2.75,1.5){};
	\graph[use existing nodes]{
		a1--[fermion]b;
		d--[cscalar]b;
		mid2--[photon]mid3;
		d--[fermion]e1;
		f1--[fermion]mid3;
		mid3--[fermion]f2;
	};
	
    \draw[postaction={decorate},thick,decoration = {markings,
    mark=at position 0.5 with {\arrow {latex[reversed]}}}] (2,3) arc (90:0:1.5);    
   \draw[postaction={decorate},thick,decoration = {markings,
    mark=at position 0.5 with {\arrow {latex}}}] (2,3) arc (90:180:1.5);
\end{tikzpicture}
\caption{}
\end{subfigure}
\begin{subfigure}[c]{0.5\textwidth}
\centering
\begin{tikzpicture}[thick,scale=1]
	\coordinate[] (a1) at (-1,1.5) {};
		\node[vtx, label=0:$\mu$] at (-0.5,1.2) {};
	\coordinate[] (a2) at (-0.5,3) {};
	
		\node[vtx, label=0:$S$] at (0,2.5) {};
	\coordinate[] (b) at (0.5,1.5) {};
		
	\coordinate[] (mid) at (2,1.5) {};
	\coordinate[] (mid2) at (2,3) {};
	\coordinate[] (d) at (3.5,1.5) {};
		\node[vtx, label=180:$q_i$] at (2.2,1.2) {};

	\coordinate[] (e1) at (5,1.5) {};
		\node[vtx, label=180:$e$] at (4.7,1.2) {};
	\coordinate[] (e2) at (4.5,3) {};
		\node[vtx, label=180:$S$] at (4,2.5) {};
	
	\coordinate[] (mid3) at (2,4.2) {};
	\node[vtx, label=180:$\gamma$ or $Z$] at (2,3.5) {};
	
	\coordinate[] (f1) at (3.5,4.7) {};
	\node[vtx, label=180:$e$] at (3.5,4.4) {};
	\coordinate[] (f2) at (3.5,3.7) {};
	\node[vtx, label=180:$e$] at (3.5,3.5) {};

	\coordinate[](quarter) at (2.75,1.5){};
	\graph[use existing nodes]{
		a1--[fermion]b;
		b--[antifermion]d;
		mid2--[photon]mid3;
		d--[fermion]e1;
		f1--[fermion]mid3;
		mid3--[fermion]f2;
	};
	
    \draw[postaction={decorate},thick,dashed,decoration = {markings,
    mark=at position 0.5 with {\arrow {latex[reversed]}}}] (2,3) arc (90:0:1.5);    
   \draw[postaction={decorate},thick,dashed,decoration = {markings,
    mark=at position 0.5 with {\arrow {latex}}}] (2,3) arc (90:180:1.5);
\end{tikzpicture}
\caption{}
\end{subfigure}\\\\\\
\hspace{0.5 cm}
\begin{subfigure}[c]{1.0\textwidth}
\centering
\begin{tikzpicture}[thick,scale=1.2]
	\coordinate[] (a1) at (0,2) {};
		\node[vtx, label=0:$\mu$] at (-0.75,2) {};
	\coordinate[] (a2) at (1,2) {};
	\coordinate[] (a3) at (3,2) {};
	\coordinate[] (a4) at (4,2) {};
		\node[vtx, label=0:$e$] at (4,2) {};

	\coordinate[] (b1) at (0,0) {};
	\node[vtx, label=0:$e$] at (-0.75,0) {};
	\coordinate[] (b2) at (1,0) {};
	\coordinate[] (b3) at (3,0) {};
	\coordinate[] (b4) at (4,0) {};
		\node[vtx, label=0:$e$] at (4,0) {};

	\node[vtx, label=0:$q_i$] at (1,1) {};		
	\node[vtx, label=0:$q_i$] at (3,1) {};	
	\node[vtx, label=0:$S{*}$] at (1.7,2.3) {};
	\node[vtx, label=0:$S$] at (1.7,0.3) {};
	\graph[use existing nodes]{
		a1--[antifermion]a2;
		b1--[fermion]b2;
		a3--[antifermion]a4;
		b3--[fermion]b4;
		a2--[fermion]b2;
		a3--[antifermion]b3;
		a2--[scalar]a3;
		b2--[scalar]b3
	};
\end{tikzpicture}
\caption{}
\end{subfigure}
\caption{There are three types of contributing diagrams to the $\mu \rightarrow eee$ process: two types of $\gamma-$ and $Z-$ penguin diagrams, and box Feynman diagrams . For our model, each penguin diagram could have either the $S_3^{4/3}$ leptoquark and down-type quarks, or $r_1$ and $r_2$ (electric charge 1/3) and up-type quarks running through the loop. The situation is similar for the box diagrams, except there will also be contributions where $r_1$ and $r_2$ are simultaneously present in the loop.}
\label{Fig: mu to eee contributions}
\end{figure}
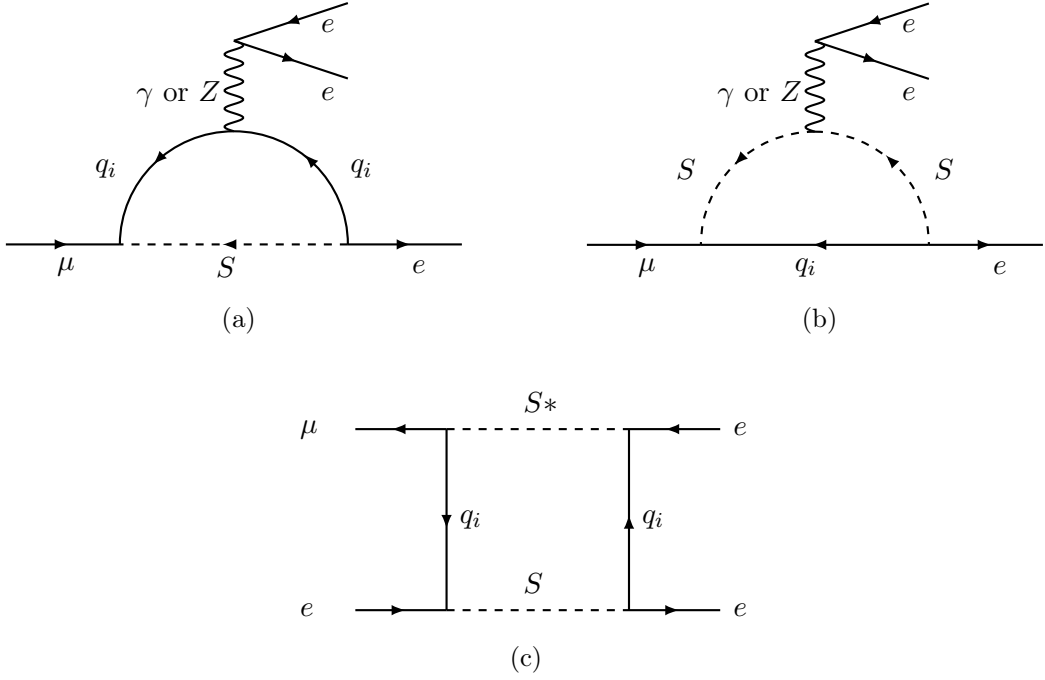}%
The $\mathbf{u}_f$ and $\mathbf{v}_f$ are the usual free-particle spinors. The variables in the loop functions are ratios of the squared masses of the quarks and the leptoquarks: $x_i = m^2_{d_i}/m^2_{S_3^{4/3}}$ and $t_{ji}= m^2_{u_i}/m^2_{r_j}$ and the loop functions $F_1(x)$ and $F_2(x)$ are
\begin{equation}
    \begin{split}
        F_1(x) &= \frac{8- 27x +36x^2 -17x^3 + (4-6x+8x^3)\, \ln x}{(x-1)^4}, \\
        F_2(x) &= \frac{-10 +27x -18x^2 +x^3 + (-8+12x+x^3)\, \ln x}{(x-1)^4}.
    \end{split} 
\end{equation} 
These loop functions are not necessarily negligible for the smaller values of $x$ generated by the first and second generations quarks. Thus, since we do not impose a priori restrictions on the leptoquark couplings, $y_3$, we cannot neglect the first and second generation contributions here. This also applies to the $Z$-penguin diagrams and the box diagrams where we must also consider all possible combinations of leptoquark mass states and quarks running through the loop.\\\\
\textbf{$Z$-penguins} --- The amplitude of the $Z$-penguin diagrams is 
\begin{equation} \begin{split}
  \mathcal{A}_{Z-\mathrm{penguin}} =& \frac{1}{m_Z^2} \overline{\mathbf{u}}_e(p_2)[ \gamma_\mu(F_L P_L + F_R P_R)]\mathbf{u}_\mu (p_1)\times \\
  \phantom{+}& \overline{\mathbf{u}}_e(p_4)[\gamma_\mu(Z_L P_L + Z_R P_R)]\mathbf{v}_e (p_3) - (p_2 \leftrightarrow p_4),
\end{split} \end{equation} 
with the Wilson coefficients:
\begin{equation}
\begin{split}
    F_L =& \frac{3}{16\pi^2}\sum_{i=1}^3 \Bigg(\frac{y^*_{i1}y_{i2}}{2~\sin\theta_W \, \mathrm{cos}\theta_W}F_3(x_i)
   \\
    \phantom{+}& + \frac{(V_{\text{CKM}}^*y)^*_{i1}(V_{\text{CKM}}^*y)_{i2}}{32\,\sin\theta_W\, \cos\theta_W(t_{1i}-1)^2(t_{1i}-t_{2i})(t_{2i}-1)^2}F_4(t_{1i},t_{2i})\Bigg),\\
    F_R =& 0,\\
    Z_L =& -\frac{e}{\sin\theta_W\, \cos\theta_W}\left(-\frac{1}{2}+ \sin^2\theta_W    \right), \\
    Z_R =& -e~\tan\theta_W.
\end{split}
\label{eq: Z-penguin FL}
\end{equation} 
Here $\theta_{W}$ is the weak angle, and $x_i$ and $t_{ji}$ are as above. We also note that the gauge coupling between the $Z$-boson and leptoquark mass states includes flavour changing contributions. Consequently, the mass states which involve mixing, specifically $r_1$ and $r_2$, must be treated together when considering the coupling between the Z-boson and the leptoquark $S_3$ in the mass basis. The loop functions in \ref{eq: Z-penguin FL} are:
\begin{subequations}
\begin{equation}
    F_3(x) = \frac{x-x^2 + x\, \ln x}{(x-1)^2}, 
\end{equation}
\begin{equation}
\begin{split}
    F_4(x_1,x_2) &= x_1^3 \Bigg[4 \cos ^2\theta  \bigg(\cos 2 \theta-2 \sin ^2\theta\, \ln x_2-3\bigg)\\
    &+x_2 \bigg(4 \cos 2 \theta-\cos 4 \theta+8 \sin ^2\theta \,\ln x_2-8 x_2+13\bigg)\Bigg]\\
    &+x_1^2 \Bigg[4 \cos ^2\theta \bigg(\cos 2 \theta \Big(2\,\ln x_1 -2\, \ln x_2-3\Big)+2\, \ln x_2+5\bigg)\\
    &+x_2 \bigg(3 \Big(\cos 4 \theta-5\Big)+\cos 2 \theta \Big(4-16 \cos ^2\theta\,\ln x_1\Big)\\
    &+ 4 x_2\, \cos 2 \theta \Big(\cos 2 \theta \big(\ln x_1-\ln x_2\big)+\ln x_1+\ln x_2-2\Big)\\
    &-4 \sin ^2\text{$\theta$} \Big(\cos 2 \theta+5\Big) \ln x_2+8 x_2^2\bigg)\Bigg]\\
    &+x_1 \Bigg[2 \sin ^2 2 \theta \bigg(\ln x_1-\ln x_2-2\bigg)+x_2^3 \bigg(4 \cos 2 \theta+\cos 4 \theta\\
    &-8 \cos ^2 \theta\, \ln x_1-13\bigg)+x_2^2 \bigg(-\cos 4 \theta \Big(\ln x_1-4 \,\ln x_2+3\Big)\\
    &+\cos 2 \theta \Big(8 \ln x_1-8 \,\ln x_2+4\Big)+9 \, \ln x_1+4 \, \ln x_2+15\bigg)\\
    &+2 x_2 \bigg(\Big(\cos 4 \theta-3\Big) \Big(\ln x_1-\ln x_2\Big)-2 \cos 2 \theta \Big(\ln x_1+\ln x_2+2 \Big) \bigg)\Bigg]\\ &+4 x_2 \, \sin ^2 \theta \Bigg[2 \cos ^2\theta \bigg(\ln x_1-\ln x_2+2\bigg)+x_2 \bigg(x_2 \Big(\cos 2 \theta\\
    &+2 \cos ^2\theta\, \ln x_1+3\Big)+\cos 2 \theta \Big(-2 \, \ln x_1+2 \, \ln x_2-3\Big)-2 \, \ln x_1-5\bigg)\Bigg].
    \end{split}
    \label{eq: Z penguin flavour changing couplings}
    \end{equation}
    \end{subequations}
In the limit of vanishing mixing angle and $x_1\rightarrow x_2$ we have the following simplification 
\begin{equation}
    \begin{split}
        \lim_{\theta \rightarrow 0, x_1 \rightarrow x_2} \frac{ F_4(x_1,x_2)}{(x_1-1)^2(x_1-x_2)(x_2-1)^2}= F_3(x). 
    \end{split}
\end{equation}
\textbf{Box diagrams} --- For our model, the non-vanishing amplitude from the contribution of box diagrams to the $\mu \rightarrow eee$ decay is
\begin{equation} \begin{split}
     \mathcal{A}_{\mathrm{Box}} =& e^2 B_1^L \overline{\mathbf{u}}_e(p_2)[\gamma^\mu P_L ]\mathbf{u}_\mu (p_1) \overline{\mathbf{u}}_e(p_4)[\gamma_\mu P_L]\mathbf{v}_e (p_3),
\end{split} 
\end{equation} 
with 
\begin{equation} \begin{split}
    B_1^L &= \frac{3}{16\pi^2 e^2}\Bigg\{\sum_{i,j = 1}^3 2 y_{i1}^*y_{i2}y_{j1}^*y_{j1}D_{00}(m^2_{S_3^{4/3}},m^2_{S_3^{4/3}},m^2_{d_i},m^2_{d_j})  \\
    &+  (V^*_{\text{CKM}}y)_{i1}^*(V^*_{\text{CKM}}y)_{i2}(V^*_{\text{CKM}}y)_{j1}^*(V^*_{\text{CKM}}y)_{j1}\bigg[\cos^4\theta \, D_{00}(m^2_{r_1},m^2_{r_1},m^2_{u_i},m^2_{u_j})\\
    & +2\cos^2\theta \, \sin^2\theta \, D_{00}(m^2_{r_1},m^2_{r_2},m^2_{u_i},m^2_{u_j}) \\
    &+  \sin^4\theta\,D_{00}(m^2_{r_2},m^2_{r_2},m^2_{u_i},m^2_{u_j})\bigg]\Bigg\}.
\end{split} \end{equation} 
The loop function for the box diagrams is defined as 
\begin{equation} \begin{split}
    m^2_4 D_{00}(m^2_1,m^2_2, m^2_3,m^2_4) = &- \frac{x_1^2 \, \ln x_1}{(x_1-1)(x_1 - x_2)(x_1-x_3)} + \frac{x_2^2 \, \ln x_2 }{(x_1-x_2)(x_2-1)(x_2-x_3)} \\
    &+ \frac{x_3^2\, \ln x_3}{(x_1-x_3)(x_3-1)(x_3-x_2)},
\end{split} \end{equation} 
where $x_i = \frac{m^2_i}{m^2_4}$ for $i=1,2,3$.\\\\
\textbf{$\mu \rightarrow eee$ Amplitude} --- Using the form factors defined above, we calculate the $\mu \rightarrow eee $ decay rate to be 
\begin{equation} \begin{split}
    \Gamma(\mu \rightarrow e e e) &= \frac{e^4}{512\pi^3}m^5_{\mu}\Bigg[|A_1^L|^2 + 2(A_1^L A_2^{R*} + \mathrm{h.c.})  \\
    &+ \bigg(|A_2^L|^2 + |A_2^R|^2\bigg)\bigg(\frac{16}{3}\, \ln \frac{m_\mu}{m_e} - \frac{22}{3}\bigg) \\
    &+\frac{1}{6}|B_1^L|^2  +\frac{1}{3}(A_1^L B_1^{L*}+ \mathrm{h.c.}) - \frac{2}{3}(A_2^R B_1^{L*} + \mathrm{h.c.})  \\
    &+ \frac{1}{3} \bigg\{2 |F_{LL}|^2 +|F_{LR}|^2 + (B_1^L F_{LL}^* + \mathrm{h.c.}) \\
    &+ 2(A_1^L F_{LL}^* + \mathrm{h.c.} ) + (A_1^L F_{LR}^* + \mathrm{h.c.} ) - 4(A_2^RF_{LL}^* + \mathrm{h.c.}) \\
    &- 2(A_2^R F_{LR}^* + \mathrm{h.c.}) \bigg\}\Bigg], 
\end{split} \end{equation} 
with 
\begin{equation} \begin{split}
    F_{LL} = \frac{F_L Z_L}{g^2 \sin^2(\theta_W) m_Z^2},\qquad F_{LR} = \frac{F_L Z_R}{g^2 \sin^2(\theta_W) m_Z^2}.
\end{split} \end{equation} 
Thus, we obtain a strong constraint on the leptoquark coupling constants to first and second generation leptons, given by
\begin{equation}
    \begin{split}
        \text{Br}(\mu\rightarrow eee) = \frac{\Gamma (\mu \rightarrow eee)}{\Gamma^{tot}_\mu},
    \end{split}
\end{equation}
where $\Gamma^{tot}_\mu = 2.99 \times 10^{-19}$ GeV.

\subsection{Rare meson decays}
\label{sec: Rare meson decays}
In the SM, rare meson decays in the form of flavour-changing neutral currents (FCNCs) arise at loop level and are thus heavily suppressed, leading them to be highly sensitive to BSM contributions. A plethora of precision experiments have placed stringent bounds on these rare decays. These processes occur at tree-level in our model, thus the couplings involved are severely constrained.\\\\
Carpentier and Davidson \cite{Carpentier:2010ue} published a comprehensive list of (order of magnitude) constraints on two-lepton--two-quark ($2l2q$) operators. They work with the effective Lagrangian
{\begin{equation} \begin{split}
\mathcal{L}^{\rm eff}_{llqq} \supset \frac{1}{2}\sum^{3}_{i,j,k,n = 1}\frac{C_{ijkn}O^{ijkn}_{\alpha}}{m_{\rm NP}^{2}} + \mathrm{h.c.},
\label{Fig: 2l2q operator lagrangian}
\end{split} \end{equation} }%
where $C_{ijkn}/2 m_{\rm NP}$ are the Wilson coefficients. The coefficient relevant to our model is that accompanying a dimension six, left-handed chiral vector effective operator
{\begin{equation} \begin{split}
O^{ijkn}_{(1)lq} &= {(\overline{l_{i}}\gamma^{\mu}P_{L}l_{j})(\overline{q_{k}}\gamma_{\mu}P_{L}q_{n})}\\
&= (\overline{\nu_{i}}\gamma^{\mu}P_{L}\nu_{j}+\overline{e_{i}}\gamma^{\mu}P_{L}e_{j})(\overline{u_{k}}\gamma_{\mu}P_{L}u_{n}+\overline{d_{k}}\gamma_{\mu}P_{L}d_{n}).
\end{split} \end{equation} }%
The bounds are set on dimensionless coefficients, $\epsilon^{ijkn}_{(n)lq}$, related to the respective Wilson coefficients by 
{\begin{equation} \begin{split}
\frac{C^{ijkn}_{(1)lq}}{m^{2}_{\rm NP}} = - \frac{4G_{F}}{\sqrt{2}}\epsilon^{ijkn}_{(1)lq}.
\end{split} \end{equation} }%
Bounds are placed on $\epsilon^{ijkn}$ by analysing the contribution of a relevant effective operator to the branching ratios of rare meson decays, one effective operator at a time. In doing so, there is a risk of overlooking possible destructive interference effects. Therefore, the constraints in this section are only order of magnitude estimates. The analysis discussed in this section allows us to place constraints on all nine leptoquark couplings, $y_{ij}$, through simply calculating the contribution of leptonic rare meson decays and semi-leptonic neutral current decays to $2l2q$ effective operators. The results are summarised in Table \ref{table: constraints on 2l2q processes}. 
{ \renewcommand{\arraystretch}{2.5}
\begin{table}[t]
		\begin{center}
	\begin{tabu}to 0.8\linewidth{X[c,m] |X[0.8,c,m] |X[0.7,c,m] | X[0.8,c,m]}
	\hline
	\hline
	{\fontsize{10}{12}\selectfont $(\overline{l_{i}}\gamma^{\mu}P_{L}l_{j})(\overline{q_{k}}\gamma^{\mu}P_{L}q_{n})$} & {\fontsize{10}{12}\selectfont Constraint on $\epsilon^{ijkn}$} &{\fontsize{10}{12}\selectfont Observable}  & {\fontsize{10}{12}\selectfont Experimental value}\\
	\hline
	\hline
	$e\mu ds$&$3.0\times 10^{-7}$&${\fontsize{8}{12}\selectfont \text{Br}(K^{0}_{L}\rightarrow \overline{e}\mu)}$&$ < 4.7\times 10^{-12}$\\
	\hline
	$\mu\mu ds$ &$7.8 \times 10^{-6}$&${\fontsize{8}{12}\selectfont \text{Br}(K^{0}_{L}\rightarrow \overline{\mu}\mu)}$& $6.84\times 10^{-9}$\\
	\hline
	$\nu_{i}\nu_{j}ds$ & $9.4 \times 10^{-6}$ & $\frac{\text{Br}(K^{+}\rightarrow \pi^{+}\overline{\nu}\nu)}{\text{Br}(K^{+}\rightarrow \pi^{0}\overline{e}\nu_{e})}$ & $\frac{1.5 \times 10^{-10}}{5.08\times 10^{-2}}$  \\
		\hline
	$\nu_{i}\nu_{j}bs$ & $1.0 \times 10^{-3}$ & $\frac{\text{Br}(B^{+}\rightarrow K^{+}\overline{\nu}\nu)}{\text{Br}(B^{+}\rightarrow D^{0}\overline{e}\nu_{e})}$ & $\frac{1.4 \times 10^{-5}}{2.2\times 10^{-2}}$  \\
	\hline
	$eesb$&$1.8\times10^{-4}$&$\frac{\mathrm{Br}(B^{+}\rightarrow K^{+}\overline{e}e)}{\mathrm{Br}(B^{+}\rightarrow D^{0}\overline{e}\nu_{e})}$&$<\frac{4.9\times10^{-7}}{1.34\times 10^{-4}}$\\
							\hline
	\hline
		\end{tabu}
		\end{center}
		\caption{{Constraints on the dimensionless coefficient $\epsilon^{ijkn}$ arising from effective operator $2\sqrt{2}G_{F}(\overline{l_{i}}\gamma^{\mu}P_{L}l_{j})(\overline{q_{k}}\gamma^{\mu}P_{L}q_{n})$ \cite[Tab. 2 and Tab. 12]{Carpentier:2010ue}. The left most column specifies the generation indices $ijkn$, the second column gives the best constraint on $\epsilon^{ijkn}$, obtained from the observable indicated in the third column and the experimental bounds given in the last column. The bounds also apply under permutation of lepton and/or quark indices. } }
	\label{table: constraints on 2l2q processes}
\vspace{-4mm}
	\end{table}}

\subsubsection{Leptonic meson decays}
\begin{figure}[b]
\centering
\begin{minipage}[c]{\textwidth}
\centering
\begin{tikzpicture}[thick,scale=0.8]
	\coordinate[] (a1) at (1,4) {};
		\node[vtx, label=0:$\overline{e_{L}}(\overline{\mu_{L}}$)] at (1,4) {};
	\coordinate[] (a2) at (0,3) {};
	\coordinate[] (a3) at (-1,4) {};
		\node[vtx, label=180:$d_{L}$] at (-1,4) {};

	\coordinate[] (b1) at (1,0) {};
		\node[vtx, label=0:$\mu_{L}$] at (1,0) {};
	\coordinate[] (b2) at (0,1) {};
	\coordinate[] (b3) at (-1,0) {};
		\node[vtx, label=180:$\overline{s_{L}}$] at (-1,0) {};

		\node[vtx, label=0:$S_{3}^{4/3*}$] at (0,2) {};

	\graph[use existing nodes]{
		a1--[antifermion]a2;
		a2--[fermion]a3;
		a2--[scalar]b2;
		b1--[fermion]b2;
		b2--[fermion]b3;
	};
\end{tikzpicture}
\\
\end{minipage}\\
\caption{Dominant contributions to $K^{0}_{L} \rightarrow \overline{e}\mu$ ($K^{0}_{L} \rightarrow \overline{\mu}\mu$).}
\label{Fig: leptonic decays}
\end{figure}
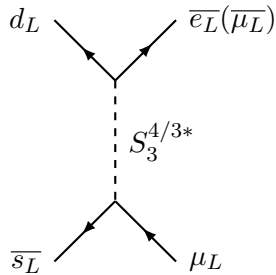

Starting with the leptonic meson decays $K^{0}_{L}\rightarrow \overline{e}\mu$ and $K^{0}_{L}\rightarrow \overline{\mu}\mu$ we place bounds on products of the first two generations of leptoquark couplings. The Wilson coefficient associated with the operator $O^{1212}_{(1)lq}=(\overline{e}\gamma^{\mu}P_{L}\mu)(\overline{s}\gamma^{\mu}P_{L}d)$ is 
\begin{equation} \begin{split}
\frac{C_{(1)lq}^{1221}}{m_{LQ}^{2}} =\frac{y_{11}y_{22}}{m_{S_{3}^{4/3}}^{2}} \leq 7.0\times 10^{-12}~\mathrm{GeV}^{-2}.
\label{eqn: Ktoemu}
\end{split} \end{equation} 
The relevant Feynman diagram, which contributes to the process $K^{0}_{L} \rightarrow \overline{e}\mu$, is depicted in Figure~\ref{Fig: leptonic decays}. This process places constraints on the first and second generation diagonal leptoquark couplings. Since the constraints on $\epsilon^{ijkn}$ are equivalent under the exchange of quark or lepton indices, the same constraints also apply to the product $y_{12}y_{21}$. Similar bounds can be placed on $y_{12}y_{22}$ from the process $K^{0}_{L}\rightarrow \overline{\mu}\mu$, with Feynman diagram depicted in Figure~\ref{Fig: leptonic decays}. 

\subsubsection{Semi-leptonic meson decays}
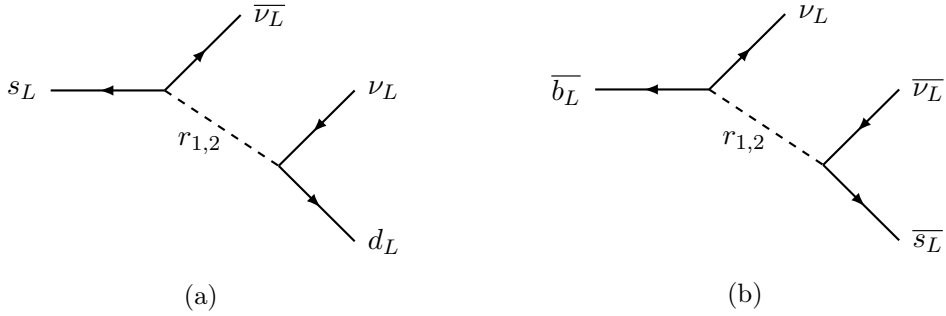
\begin{figure}[t]
\centering
\begin{subfigure}[l]{0.45\textwidth}
\centering
\begin{tikzpicture}[thick,scale=1]
	\coordinate[] (a1) at (0,2) {};
		\node[vtx, label=0:$s_{L}$] at (-0.75,2) {};
	\coordinate[] (a2) at (1.5,2) {};
	\coordinate[] (a3) at (2.5,3) {};
		\node[vtx, label=0:$\overline{\nu_{L}}$] at (2.5,3) {};

	\coordinate[] (b1) at (4,2) {};
		\node[vtx, label=0:$d_{L}$] at (4,0) {};
	\coordinate[] (b2) at (3,1) {};
	\coordinate[] (b3) at (4,0) {};
		\node[vtx, label=0:$\nu_{L}$] at (4,2) {};

		\node[vtx, label=0:$r_{1,2}$] at (1.5,1.3) {};

	\graph[use existing nodes]{
		a1--[antifermion]a2;
		a2--[fermion]a3;
		a2--[scalar]b2;
		b1--[fermion]b2;
		b2--[fermion]b3;
	};
\end{tikzpicture}
\\
\caption{}
\label{Fig: semi-leptonic decays a}
\end{subfigure}
\hspace{0.1cm}
\begin{subfigure}[c]{0.45\textwidth}
\centering
\begin{tikzpicture}[thick,scale=1]
	\coordinate[] (a1) at (0,2) {};
		\node[vtx, label=0:$\overline{b_{L}}$] at (-0.75,2) {};
	\coordinate[] (a2) at (1.5,2) {};
	\coordinate[] (a3) at (2.5,3) {};
		\node[vtx, label=0:$\nu_{L}$] at (2.5,3) {};

	\coordinate[] (b1) at (4,2) {};
		\node[vtx, label=0:$\overline{s_{L}}$] at (4,0) {};
	\coordinate[] (b2) at (3,1) {};
	\coordinate[] (b3) at (4,0) {};
		\node[vtx, label=0:$\overline{\nu_{L}}$] at (4,2) {};

		\node[vtx, label=0:$r_{1,2}$] at (1.5,1.3) {};

	\graph[use existing nodes]{
		a1--[antifermion]a2;
		a2--[fermion]a3;
		a2--[scalar]b2;
		b1--[fermion]b2;
		b2--[fermion]b3;
	};
\end{tikzpicture}
\\
\caption{}
\label{Fig: semi-leptonic decays b}
\end{subfigure}\\
\caption{Dominant contributions to (a) $K^{+} \rightarrow \pi^{+} \overline{\nu}\nu$ and (b) $B^{+} \rightarrow K^{+}\overline{\nu}\nu$.}
\label{Fig: semi-leptonic decays.}
\end{figure}

The semi-leptonic meson decays, the most tightly constrained being $K^{+}\rightarrow \pi^{+}\overline{\nu}\nu$ and $B^{+}\rightarrow K^{+}\overline{\nu}\nu$, place bounds on third generation leptoquark couplings, as well as additional bounds on the couplings already discussed. The process $K^{+}\rightarrow \pi^{+}\overline{\nu}\nu$, depicted in Figure \ref{Fig: semi-leptonic decays a}, induces the following Wilson coefficient, associated with the operator $O^{ij12}_{(1)lq}=(\overline{\nu_{i}}\gamma^{\mu}P_{L}\nu_{j})(\overline{d}\gamma^{\mu}P_{L}s)$:
{\begin{equation} \begin{split}
\frac{C_{(1)lq}^{ij12}}{m_{LQ}^{2}} =\sum_{i,j=1}^{3} \frac{y_{1i}y_{2j}}{2}\left( \frac{\cos{\theta}}{m_{r_{1}}^{2}} + \frac{\sin{\theta}}{m_{r_{2}}^{2}} \right),
\label{eqn: semileptonic wilson coeff}
\end{split} \end{equation} }%
where we sum over the neutrino flavour. The bounds on $\epsilon^{ijkn}$, found in Table~\ref{table: constraints on 2l2q processes}, are calculated one flavour at a time, with all other contributions set to zero. After this analysis, the only unconstrained leptoquark parameters are those involving third generation quarks. \\\\
The leptoquark couplings to third generation quarks can be constrained via the process $B^{+}\rightarrow K^{+}\overline{\nu}\nu$, which induces the effective operator $O^{ij23}_{(1)lq}=(\overline{\nu_{i}}\gamma^{\mu}P_{L}\nu_{j})(\overline{s}\gamma^{\mu}P_{L}b)$. The Feynman diagram associated with this operator is pictured in Figure \ref{Fig: semi-leptonic decays b}, and the Wilson coefficients $\frac{C^{ij23}}{m_{LQ}^{2}}$, are identical to Equation \ref{eqn: semileptonic wilson coeff} apart from the quark indices.\\\\
The constraints applied to each set of parameters are summarised in Table \ref{table: constraints on parameters from 2q2l operators}. Note that we do not include constraints from the anomalous decay $B_s \rightarrow \overline{\mu}\mu$, which will be discussed in Section \ref{Rk anomalies}.

	{ \renewcommand{\arraystretch}{2.5}	\begin{table}[b]
		\begin{center}
	\begin{tabu}to 0.8\linewidth{X[0.33,c,m] |X[0.33,c,m]| X[0.33,c,m]}
	\hline
	\hline
	{\small Process} & {\small Parameters Constrained}& {\small Constraints $\mathrm{(GeV)^{-2}}$}\\
	\hline
	\hline
	$K^{0}_{L}\rightarrow \overline{e}\mu$&$\frac{y_{\scaleto{11\mathstrut}{4pt}}y_{\scaleto{22\mathstrut}{4pt}}}{m_{ \scaleto{S_{\scaleto{3\mathstrut}{5pt}}^{\scaleto{4/3}{5pt}}}{8pt}}^{2}}$  {\small and}  $\frac{y_{\scaleto{12\mathstrut}{4pt}}y_{\scaleto{21\mathstrut}{4pt}}}{m_{ \scaleto{S_{\scaleto{3\mathstrut}{5pt}}^{\scaleto{4/3}{5pt}}}{8pt}}^{2}}$& $7.1 \times 10^{-12}$\\
	\hline
	$K^{0}_{L}\rightarrow \overline{\mu}\mu$& $\frac{y_{\scaleto{12\mathstrut}{4pt}}y_{\scaleto{22\mathstrut}{4pt}}}{m_{ \scaleto{S_{\scaleto{3\mathstrut}{5pt}}^{\scaleto{4/3}{5pt}}}{8pt}}^{2}}$&$1.8\times 10^{-10}$\\
	\hline
	$K^{+}\rightarrow \pi^{+}\overline{\nu}\nu$&$\frac{y_{\scaleto{11\mathstrut}{4pt}}y_{\scaleto{22\mathstrut}{4pt}}}{2}\left( \frac{\cos\theta}{m_{r_{\scaleto{1\mathstrut}{4pt}}}^{2}} + \frac{\sin\theta}{m_{r_{\scaleto{2\mathstrut}{4pt}}}^{2}} \right)$, $\frac{y_{\scaleto{12\mathstrut}{4pt}}y_{\scaleto{21\mathstrut}{4pt}}}{2}\left( \frac{\cos\theta}{m_{r_{\scaleto{1\mathstrut}{4pt}}}^{2}} + \frac{\sin\theta}{m_{r_{\scaleto{2\mathstrut}{4pt}}}^{2}} \right)$, {\small and} $\frac{y_{\scaleto{13\mathstrut}{4pt}}y_{\scaleto{23\mathstrut}{4pt}}}{2}\left( \frac{\cos\theta}{m_{r_{\scaleto{1\mathstrut}{4pt}}}^{2}} + \frac{\sin\theta}{m_{r_{\scaleto{2\mathstrut}{4pt}}}^{2}} \right)$& $2.3\times 10^{-10}$ \\
		\hline
	$B^{+}\rightarrow K^{+}\overline{\nu}\nu$&$\frac{y_{\scaleto{22\mathstrut}{4pt}}y_{\scaleto{33\mathstrut}{4pt}}}{2}\left( \frac{\cos\theta}{m_{r_{\scaleto{1\mathstrut}{4pt}}}^{2}} + \frac{\sin\theta}{m_{r_{\scaleto{2\mathstrut}{4pt}}}^{2}} \right)$, $\frac{y_{\scaleto{23\mathstrut}{4pt}}y_{\scaleto{32\mathstrut}{4pt}}}{2}\left( \frac{\cos\theta}{m_{r_{\scaleto{1\mathstrut}{4pt}}}^{2}} + \frac{\sin\theta}{m_{r_{\scaleto{2\mathstrut}{4pt}}}^{2}} \right)$, {\small and} $\frac{y_{\scaleto{23\mathstrut}{4pt}}y_{\scaleto{31\mathstrut}{4pt}}}{2}\left( \frac{\cos\theta}{m_{r_{\scaleto{1\mathstrut}{4pt}}}^{2}} + \frac{\sin\theta}{m_{r_{\scaleto{2\mathstrut}{4pt}}}^{2}} \right)$& $2.4\times 10^{-8}$ \\
	\hline
	$B^{+}\rightarrow K^{+}\overline{e}e$&$\frac{y_{\scaleto{21\mathstrut}{4pt}}y_{\scaleto{31\mathstrut}{4pt}}}{m_{ \scaleto{S_{\scaleto{3\mathstrut}{5pt}}^{\scaleto{4/3}{5pt}}}{8pt}}^{2}}$&$4.2 \times 10^{-9}$\\
							\hline
	\hline
		\end{tabu}
		\end{center}
		\caption{ Constraints applied to expression involving leptoquark couplings and masses, derived from leptonic and semi--leptonic rare meson decays. }
	\label{table: constraints on parameters from 2q2l operators}
			\end{table}}

\subsection{Neutral meson anti-meson mixing}
Mixing of neutral mesons occurs in the SM through box diagrams with $W$--bosons and top quarks as the propagators. Since neither $S_{3}$ nor $S_{1}$ have any restrictions with respect to the generation of the SM fermions they couple to, meson mixing gets contributions from diagrams with leptoquark couplings as well as diquark couplings. Another consequence of unrestrained flavour couplings is that there are contributions to the mixing of all neutral-meson species. We focus on neutral kaon mixing, $B_{s}-\overline{B}_{s}$ and $B_{d}-\overline{B}_{d}$ mixing.  The most general effective Hamiltonian for neutral meson mixing is 
{\begin{equation} \begin{split}
\mathcal{H}^{\rm eff}_{M-\overline{M}} = \sum_{m=1}^{5}C^{ij}_{m} O_{m}^{ij} +  \sum_{m=1}^{3}\tilde{C}^{ij}_{m} \tilde{O}_{m}^{ij},
\label{eqn: effective operator kaon mixing 1}
\end{split} \end{equation} }%
with the effective operator relevant to meson mixing both in the SM and in our model being 
{\begin{equation} \begin{split}
O_{1}^{ij} = (\overline{q}_{i}^{\alpha}\gamma_{\mu}P_{L}q_{j}^{\alpha})(\overline{q}_{i}^{\beta}\gamma^{\mu}P_{L}q_{j}^{\beta}).
\label{eqn: effective operator kaon mixing 2}
\end{split} \end{equation} }%
(The other operators in the above equation are listed in \cite{Bona:2007vi}). The operators $\tilde{O}_{m}^{ij}$ are identical to $O_{m}^{ij}$ except for the exchange $L \leftrightarrow R$. Latin letters indicate fermion flavour indices, while the Greek letters $\alpha$ and $\beta$ indicate colour indices. \\\\
The SM Wilson coefficient for meson mixing is 
\begin{equation} \begin{split}
C_{1,SM}^{ij}=-\frac{G_{F}^{2}m_{W}}{12\pi^{2}}V_{ti}^{2}V_{tj}^{*2}S_{0}\left(\frac{m_{t}^{2}}{m_{W}^{2}}\right),
\end{split} \end{equation}  
where $m_{t}$ is the mass of the top quark, $m_{W}$ is the mass of the $W$ boson and $S_{0}(x)$ is the Inami-Lim function~\cite{Inami:1980fz},
\begin{equation} \begin{split}
S_{0} (x)= x\left(\frac{1}{4}+\frac{9}{4}\frac{1}{1-x}\right) - \frac{3}{2}\left( \frac{x}{1-x}\right)^{3}.
\end{split} \end{equation} 
{\begin{figure}[t]
\centering
\begin{minipage}[l]{0.45\textwidth}
\centering
\begin{tikzpicture}[thick,scale=1]
	\coordinate[] (a1) at (0,2) {};
		\node[vtx, label=0:$\overline{s_{L}}$] at (-0.75,2) {};
	\coordinate[] (a2) at (1,2) {};
	\coordinate[] (a3) at (3,2) {};
	\coordinate[] (a4) at (4,2) {};
		\node[vtx, label=0:$\overline{d_{L}}$] at (4,2) {};

	\coordinate[] (b1) at (0,0) {};
	\node[vtx, label=0:$d_{L}$] at (-0.75,0) {};
	\coordinate[] (b2) at (1,0) {};
	\coordinate[] (b3) at (3,0) {};
	\coordinate[] (b4) at (4,0) {};
		\node[vtx, label=0:$s_{L}$] at (4,0) {};

	\node[vtx, label=0:$\overline{t^{C}_{L}}$] at (1,1) {};		
	\node[vtx, label=0:$t_{L}^{C}$] at (3,1) {};	
	\node[vtx, label=0:$S_{1}^{*}$] at (1.7,2.3) {};
	\node[vtx, label=0:$S_{1}$] at (1.7,0.3) {};
	\graph[use existing nodes]{
		a1--[antifermion]a2;
		b1--[fermion]b2;
		a3--[antifermion]a4;
		b3--[fermion]b4;
		a2--[fermion]b2;
		a3--[antifermion]b3;
		a2--[scalar]a3;
		b2--[scalar]b3
	};
\end{tikzpicture}
\end{minipage}
\hspace{0.1cm}
\begin{minipage}[c]{0.45\textwidth}
\centering
\begin{tikzpicture}[thick,scale=1]
	\coordinate[] (a1) at (0,2) {};
		\node[vtx, label=0:$\overline{s_{L}}$] at (-0.75,2) {};
	\coordinate[] (a2) at (1,2) {};
	\coordinate[] (a3) at (3,2) {};
	\coordinate[] (a4) at (4,2) {};
		\node[vtx, label=0:$\overline{d_{L}}$] at (4,2) {};

	\coordinate[] (b1) at (0,0) {};
	\node[vtx, label=0:$d_{L}$] at (-0.75,0) {};
	\coordinate[] (b2) at (1,0) {};
	\coordinate[] (b3) at (3,0) {};
	\coordinate[] (b4) at (4,0) {};
		\node[vtx, label=0:$s_{L}$] at (4,0) {};

	\node[vtx, label=0:$S_{1}^{*}$] at (1,1) {};		
	\node[vtx, label=0:$S_{1}$] at (3,1) {};	
	\node[vtx, label=0:$\overline{t^{C}_{L}}$] at (1.7,2.3) {};
	\node[vtx, label=0:$t_{L}^{C}$] at (1.7,0.3) {};
	\graph[use existing nodes]{
		a1--[antifermion]a2;
		b1--[fermion]b2;
		a3--[antifermion]a4;
		b3--[fermion]b4;
		a2--[scalar]b2;
		a3--[scalar]b3;
		a2--[fermion]a3;
		b2--[antifermion]b3
	};
\end{tikzpicture}
\end{minipage}\\
\begin{minipage}[l]{0.45\textwidth}
\centering
\begin{tikzpicture}[thick,scale=1]
	\coordinate[] (a1) at (0,2) {};
		\node[vtx, label=0:$\overline{s_{L}}$] at (-0.75,2) {};
	\coordinate[] (a2) at (1,2) {};
	\coordinate[] (a3) at (3,2) {};
	\coordinate[] (a4) at (4,2) {};
		\node[vtx, label=0:$\overline{d_{L}}$] at (4,2) {};

	\coordinate[] (b1) at (0,0) {};
	\node[vtx, label=0:$d_{L}$] at (-0.75,0) {};
	\coordinate[] (b2) at (1,0) {};
	\coordinate[] (b3) at (3,0) {};
	\coordinate[] (b4) at (4,0) {};
		\node[vtx, label=0:$s_{L}$] at (4,0) {};

	\node[vtx, label=0:$\overline{\nu^{C}_{L}}$] at (1,1) {};		
	\node[vtx, label=0:$\nu_{L}^{C}$] at (3,1) {};	
	\node[vtx, label=0:$r_{1,2}$] at (1.7,2.3) {};
	\node[vtx, label=0:$r_{1,2}^{*}$] at (1.7,0.3) {};
	\graph[use existing nodes]{
		a1--[antifermion]a2;
		b1--[fermion]b2;
		a3--[antifermion]a4;
		b3--[fermion]b4;
		a2--[fermion]b2;
		a3--[antifermion]b3;
		a2--[scalar]a3;
		b2--[scalar]b3
	};
\end{tikzpicture}
\end{minipage}
\hspace{0.1cm}
\begin{minipage}[c]{0.45\textwidth}
\centering
\begin{tikzpicture}[thick,scale=1]
	\coordinate[] (a1) at (0,2) {};
		\node[vtx, label=0:$\overline{s_{L}}$] at (-0.75,2) {};
	\coordinate[] (a2) at (1,2) {};
	\coordinate[] (a3) at (3,2) {};
	\coordinate[] (a4) at (4,2) {};
		\node[vtx, label=0:$\overline{d_{L}}$] at (4,2) {};

	\coordinate[] (b1) at (0,0) {};
	\node[vtx, label=0:$d_{L}$] at (-0.75,0) {};
	\coordinate[] (b2) at (1,0) {};
	\coordinate[] (b3) at (3,0) {};
	\coordinate[] (b4) at (4,0) {};
		\node[vtx, label=0:$s_{L}$] at (4,0) {};

	\node[vtx, label=0:$r_{1,2}$] at (1,1) {};		
	\node[vtx, label=0:$r_{1,2}^{*}$] at (3,1) {};	
	\node[vtx, label=0:$\overline{\nu^{C}_{L}}$] at (1.7,2.3) {};
	\node[vtx, label=0:$\nu_{L}^{C}$] at (1.7,0.3) {};
	\graph[use existing nodes]{
		a1--[antifermion]a2;
		b1--[fermion]b2;
		a3--[antifermion]a4;
		b3--[fermion]b4;
		a2--[scalar]b2;
		a3--[scalar]b3;
		a2--[fermion]a3;
		b2--[antifermion]b3
	};
\end{tikzpicture}
\end{minipage}\\
\begin{minipage}[l]{0.45\textwidth}
\centering
\begin{tikzpicture}[thick,scale=1]
	\coordinate[] (a1) at (0,2) {};
		\node[vtx, label=0:$\overline{s_{L}}$] at (-0.75,2) {};
	\coordinate[] (a2) at (1,2) {};
	\coordinate[] (a3) at (3,2) {};
	\coordinate[] (a4) at (4,2) {};
		\node[vtx, label=0:$\overline{d_{L}}$] at (4,2) {};

	\coordinate[] (b1) at (0,0) {};
	\node[vtx, label=0:$d_{L}$] at (-0.75,0) {};
	\coordinate[] (b2) at (1,0) {};
	\coordinate[] (b3) at (3,0) {};
	\coordinate[] (b4) at (4,0) {};
		\node[vtx, label=0:$s_{L}$] at (4,0) {};

	\node[vtx, label=0:$\overline{e^{C}_{L}}$] at (1,1) {};		
	\node[vtx, label=0:$e_{L}^{C}$] at (3,1) {};	
	\node[vtx, label=0:$S_{3}^{4/3}$] at (1.7,2.3) {};
	\node[vtx, label=0:$S_{3}^{4/3*}$] at (1.7,0.3) {};
	\graph[use existing nodes]{
		a1--[antifermion]a2;
		b1--[fermion]b2;
		a3--[antifermion]a4;
		b3--[fermion]b4;
		a2--[fermion]b2;
		a3--[antifermion]b3;
		a2--[scalar]a3;
		b2--[scalar]b3
	};
\end{tikzpicture}
\end{minipage}
\hspace{0.1cm}
\begin{minipage}[c]{0.45\textwidth}
\centering
\begin{tikzpicture}[thick,scale=1]
	\coordinate[] (a1) at (0,2) {};
		\node[vtx, label=0:$\overline{s_{L}}$] at (-0.75,2) {};
	\coordinate[] (a2) at (1,2) {};
	\coordinate[] (a3) at (3,2) {};
	\coordinate[] (a4) at (4,2) {};
		\node[vtx, label=0:$\overline{d_{L}}$] at (4,2) {};

	\coordinate[] (b1) at (0,0) {};
	\node[vtx, label=0:$d_{L}$] at (-0.75,0) {};
	\coordinate[] (b2) at (1,0) {};
	\coordinate[] (b3) at (3,0) {};
	\coordinate[] (b4) at (4,0) {};
		\node[vtx, label=0:$s_{L}$] at (4,0) {};

	\node[vtx, label=0:$S_{3}^{4/3}$] at (1,1) {};		
	\node[vtx, label=0:$S_{3}^{4/3*}$] at (3,1) {};	
	\node[vtx, label=0:$\overline{e^{C}_{L}}$] at (1.7,2.3) {};
	\node[vtx, label=0:$e_{L}^{C}$] at (1.7,0.3) {};
	\graph[use existing nodes]{
		a1--[antifermion]a2;
		b1--[fermion]b2;
		a3--[antifermion]a4;
		b3--[fermion]b4;
		a2--[scalar]b2;
		a3--[scalar]b3;
		a2--[fermion]a3;
		b2--[antifermion]b3
	};
\end{tikzpicture}
\end{minipage}
\caption{Dominant contributions to $K-\overline{K}$ mixing.}
\label{Fig: kaon mixing.}
\end{figure}
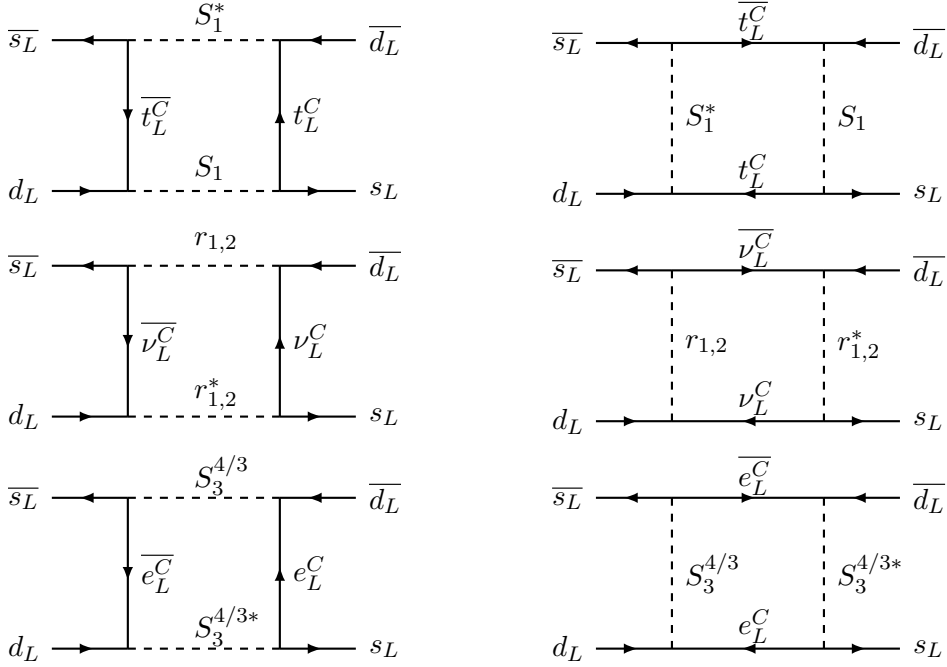}%
\noindent As can be seen in Figure \ref{Fig: kaon mixing.}, which depicts the NP Feynman diagrams for kaon mixing, our model contributes to meson mixing through box diagrams with both diquark and leptoquark propagators. When considering the contributions from the diquark, $S_{1}$, we only include contributions which include the top quark as a propagator, just as in the SM calculation. This is because the CKM matrix elements involving the top quark dominate over the others. Leptoquark contributions occur through box diagrams with either neutrinos and $r_{1,2}$ or charged leptons and $S_{3}^{4/3}$. Summing over all contributions, we have the following BSM Wilson coefficient for meson mixing 
\begin{equation} \begin{split}
C^{ij}_{1,{\rm NP}} =&- \frac{(zV_{\text{CKM}}^{\dagger})^{2}_{3i}(V_{\text{CKM}}^{*}z)^{2}_{3j}}{96\pi^{2}m_{S_{1}}^{2}}  -  \frac{1}{96\pi^{2}}\left(\sum_{k}\frac{y_{ik}y_{jk}}{m_{S_{3}^{4/3}}\pi} \right)^{2}\\
 &- \frac{1}{384\pi^2} \left(\sum_{k}y_{ik}y_{jk} \right)^{2}\left[\left(\frac{\cos{\theta}}{m_{r_{1}}}\right)^{2}+ \left(\frac{\sin{\theta}}{m_{r_{2}}}\right)^{2}+ \frac{2\sin{\theta}\,  \cos{\theta}}{m_{r_{1}}^{2}-m_{r_{2}}^{2}}\mathrm{ln} \frac{m_{r_{1}}^{2}}{m_{r_{2}}^{2}} \right] .           
\end{split} \end{equation} 
\\\\
The effective operators and Wilson coefficients depend on the renormalisation scheme and scale. However, since we are only interested in an order of magnitude estimate, we neglect the running from the scale of new physics to the top quark mass and take the ratio of the new physics contribution to the SM contribution. This ratio is independent of QCD running and is simply a ratio of the respective Wilson coefficients,
\begin{equation} \begin{split}
\frac{\bra{M}\mathcal{H}^{\rm NP}_{\rm eff}\ket{\overline{M}}}{\bra{M}\mathcal{H}^{\rm SM}_{\rm eff}\ket{\overline{M}}} = \frac{C_{1,{\rm NP}}^{ij}}{C_{1,{\rm SM}}^{ij}}.
\end{split} \end{equation}
\\\\
The UT\textit{fit} collaboration has published model-independent constraints on $\Delta F = 2$ operators \cite{Bona:2007vi}, with the results from the latest fit published in \cite{UTFitweb}:
{\begin{subequations}
\begin{equation}
\begin{split}
C_{B_{q}}e^{2i\phi_{B_{q}}}&=\frac{\bra{B_{q}}\mathcal{H}_{\rm eff}^{\rm full}\ket{\overline{B}_{q}}}{\bra{B_{q}}\mathcal{H}_{\rm eff}^{\rm SM}\ket{\overline{B}_{q}}} = 1 + \frac{\bra{B_{q}}\mathcal{H}_{\rm eff}^{\rm NP}\ket{\overline{B}_{q}}}{\bra{B_{q}}\mathcal{H}_{\rm eff}^{\rm SM}\ket{\overline{B}_{q}}}
\label{eqn: wilson B meson mixing}
\end{split} 
\end{equation} 
\begin{equation}
\begin{split}
C_{\Delta m_{K}}&=\frac{\mathrm{Re}[\bra{K}\mathcal{H}_{\rm eff}^{\rm full}\ket{\overline{K}}]}{\mathrm{Re}[\bra{K}\mathcal{H}_{\rm eff}^{\rm SM}\ket{\overline{K}}]} = 1 +\frac{\mathrm{Re}[\bra{K}\mathcal{H}_{\rm eff}^{\rm NP}\ket{\overline{K}}]}{\mathrm{Re}[\bra{K}\mathcal{H}_{\rm eff}^{\rm SM}\ket{\overline{K}}]}
\label{eqn: wilson kaon mixing real part}
\end{split} 
\end{equation} 
\begin{equation}
\begin{split}
C_{\Delta \epsilon_{K}}&=\frac{\mathrm{Im}[\bra{K}\mathcal{H}_{\rm eff}^{\rm full}\ket{\overline{K}}]}{\mathrm{Im}[\bra{K}\mathcal{H}_{\rm eff}^{\rm SM}\ket{\overline{K}}]} = 1 +\frac{\mathrm{Im}[\bra{K}\mathcal{H}_{\rm eff}^{\rm NP}\ket{\overline{K}}]}{\mathrm{Im}[\bra{K}\mathcal{H}_{\rm eff}^{\rm SM}\ket{\overline{K}}]}. \label{eqn: wilson kaon mixing imaginary part}
\end{split} 
\end{equation} 
\end{subequations}}%
The current best fit values for these parameters are given by \cite{UTFitweb}
{\begin{equation} \begin{split}
\begin{tabular}{lll}
$C_{B_{d}}=1.03 \pm 0.11$, &\phantom{blankspaces} & $\phi_{B_{d}}= -1.8^{\circ} \pm	1.7^{\circ}$,\\
$C_{B_{s}}= 1.070\pm 0.088$, & \phantom{blankspaces}& $\phi_{B_{s}}= 0.054^{\circ} \pm 0.951^{\circ}$,\\
$C_{\Delta m_{K}}= 0.978 \pm 0.331$,& \phantom{blankspaces}& $C_{\epsilon_{K}}=1.04 \pm 0.11$.
\end{tabular}
\label{eqn: UTfit values}
\end{split} \end{equation} }%
As can be seen in Equation \ref{eqn: UTfit values}, $\phi_{B_{d}}$ and $\phi_{B_{s}}$ are not measured precisely, thus in placing constraints on B--meson mixing we simply require that the magnitude of the right-handed side of Equation \ref{eqn: wilson B meson mixing} be within the one standard deviation range of $C_{B_{d}}$ and $C_{B_{s}}$, as per the values quoted in Equation \ref{eqn: UTfit values}. Specifically
{\begin{equation} \begin{split}
0.92 < \left|1 + \frac{C_{1,{\rm NP}}^{13}}{C_{1,{\rm SM}}^{13}}\right| < 1.14
\end{split} \end{equation} }%
for $B_{d}$ mixing, and 
{\begin{equation} \begin{split}
0.982 < \left|1 + \frac{C_{1,{\rm NP}}^{23}}{C_{1,{\rm SM}}^{23}}\right| < 1.158
\end{split} \end{equation} }%
for $B_{s}$ mixing. Concerning the bounds placed on couplings involved in kaon mixing, only the imaginary part of the ratio of NP to SM Wilson coefficients, seen in Equation \ref{eqn: wilson kaon mixing imaginary part}, was constrained\footnote{The SM contribution for $\Delta m_{K}$ cannot be measured reliably, thus this parameter was not used to constrain the NP contributions to kaon mixing.}:
\begin{equation} \begin{split}
\frac{C_{1,{\rm NP}}^{ij}}{C_{1,{\rm SM}}^{ij}} < 0.15.
\end{split} \end{equation} 
\subsection{Solving the $R_{K^{(*)}}$ anomaly}
\label{Rk anomalies}
While not the focus of this paper, another key feature of our model is its ability to explain the $R_{K{(*)}}$ flavour anomalies due to the presence of the $S_3$ scalar leptoquark. The $R_{K^{(*)}}$ flavour anomalies are a set of deviations from SM predictions in the decays of B-mesons. The anomalous quantities are the ratios of branching fractions 
\begin{equation} \begin{split}
    R_{K^{(*)}}= \frac{\text{Br}(\overline{B} \rightarrow \overline{K}^{(*)}\mu \mu)}{\text{Br}(\overline{B} \rightarrow \overline{K}^{(*)}ee)}.
\end{split} \end{equation} 
The SM prediction of $R_{K^{(*)}} =1.0003 \pm 0.0001$  \cite{Bobeth:2007dw} is close to unity due to lepton flavour universality, with the only difference in the measured branching fractions coming from their dependence on the masses of the final state leptons. Most recently, the LHCb collaboration has updated the measurement of $R_K$ by combining Run-1 data with 2 ${\rm fb}^{-1}$ of Run-2 data \cite{Aaij:2019wad}. They found
\begin{equation} \begin{split}
R_K = 0.846^{+0.060~+ 0.016}_{-0.054~-0.014},~~~~~~~q^{2} = [1.1, 6]\ \mathrm{GeV}^2,
\end{split} \end{equation} 
where $q^2$ is the dilepton invariant mass squared, the first uncertainty is statistical and the second is systematic. This result continues to be in tension with theory at $2.5 \sigma.$ The $R_{K^*}$ results have also been updated, with preliminary measurements by Belle \cite{Abdesselam:2019wac} of 
\begin{equation} \begin{split}
R_{K^{*}}&=
\begin{dcases}
0.90^{+0.27}_{-0.21}\pm 0.10~~q^{2} = [0.1, 8]\ \mathrm{GeV}^{2}\\
1.18^{+0.52}_{-0.32}\pm 0.10~~q^{2} = [15, 19]\ \mathrm{GeV}^{2},
\end{dcases}
\end{split} \end{equation} 
in agreement with the existing measurements by LHCb \cite{Aaij:2017vbb} of
\begin{equation} \begin{split}
R_{K^{*}}&=
\begin{dcases}
0.660^{+0.110}_{-0.070}\pm 0.024~~q^{2} = [0.045, 1.1]~\mathrm{GeV}^{2}\\
0.685^{+0.113}_{-0.069}\pm 0.047~~q^{2} = [1.1, 6]~\mathrm{GeV}^{2}.
\end{dcases}
\end{split} \end{equation} 
Interestingly, the large uncertainties in the Belle measurement allow it to be in agreement with both the SM and the LHCb measurement, which deviates from the SM prediction by $\sim 2.5 \sigma$. \\\\
Our model contributes to $R_{K^{(*)}}$ via semi-leptonic B-meson decays mediated by $S_3$. The $b\rightarrow sll$ transition can be described by the effective Hamiltonian 
\begin{equation} \begin{split}
\mathcal{H}_{\rm eff} = -\frac{4G_{F}}{\sqrt{2}}(V_{\text{CKM}})_{33}(V_{\text{CKM}})_{32}^{*}\frac{\alpha_{e}}{4\pi}\sum_{i} C_{i}(\mu)O_{i}(\mu),
\end{split} \end{equation} 
where $\alpha_e$ is the fine structure constant. The relevant effective operator for our model, presented here in the chiral basis, is 
\begin{equation} \begin{split}
    O^{\mu \mu}_{LL} = (\overline{\mu}\gamma_\nu P_L \mu)(\overline{s}\gamma_\nu P_L b). 
\end{split} \end{equation} 
Reference \cite{Aebischer:2019mlg} argues that contributions to the effective operator $O_{LL}^{\mu \mu}$ with Wilson coefficient 
\begin{equation} \begin{split}
C^{\mu \mu}_{LL} = \frac{\pi}{\alpha_{e}}\frac{y_{22}y_{32}}{(V_{\text{CKM}})_{33}(V_{\text{CKM}})^{*}_{32}}\frac{\sqrt{2}}{2m_{S_{3}^{4/3}}^{2}G_{F}} \approx -0.53 \pm 0.08 (0.16),
\end{split} \end{equation} 
ameliorates the $R_{K^{(*)}}$ anomalies, with the uncertainties giving the $1 (2)\sigma$ bounds. In our model, this corresponds to a central value of 
\begin{equation} \begin{split}
    y_{22}y_{32} \approx \Big(\frac{m_{S_3^{4/3}}}{35~\mathrm{TeV}}\Big)^2.
\end{split} \end{equation} 
\section{Results}
\label{sec:results}
We will now present the results of our random parameter scans, and discuss the predictions and limitations of the model. The discussion is broken up into three sections, each discussing one of the three leptoquark coupling matrix textures introduced in Table \ref{table: parameters}. The strongest constraint on the model comes from $\mu \rightarrow e$ conversion in gold nuclei,  which is mediated by leptoquark $S_3$ at tree level. Consequently, we will also explore the potential consequences for the model if future $\mu \rightarrow e$ experiments fail to measure a signal at the promised prospective sensitivities. \\\\
The scans were performed over leptoquark couplings $y_{11},$ $y_{12}$, $y_{21}$ and $y_{22}$, and the leptoquark mass parameter $\mu_{S_3}$, with all other couplings fixed. In 2018 the CMS experiment at the LHC set an exclusion on diquark masses below $7.2$ TeV at $95\%$ confidence interval \cite{Sirunyan:2018xlo}. While the exclusion was calculated for a particular diquark which features in superstring inspired $E_6$ models \cite{HEWETT1989193}, similar bounds are expected to apply to other diquarks. The CMS experiment has also set limits on leptoquarks masses, with masses below $1.1$ TeV being excluded at $95 \%$ confidence interval for third generation leptoquarks decaying to $b\nu$ \cite{Sirunyan:2018kzh}. We are not aware of current limits on exotic scalars that only couple to other scalars and gauge bosons, such as $\phi_3$. With these considerations in mind, we stick to conservative lower bounds that have the potential to be directly probed at the LHC and indirectly probed at precision- or luminosity-frontier experiments. We fix the diquark mass parameter in our model to a benchmark value of $\mu_{S_1}= 7.5$ TeV, the mass parameter for the scalar $\phi_3$ to $\mu_{\phi_3}= 1.5$ TeV, and scan over the leptoquark mass parameter such that $1.1 \leq \mu_{S_3} \leq 10~\text{TeV}$. The limits on leptoquark masses are set assuming sufficiently large leptoquark couplings of $y \geq 10^{-7}$ to guarantee prompt decay of leptoquarks in the detector. We also found that due to the relationship between $y_3^{LL}$ and $z_1^{LL}$ displayed in Equation \ref{eqn: casas Ibarra for z}, leptoquark couplings below $10^{-5}$ had a high likelihood of requiring large, non-perturbative diquark couplings in order to guarantee the desired neutrino masses. Thus, the free leptoquark couplings: $y_{11},$ $y_{12}$, $y_{21}$ and $y_{22}$ are allowed to vary between $10^{-5}$ and $1$. The other five leptoquark couplings are set to benchmark values. We investigate three leptoquark matrix textures, as indicated in Table \ref{table: parameters} and detailed below.
\newpage
\subsection{Texture A}
\begin{figure}[!t]
\centering
\includegraphics[width=0.9\linewidth]{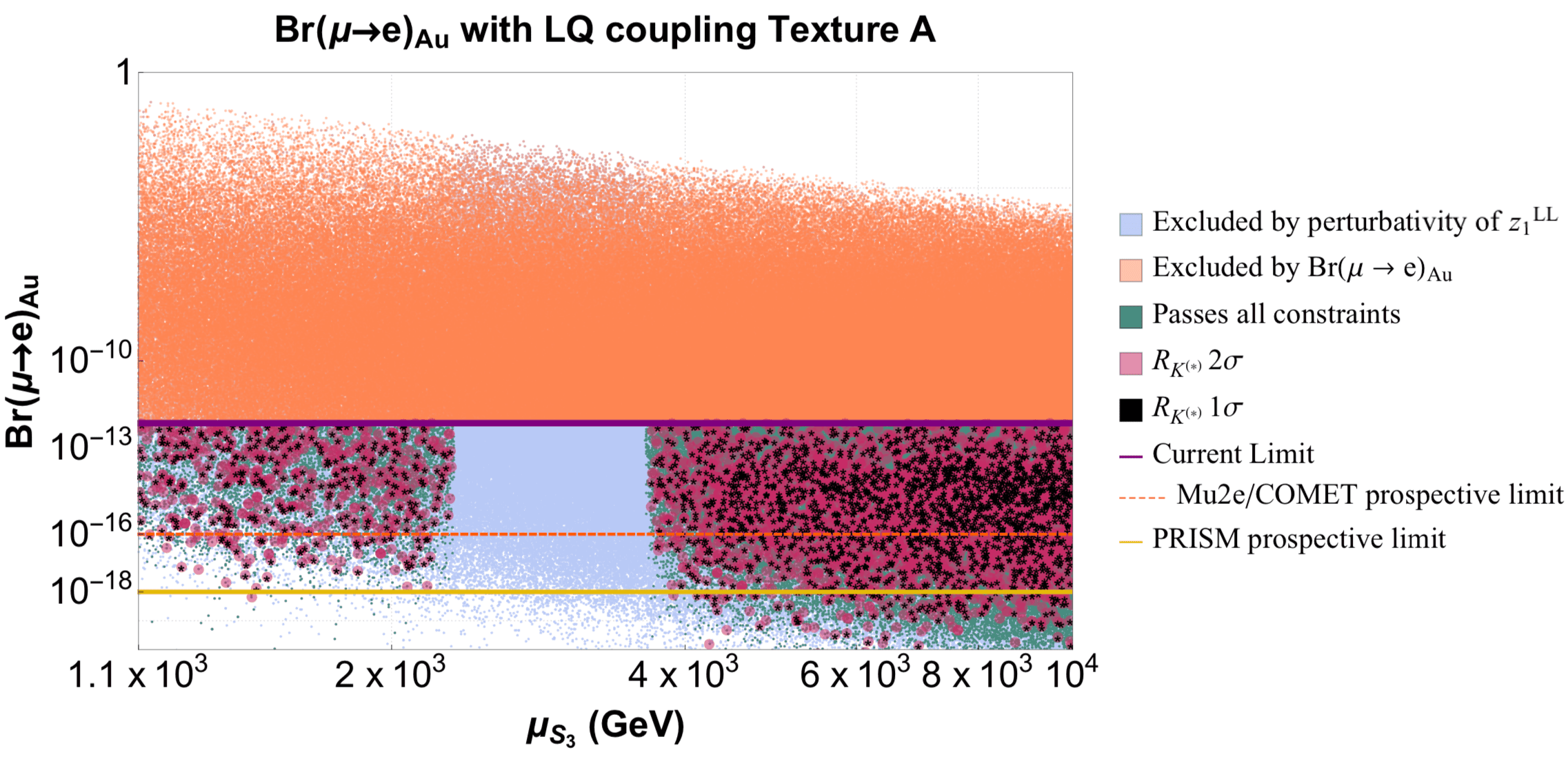}
\caption{Indicative plot of the allowed parameter space scans showing $\mu_{S_3}$ vs. $\text{Br}(\mu~\rightarrow~e)_{\text{Au}}$, for leptoquark coupling matrix Texture A. The purple line indicates the current $\text{Br}(\mu~\rightarrow~e)_{\text{Au}}$ cut-off of $7.0 \times 10^{-13}$, and the dotted orange (yellow) line indicates the prospective cut-off from the Mu2e/COMET (PRISM) experiment which is of order $10^{-16}$ ($10^{-18}$). The orange region of the parameter space is constrained by $\text{Br}(\mu~\rightarrow~e)_{\text{Au}}$, the teal region is allowed; specifically it passes all constraints discussed in Section \ref{Sec: Constraints} as well as perturbativity constraints on the diquark coupling constants $z^{LL}_{1ij}$. The light blue region is excluded by the perturbativity constraints on diquark couplings. The black (pink) region solves the $R_{K^{(*)}}$ anomalies to $1(2)\sigma$.}
\label{Fig: parameter muS3 vs mutoe Texture A}
\end{figure}
\begin{figure}[!t]
\centering
\includegraphics[width=0.9\linewidth]{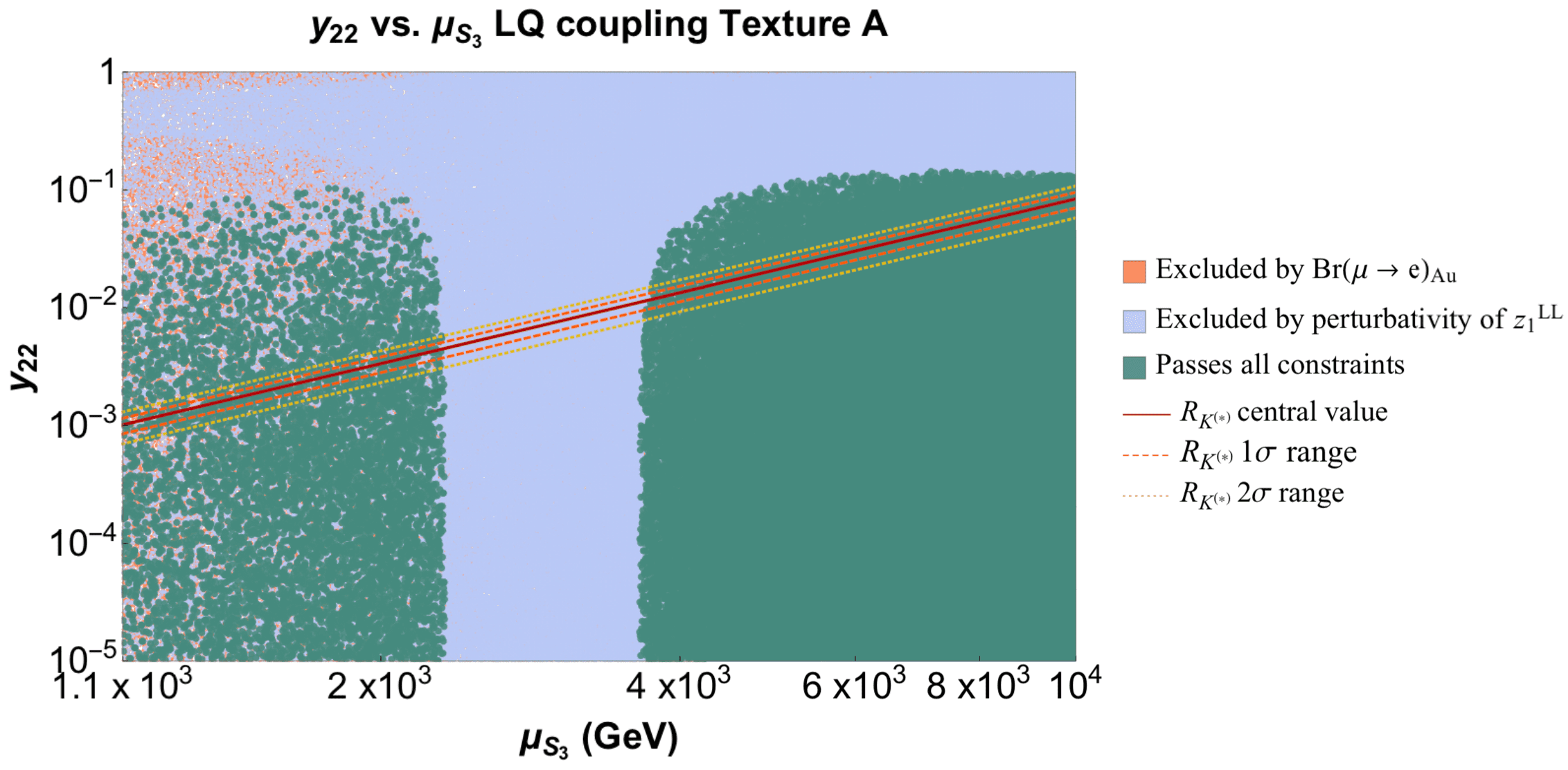}
\caption{Indicative plot of the allowed parameter space in the $y_{22}$ vs. $\mu_{S_3}$ plane, for leptoquark coupling matrix Texture A. The red line indicates the central value needed to explain the $R_{K^{(*)}}$ anomalies with the dotted orange (yellow) lines indicating the $1(2)~\sigma$  bounds. The orange region of the parameter space is constrained by $\text{Br}(\mu \rightarrow e)_{\text{Au}}$, the teal region is allowed; specifically it passes all constraints discussed in Section \ref{Sec: Constraints} as well as perturbativity constraints on the diquark coupling constants $z^{LL}_{1ij}$. The light blue region is excluded by perturbativity constraints on diquark couplings.}
\label{Fig: parameter muS3 vs y22 Texture A}
\end{figure}
\begin{figure}[!b]
\centering
\includegraphics[width=0.9\linewidth]{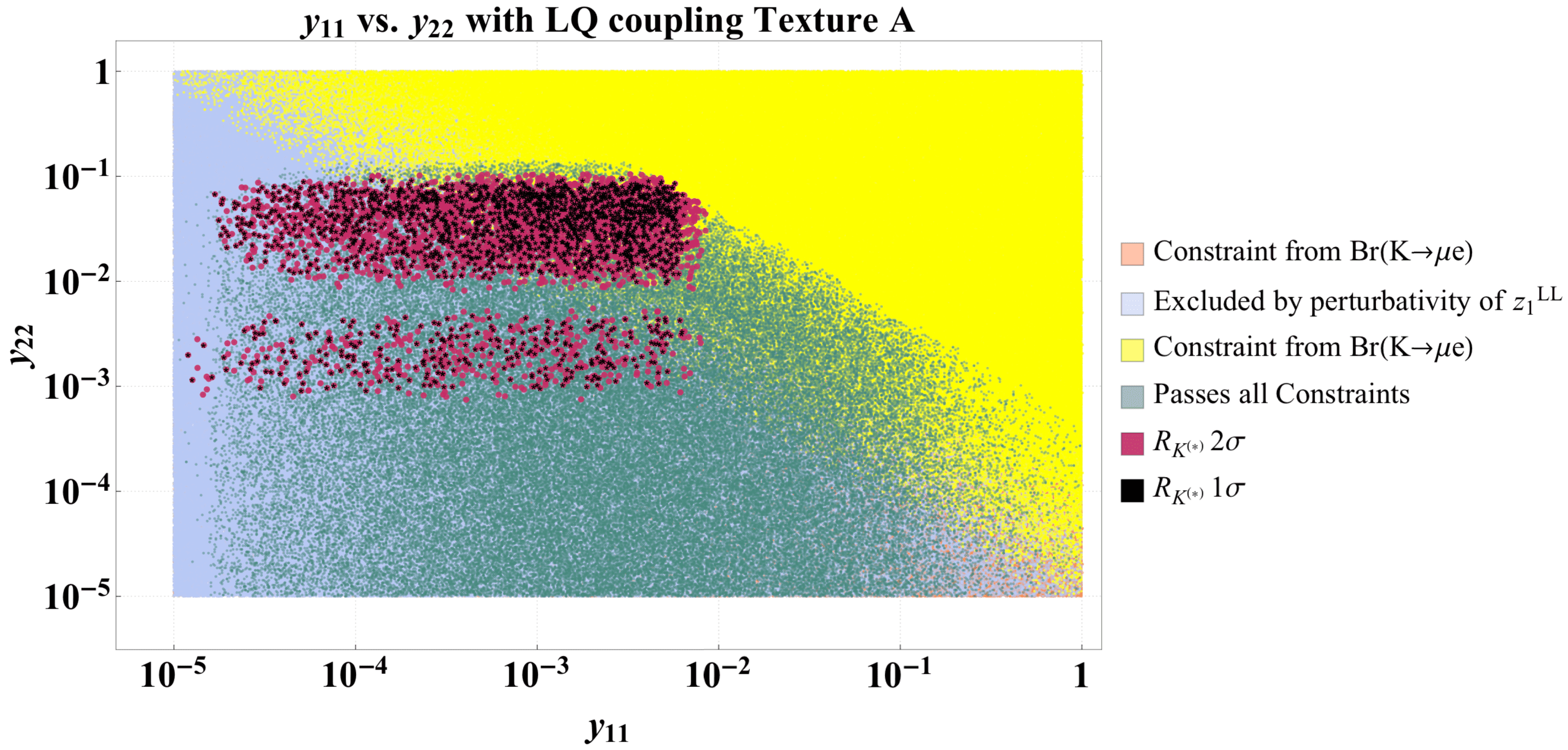}
\caption{Indicative plot of the allowed parameter space in the $y_{22}$ vs. $y_{11}$ plane, for leptoquark coupling matrix Texture A. The yellow section is ruled out by $\text{Br}(K \rightarrow \mu e)$, the orange section is ruled out by $\text{Br}(\mu \rightarrow e)_{\text{Au}}$, while the teal region is allowed parameter space; specifically it passes all constraints discussed in Section \ref{Sec: Constraints} as well as perturbativity constraints on the diquark coupling constants $z^{LL}_{1ij}$. The light blue region is excluded by perturbativity constraints on diquark couplings.  }
\label{Fig: parameter y11 vs y22 Texture A}
\end{figure}

The leptoquark matrix texture investigated first is Texture A 
\begin{equation}
    \begin{split}
y=\begin{pmatrix}
y_{11}&y_{12}&\epsilon_A
\\
y_{21}&y_{22}&5\epsilon_A\\
\epsilon_A&1&10\epsilon_A
\end{pmatrix},
    \end{split}
\end{equation}
with $\epsilon_A = 10^{-5}$.
If leptoquark couplings to third generation quarks and leptons are very weak, as they are for Texture A, ample parameter space is available when accounting for current constraints, including regions which allow the model to explain the $R_{K^{(*)}}$ flavour anomalies, as summarised in Figures \ref{Fig: parameter muS3 vs mutoe Texture A}, \ref{Fig: parameter muS3 vs y22 Texture A} and \ref{Fig: parameter y11 vs y22 Texture A}. In fact, with leptoquark couplings of the order of $\sim 10^{-5}$, this neutrino mass model is able to explain $R_{K^{(*)}}$ at a $1 \sigma$ level even if NP is not discovered by future experiments, such as Mu2e/COMET and PRISM. This is evident in Figure \ref{Fig: parameter muS3 vs mutoe Texture A}, where viable parameter space is plotted in teal, points that solve the $R_{K^{(*)}}$ anomalies to $1(2)\sigma$ are black (pink) and the dotted orange (yellow) lines represent prospective constraints for $\text{Br}(\mu \rightarrow e)_{\text{Au}}$ from the Mu2e/COMET (PRISM) experiments.
\newpage
\FloatBarrier
\noindent Figures \ref{Fig: parameter muS3 vs mutoe Texture A} and \ref{Fig: parameter muS3 vs y22 Texture A} both contain a curious feature: a region of parameter space (in the form of a band in $\mu_{S_3}$) which is excluded due to perturbativity constraints placed on the diquark coupling constants. This feature can be understood as follows. Small leptoquark couplings lead to a higher probability of non-perturbative diquark couplings. This is a consequence of Equation \ref{eqn: casas Ibarra for z}, which parametrises the diquark couplings in terms of the leptoquark couplings to ensure the desired neutrino masses are computed. The value of $z_1^{LL}$ is also inversely proportional to the value of the integral. The curious feature of the band of excluded parameter space in the $\mu_{S_3} \sim 2-4$ TeV, displayed in Figures \ref{Fig: parameter muS3 vs mutoe Texture A} and \ref{Fig: parameter muS3 vs y22 Texture A}, corresponds to the sign change in the integral. This can be verified in Figure \ref{Fig: integral vs mS3}. Simply put, there exits a region of $\mu_{S_3}$ for which the value of the leptoquark coupling and the integral are sufficiently small, making the diquark couplings $z_{1ij}^{LL}$ too large to be perturbative.\\\\
The $R_{K^{(*)}}$ anomalies can be solved for a sub-region of the allowed parameter space, for $y_{11} \lesssim 10^{-2}$ and $10^{-3}\lesssim y_{22}\lesssim 10^{-1}$. There is a band around $y_{22} \simeq 10^{-2}$ for which the model cannot explain the $R_{K^{(*)}}$ anomalies, which corresponds to the non-perturbative band in Figure \ref{Fig: parameter muS3 vs y22 Texture A}. \\\\
Figure \ref{Fig: parameter y11 vs y22 Texture A} also shows that while $\mu \rightarrow e$ conversion is found to be the most constraining process for this model in general as it appears at tree level, there are other important signals. Since the neutrino mass of this model does not depend on quark masses, there is no reason to set couplings to first generation quarks to zero, meaning our model is also sensitive to probes involving first generation quarks, such as $K^0_L \rightarrow \bar{e}\mu$ and $K^+ \rightarrow \pi^+\bar{\nu}\nu$. The process $K^0_L \rightarrow \bar{e}\mu$ is the most constraining for our model, with Figure \ref{Fig: parameter y11 vs y22 Texture A} showing that the leptoquark couplings $y_{11}$ and $y_{22}$ cannot be simultaneously close to unity. There exists a trade off due largely to constraints coming from $\text{Br}(K \rightarrow \mu e)$. It should be noted that $\mu \rightarrow e$ conversion, shown in orange in Figures \ref{Fig: parameter y11 vs y22 Texture A} and \ref{Fig: parameter muS3 vs y22 Texture A}, is also strongly constraining in the regions excluded by perturbativity and $\text{Br}(\mu \rightarrow e)_{\text{Au}}$ constraints. 
\FloatBarrier
\subsection{Texture B}
\begin{figure}[!t]
\centering
\includegraphics[width=0.9\linewidth]{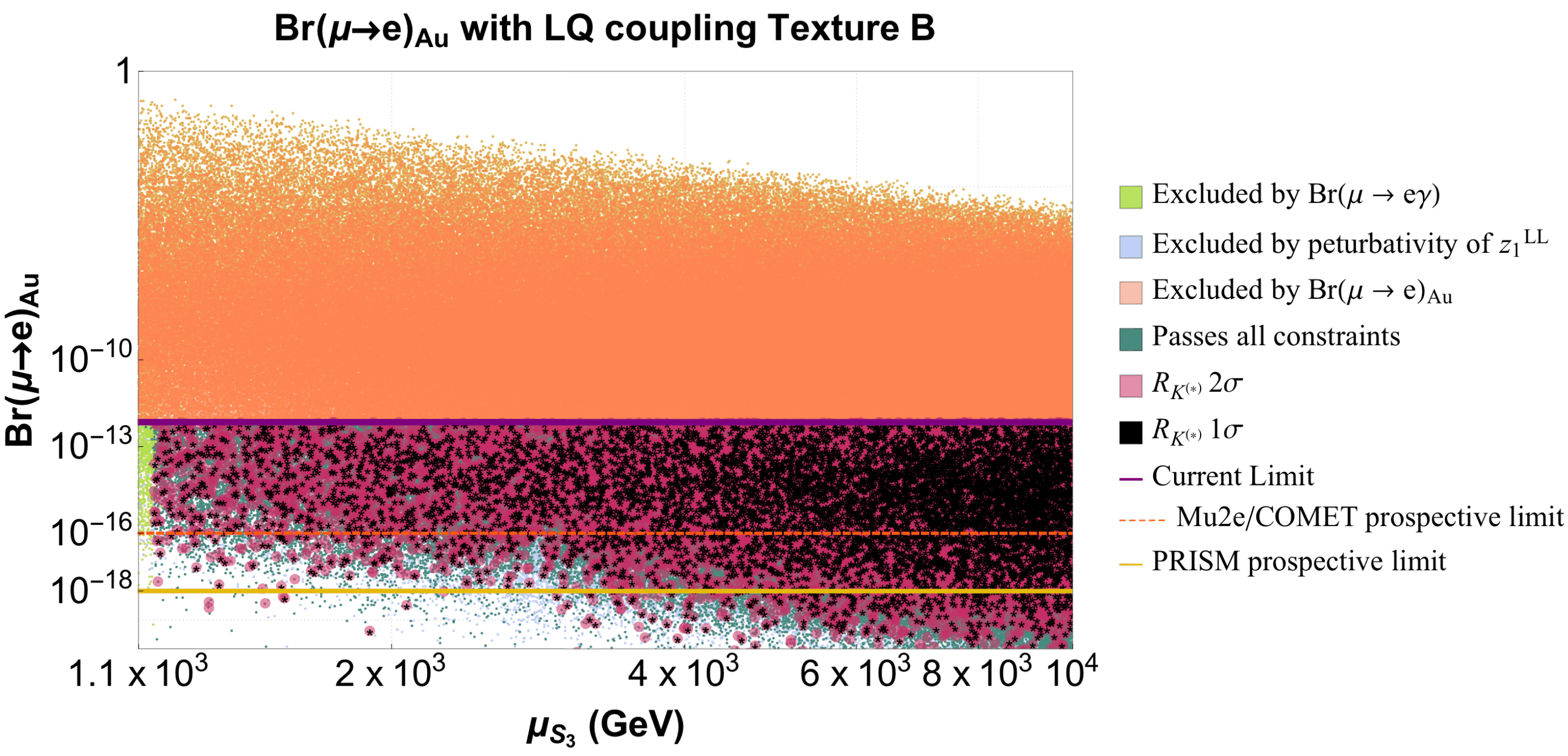}
\caption{Indicative plot of the allowed parameter space scans showing $\mu_{S_3}$ vs. $\text{Br}(\mu~\rightarrow~e)_{\text{Au}}$, for leptoquark coupling matrix Texture B. The purple line indicates the current $\text{Br}(\mu~\rightarrow~e)_{\text{Au}}$ cut-off, the dotted orange (yellow) line indicates the prospective cut-off from the Mu2e/COMET (PRISM) experiment. The light green region shows parameter space that is excluded by $\text{Br}(\mu~\rightarrow~e \gamma)$. The orange region of the parameter space is constrained by $\text{Br}(\mu~\rightarrow~e)_{\text{Au}}$, the teal region is allowed; specifically it passes all constraints discussed in Section \ref{Sec: Constraints} as well as perturbativity constraints on the diquark coupling constants $z^{LL}_{1ij}$. The black (pink) region solves the $R_{K^{(*)}}$ anomalies to $1(2)\sigma$.}
\label{Fig: parameter muS3 vs mutoe Texture B}
\end{figure}
\begin{figure}[!htbp]
\centering
\includegraphics[width=0.9\linewidth]{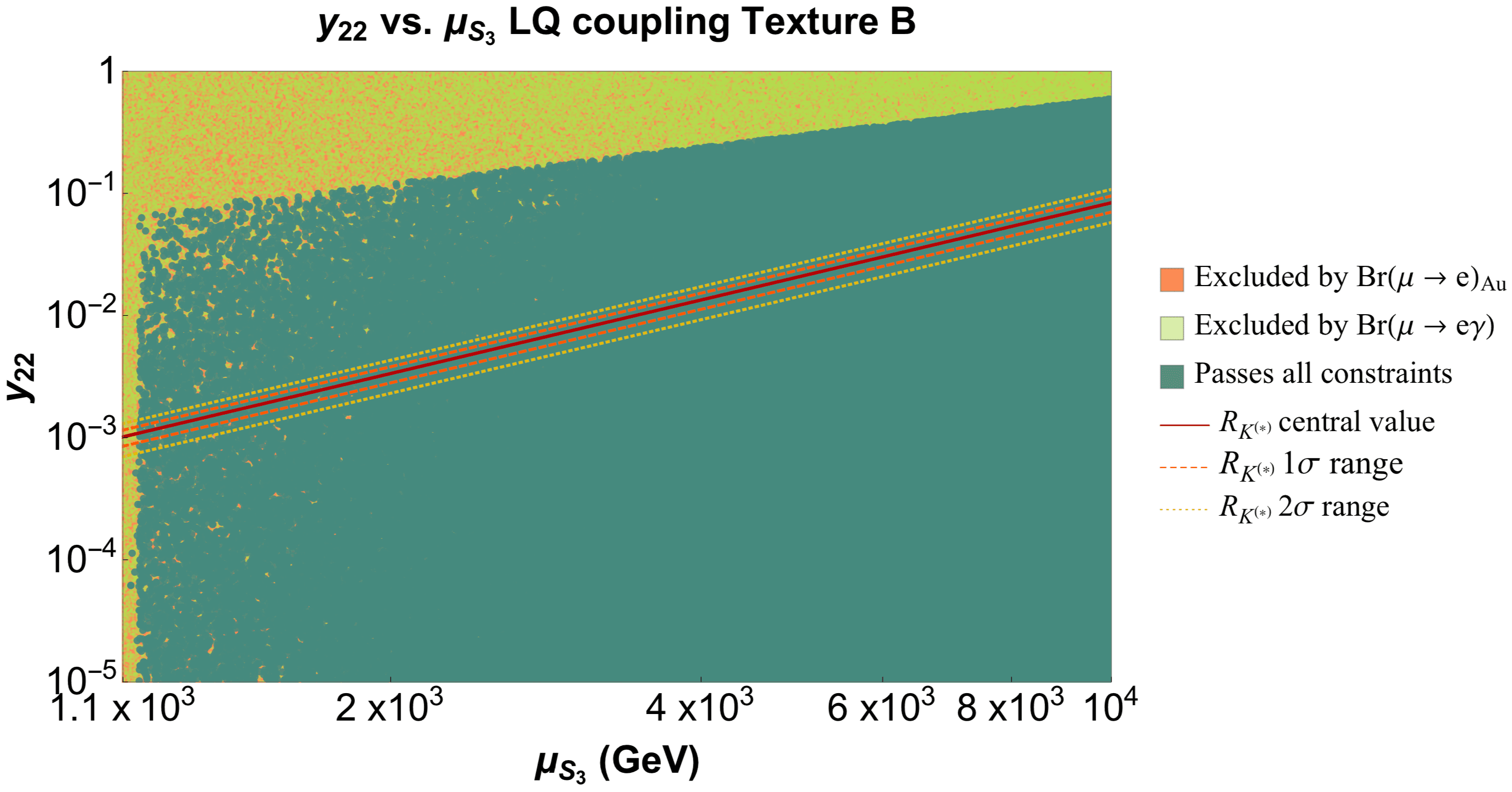}
\caption{Indicative plot of the allowed parameter space in the $y_{22}$ vs. $\mu_{S_3}$ plane, for leptoquark coupling matrix Texture B. The red line indicates the central value needed to explain the $R_{K^{(*)}}$ anomalies with the dotted orange (yellow) lines indicating the $1(2) \sigma$ bounds. The orange region of the parameter space is constrained by $\text{Br}(\mu~\rightarrow~e)_{\text{Au}}$, the teal region is allowed; specifically it passes all constraints discussed in Section \ref{Sec: Constraints} as well as perturbativity constraints on the diquark coupling constants $z^{LL}_{1ij}$. The light green region, which in this plot has been superimposed on the orange, shows parameter space that is excluded by $\text{Br}(\mu~\rightarrow~e \gamma)$. }
\label{Fig: parameter muS3 vs y22 Texture B}
\end{figure}
\begin{figure}[!htbp]
\centering
\includegraphics[width= 0.9\linewidth]{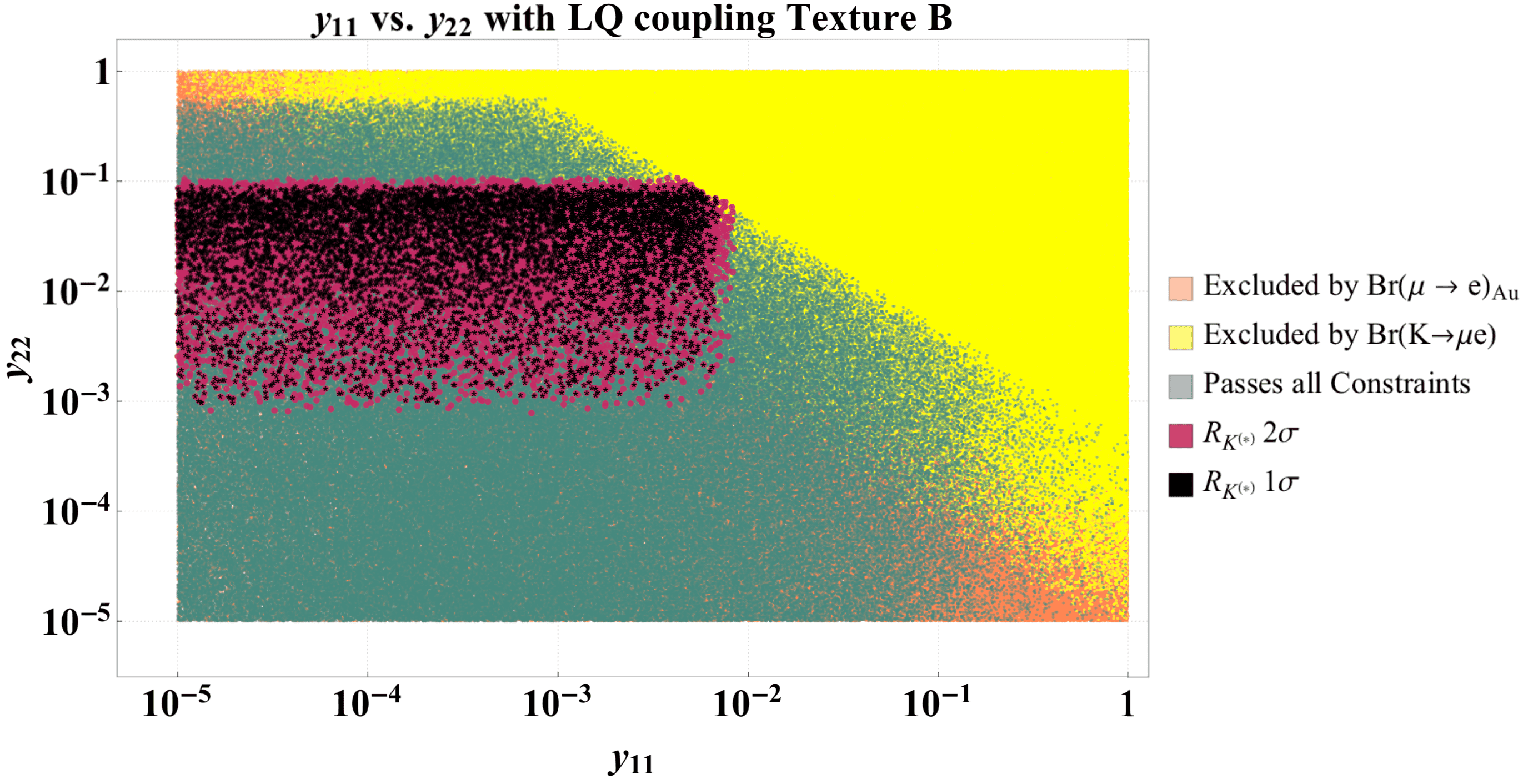}
\caption{Indicative plot of the allowed parameter space in the $y_{22}$ vs. $y_{11}$ plane, for leptoquark coupling matrix Texture B. The yellow section is ruled out by $\text{Br}(K \rightarrow \mu e)$, the orange section is ruled out by $\text{Br}(\mu \rightarrow e)_{\text{Au}}$, while the teal region is allowed parameter space; specifically it passes all constraints discussed in Section \ref{Sec: Constraints} as well as perturbativity constraints on the diquark coupling constants $z^{LL}_{1ij}$.}
\label{Fig: parameter y11 vs y22 Texture B}
\end{figure}

The leptoquark matrix texture considered next is Texture B
\begin{equation}
    \begin{split}
        y=\begin{pmatrix}
y_{11}&y_{12}&\epsilon_B
\\
y_{21}&y_{22}&5\epsilon_B\\
\epsilon_B&1&10\epsilon_B
\end{pmatrix},
    \end{split}
\end{equation}
where $\epsilon_B = 10^{-3}$. For this leptoquark matrix coupling texture, constraints due to the perturbativity of the diquark couplings are no longer a major concern. Texture B gives similar results to Texture A in that it is strongly constrained by $\text{Br}(\mu \rightarrow e)_{\text{Au}}$ and $\text{Br}(K \rightarrow \mu e)$, yet still able to solve the $R_{K^{(*)}}$ anomalies. In fact, with leptoquark couplings to third generation quarks of order $10^{-3}$, there exists parameter space which is able to solve the $R_{K^{(*)}}$ anomalies to $1 \sigma (2 \sigma)$ even if NP is not discovered by future experiments such as Mu2e/COMET and PRISM, as can be seen in Figure \ref{Fig: parameter muS3 vs mutoe Texture B} in black (pink). Figure \ref{Fig: parameter muS3 vs mutoe Texture B} also shows, in light green, that we now have a constraint from $\text{Br}(\mu \rightarrow e \gamma)$ on the leptoquark mass parameter $\mu_{S_3}$. \\\\
As for Texture A, leptoquark couplings of $y_{11} \lesssim 10^{-2}$ and $10^{-3}\lesssim y_{22}\lesssim 10^{-1}$ can currently solve the $R_{K^{(*)}}$ anomalies, as depicted in Figure \ref{Fig: parameter y11 vs y22 Texture B}. In can also be seen that $y_{22}$ must be less than unity, or the model is not viable at all due to constraints from $\mu \rightarrow e$ conversion. Constraints from $\text{Br}(K \rightarrow \mu e)$ enforces a trade off between large $y_{11}$ and $y_{22}$ couplings for Texture B. additionally, Figure \ref{Fig: parameter muS3 vs y22 Texture B} shows that $y_{22} \lesssim 10^{-1}$ is necessary for $\mu_{S_3} \simeq 1.2$ TeV.

\FloatBarrier
\subsection{Texture C}
\begin{figure}[!t]
\centering
\includegraphics[width=0.9\linewidth]{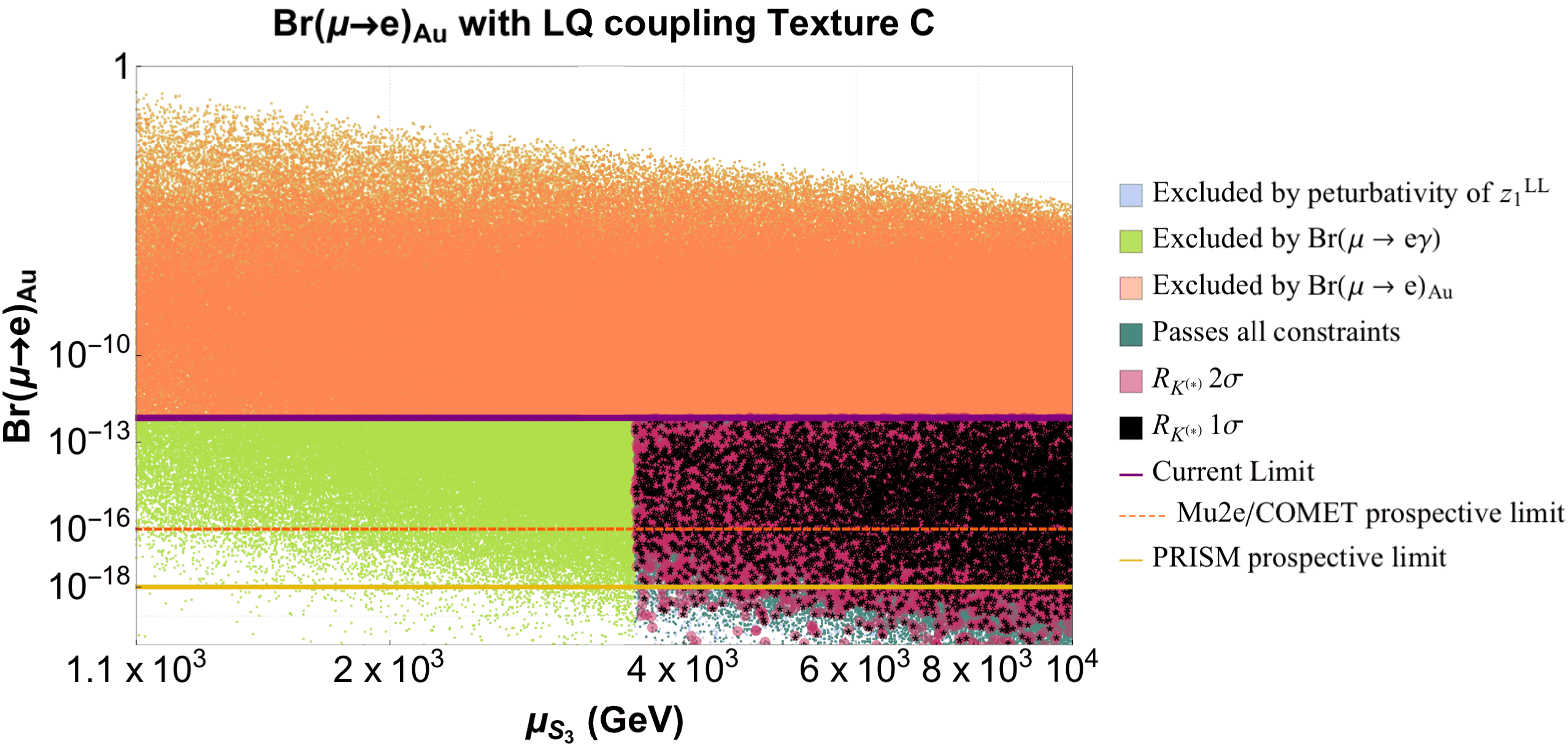}\\
\caption{Indicative plot of the allowed parameter space scans showing $\mu_{S_3}$ vs. $\text{Br}(\mu~\rightarrow~e)_{\text{Au}}$, for leptoquark coupling matrix Texture C. The purple line indicates the current $\text{Br}(\mu~\rightarrow~e)_{\text{Au}}$ cut-off, the dotted orange (yellow) line indicates the prospective cut-off from the Mu2e/COMET (PRISM) experiment. The light green region shows parameter space that is excluded by $\text{Br}(\mu~\rightarrow~e \gamma)$. The orange region of the parameter space is constrained by $\text{Br}(\mu~\rightarrow~e)_{\text{Au}}$, the teal region is allowed; specifically it passes all constraints discussed in Section \ref{Sec: Constraints} as well as perturbativity constraints on the diquark coupling constants $z^{LL}_{1ij}$. The black (pink) region solves the $R_{K^{(*)}}$ anomalies to $1(2)\sigma$.}
\label{Fig: parameter muS3 vs mutoe Texture C}
\end{figure}

\begin{figure}[!htbp]
\centering
\includegraphics[width=0.9\linewidth]{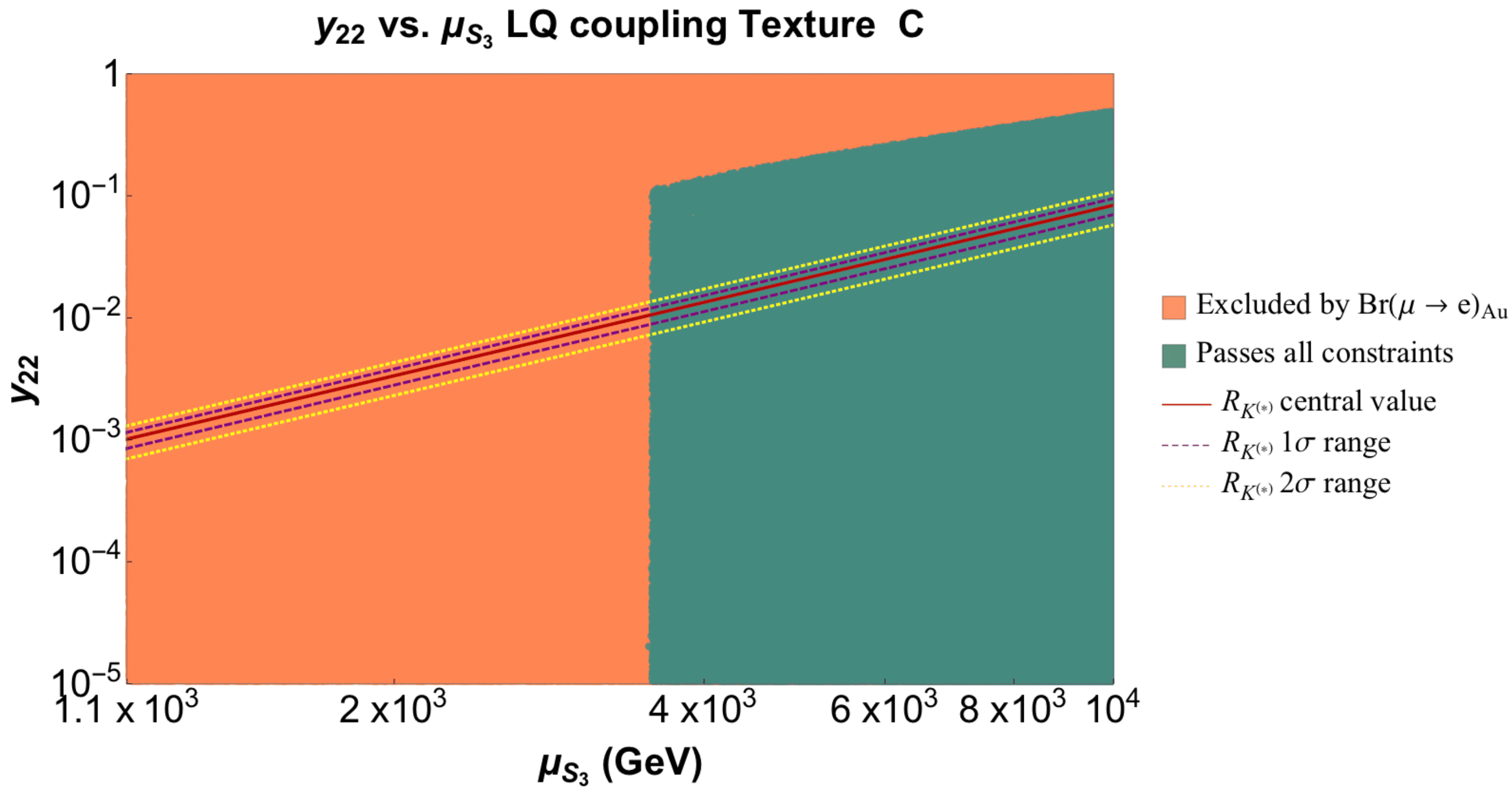}\\
\caption{Indicative plot of the allowed parameter space (in teal) in the $y_{22}$ vs. $\mu_{S_3}$ plane, for leptoquark coupling matrix Texture C. The red line indicates the central value needed to explain the $R_{K^{(*)}}$ anomalies with the purple (yellow) lines indicating the $1(2) \sigma$ bounds. Parameter space exluded by $\text{Br}(\mu \rightarrow e)$ is indicated in orange.}
\label{Fig: parameter muS3 vs y22 Texture C}
\end{figure}

\begin{figure}[!htbp]
\centering
\includegraphics[width=\linewidth]{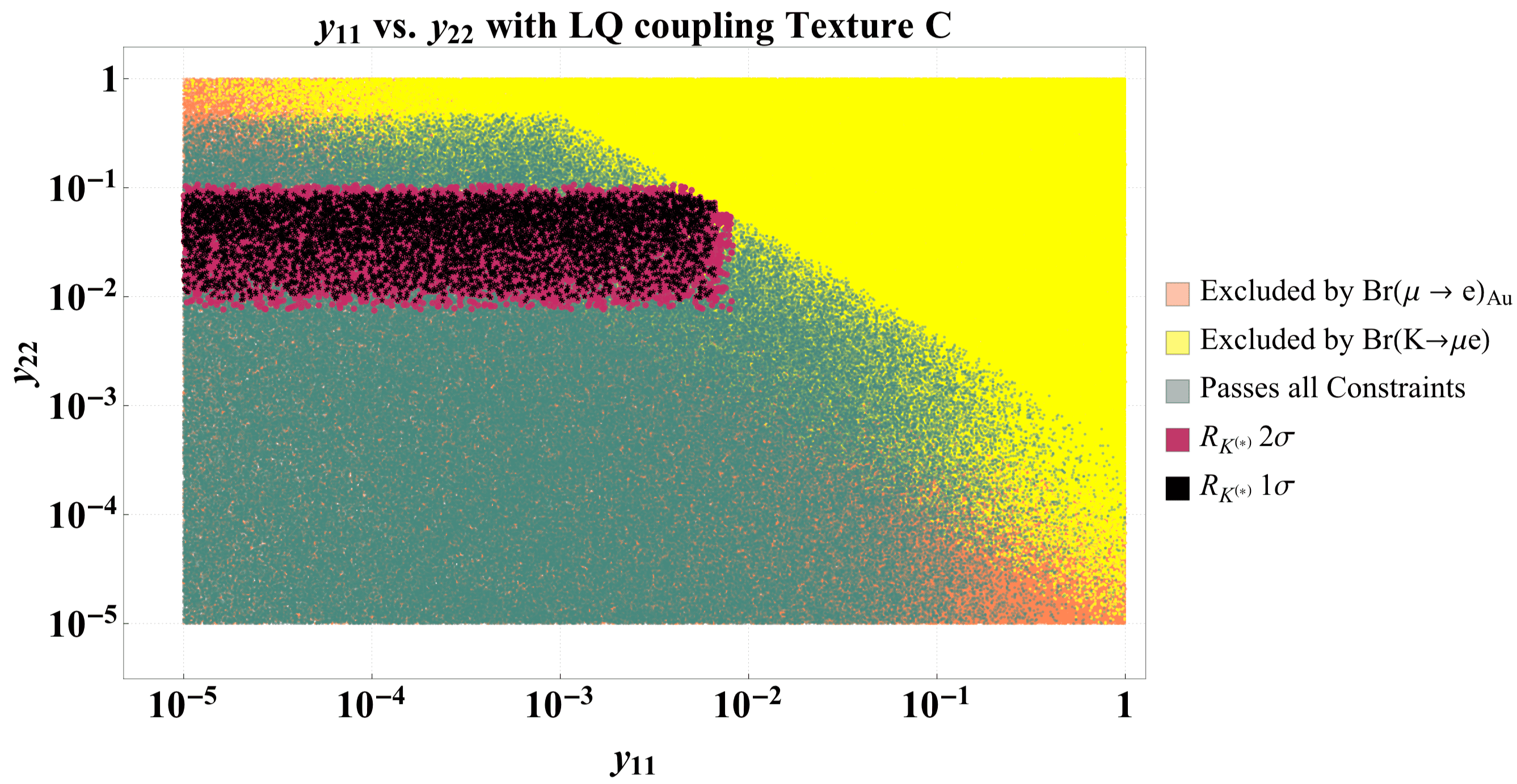}\\
\caption{Indicative plot of the allowed parameter space in the $y_{22}$ vs. $y_{11}$ plane, for leptoquark coupling matrix Texture C. The yellow section is ruled out by $\text{Br}(K \rightarrow \mu e)$, the orange section is ruled out by $\text{Br}(\mu \rightarrow e)_{\text{Au}}$, while the teal region is allowed parameter space; specifically it passes all constraints discussed in Section \ref{Sec: Constraints} as well as perturbativity constraints on the diquark coupling constants $z^{LL}_{1ij}$.}
\label{Fig: parameter y11 vs y22 Texture C}
\end{figure}

\begin{figure}[!htbp]
\centering
\includegraphics[width=\linewidth]{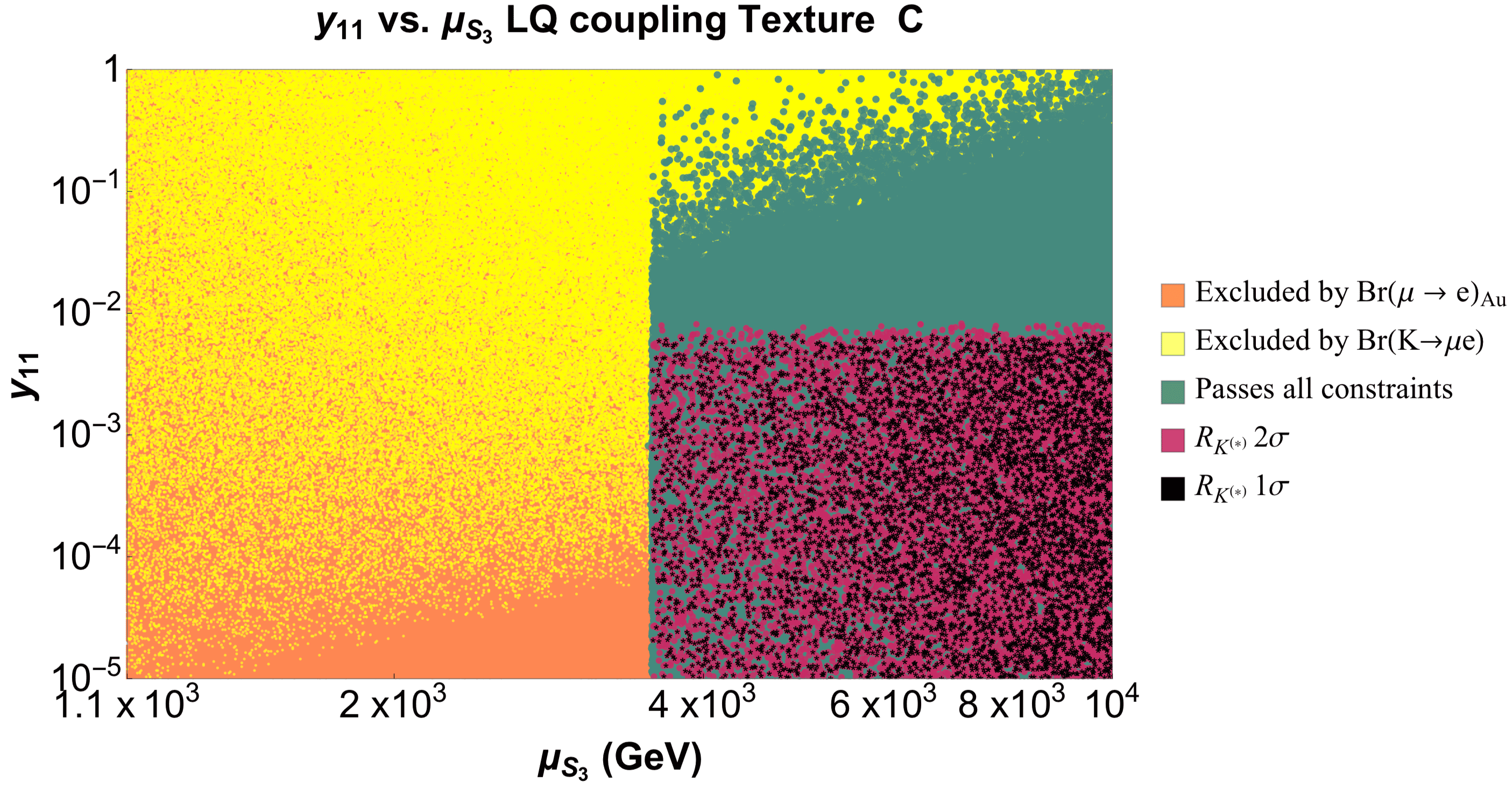}\\
\caption{Indicative plot of the allowed parameter space in the $y_{11}$ vs. $m_{S_3}$ plane, for leptoquark coupling matrix Texture C. The teal region shows allowed parameter space, while the orange region indicates parameter space excluded by constraints on $\text{Br}(\mu \rightarrow e)_{\text{Au}}$, while the yellow region show parameter space excluded by $\text{Br}(K \rightarrow \mu e)$. The black (pink) points indicate bounds on the parameters $y_{11}$ and $\mu_{S_3}$ needed to explain the $R_{K^{(*)}}$ anomalies to within $1(2) \sigma$.}
\label{Fig: parameter y11 vs mS3 Texture C}
\end{figure}

Finally, we consider leptoquark matrices of Texture C
\begin{equation}
    \begin{split}
y=\begin{pmatrix}
y_{11}&y_{12}&\epsilon_C^2
\\
y_{21}&y_{22}&\epsilon_C\\
\epsilon_C^2&1&1
\end{pmatrix},
    \end{split}
\end{equation}
where $\epsilon_C = 10^{-1}$. When couplings between electrons or muons and third generation quarks are $\geq 10^{-2}$ there is a strong constraint on the leptoquark mass parameter coming from $\text{Br}(\mu \rightarrow e \gamma)$, which excludes parameter space less than  $\mu_{S_3} \approx 4$ TeV, as seen in Figure \ref{Fig: parameter muS3 vs mutoe Texture C}. A similar bound in the $y_{22}$ versus $\mu_{S_3}$ slice of parameter space comes from the $\text{Br}(\mu \rightarrow e)_{Au}$ constraint. This can be seen in Figures \ref{Fig: parameter muS3 vs y22 Texture C} and \ref{Fig: parameter y11 vs y22 Texture C}. Constraints from $\text{Br}(K\rightarrow \mu e)$ still enforce that $y_{11}$ and $y_{22}$ cannot be simultaneously large, but it is the $\text{Br}(\mu \rightarrow e)_{\text{Au}}$ constraint that places a bound on the leptoquark couplings such that $y_{22}\lesssim 0.5$.  \\\\
The model can solve the $R_{K^{(*)}}$ anomalies to $1 (2) \sigma$ for $\mu_{S_3} \lesssim 4$ TeV, $y_{11} \lesssim 10^{-2}$ and $ 10^{-2}\lesssim y_{11} \lesssim 10^{-1}$, as can be seen in black (pink) in Figures \ref{Fig: parameter y11 vs y22 Texture C}, \ref{Fig: parameter muS3 vs mutoe Texture C} and \ref{Fig: parameter y11 vs mS3 Texture C}. Unsurprisingly, regardless of the leptoquark coupling matrix texture, constraints from processes involving first generation quarks and leptons, expecially $\text{Br}(\mu \rightarrow e) $ provide the strongest bounds on the parameter space of our model and would also be the most promising signals. 

\FloatBarrier
\section{Conclusion}
\label{Sec: Conclusion}
There are a large number of candidate radiative Majorana neutrino mass models. The $\Delta L = 2$ interactions responsible may be classified according to the dominant low-energy effective operators generated at tree-level from integrating out the massive exotic fields. The baryon-number conserving $\Delta L = 2$ operators occur at odd mass dimension, and an extensive list has been compiled up to mass dimension 11~\cite{explodingOperators}. An interesting question is: Beyond what mass dimension are phenomenologically viable models no longer possible? One generally expects that the higher the mass dimension, the higher also will be the number of vertices and loops in the neutrino self-energy graphs. Each additional vertex and loop contributes to additional suppression of the scale of neutrino mass, provided that the coupling constants at the vertices are small enough. There is also the prospect of suppression from powers of the ratio of electroweak scale to the new physics scale. At some point, the net suppression should become so strong that the $0.06$ eV lower bound on the neutrino mass scale cannot be generated for phenomenologically-acceptable exotic particle masses. Indeed, it has been argued that models constructed from opening up mass dimension 13 and higher operators are unlikely to be viable. Our findings in this paper, arrived at by a detailed examination of a specific model constructed from a dimension-11 operator, cast doubt on this tentative conclusion.\\\\
The basic reason is evident from Equation~\ref{eqn: O_{47} neutrino mass generic}: While there is a product of a few coupling constants in the numerator, and there is the $(1/16\pi^2)^2$ two-loop suppression factor, the mass suppression is only $v^2/\Lambda$, so identical with that of the usual seesaw models. Formally, this is due to the neutrino mass diagram generating the same dimension-5 Weinberg operator as underpins the seesaw models, with the main difference being that it is generated at loop- rather than tree-level. Additional insertions of $v$ are often produced in radiative models from the need to use quark or charged-lepton mass insertions, but there is no such necessity in this model: the dominantly induced Weinberg-type operator is the standard one at dimension-5, rather than a higher-dimension generalisation obtained by multiplying by powers of $H^\dagger H$. At the level of the underlying renormalisable theory, one also observes that one of the contributing vertices in the numerator is a trilinear scalar coupling, thus having the dimension of mass, and most naturally set at the scale of the new physics. One possible source of suppression is thus absent. With order one dimensionless couplings, and the scalar trilinear coupling set at the new physics scale, the masses of the exotic scalars can be pushed as high as $10^7$ TeV. With the dimensionless couplings at $0.01$, so of fine-structure constant magnitude, the scale of new physics drops to the hundreds of GeV level. Hence, while the existence of new particles at the $1$-$100$ TeV scale explorable at current and proposed colliders is consistent with this model, it is not inevitable. From this perspective, models which require some of the couplings in the neutrino mass diagram to be standard-model Yukawa couplings (other than that of the top quark) are more experimentally relevant. From the model-building perspective, however, our analysis raises the prospect that some models based on tree-level UV completions of mass dimension 13 (and possibly even higher) operators might be viable.\\\\
The specific model analysed in this paper, consisting of an isotriplet scalar leptoquark, an isosinglet diquark and a third exotic scalar multiplet that has no Yukawa interactions, successfully generates neutrino masses and mixings at two-loop level consistent with experimental bounds from a variety of processes, of which $\mu \to e$ conversion on nuclei proved to be the most stringent. It can also ameliorate the discrepancies between measurements and standard-model predictions for $R_{K^{(*)}}$ and the anomalous magnetic moment of the muon.

\acknowledgments
We acknowledge the use of Package X \cite{Patel_2015}, which was used to perform the calculations in Section \ref{sec: mu to eee}. Some Feynman diagrams were drawn with the help of the Ti\textit{k}Z-Feynman package for \LaTeX~\cite{Ellis:2016jkw}. 
This work was supported in part by the Australian Research Council.
\appendix
\section{The full Lagrangian}
\label{App: full lagrangian}

The most general $SU(3)_c \otimes SU(2)_L \otimes U(1)_Y $ gauge-invariant, renormalisable Lagrangian will consist of the SM Lagrangian, together with additional NP terms,
\begin{equation} \begin{split}
\mathcal{L}_{\text{NP}} &= \mathcal{L}_{\text{gauge}-S} + \mathcal{L}_{S-H} + \mathcal{L}_{S-F} + \mathcal{L}_{S-S^{(\prime)}} ,
\label{eqn: full lagrangian}
\end{split} \end{equation}  
that involve interactions between the following: the exotic scalars, denoted $S^{(')}$, and the SM fermions, $F$; the exotic scalars and the Higgs boson, $H$; and the exotic scalars themselves. It also includes the gauge interactions of the exotic scalars. The accidental lepton and baryon number symmetries which are conserved by the SM Lagrangian have not been imposed here. \\
\subsection{The gauge sector}
The gauge covariant derivative acting on the scalar fields is 
{\begin{equation} \begin{split}
D_{\mu} = \partial_{\mu}+ ig_{1}YB_{\mu} + ig_{2}I_{k}W^{k}_{\mu} + ig_{3}\frac{\lambda_{A}}{2}G_{\mu}^{A},
\end{split} \end{equation} }%
where $Y$ is the hypercharge of the scalar, $I_{k}$ are the $SU(2)$ representation matrices ($I_{k}$ = 0 for SU(2) singlets and $-i\epsilon_{k}$ for $SU(2)$ triplets),  $\lambda_{A}$ for $A = 1,...,8$ are the Gell-Mann matrices, $g_{1}$, $g_{2}$ and $g_{3}$ are the respective coupling constants and after electroweak symmetry breaking we have $e= g_{2}\mathrm{sin~\theta_{W}}>0$ and $g_{1}/g_{2} = \mathrm{tan~\theta_{W}}$ \cite{Dorsner:2016wpm}. \\\\
The gauge interactions of $S = S_{1}$, $S_{3}$ and $\phi_{3}$  are then
{\begin{equation} \begin{split}
\mathcal{L}_{\text{gauge}-S} \supset (D_{\mu}S)^{\dagger}(D^{\mu}S),
\label{eqn: gauge lagrangian for exotics}
\end{split} \end{equation} }%
with the appropriate selections of hypercharge and $SU(2)$ representation matrices. 
\subsection{The fermion sector}
The part of the Lagrangian involving couplings between the fermions and the scalar bosons is 
{\begin{equation} \begin{split}
\mathcal{L}_{S-F} \supset \mathcal{L}_{S_3-F}+\mathcal{L}_{S_1-F} + \mathcal{L}_{\phi_3-F},
\label{eqn: fermion full lagrandian}
\end{split} \end{equation} }%
where 
{\begin{equation}
\begin{split}
\mathcal{L}_{S_3-F} =  &+y^{LL}_{3i,j}\overline{Q^{C}}^{~i,a}\epsilon_{ab}(\tau^k S^k_3)^{bc} L_L^{j,c} + z_3^{LL} \overline{Q^{C}}^{~i,a}\epsilon_{ab}(\tau^k S^k_3)^{\dagger~ bc} Q^{j,c} + \mathrm{h.c},
\\
 = &-\frac{(y_{3}^{LL})_{ij}}{\sqrt{2}}\overline{d_{L}^{C}}^{~i}S_{3}^{1/3}\nu^{j} - y_{3 ij}^{LL}\overline{d_{L}^{C}}^{~i}S_{3}^{4/3}e_{L}^{j}\\
 &+ (V^{*}_{\text{CKM}}y_{3}^{LL})_{ij}\overline{u_{L}^{C}}^{~i}S_{3}^{-2/3}\nu^{j} - \frac{(V^{*}_{\text{CKM}}y_{3^{LL}})_{ij}}{\sqrt{2}}\overline{u_{L}^{C}}^{~i}S_{3}^{1/3}e_{L^{j}}\\
 &-\frac{(z_{3}^{LL}V_{\text{CKM}}^{\dagger})_{ij}}{\sqrt{2}}\overline{d_{L}^{C}}^{~i}S_{3}^{1/3*}u_{L}^{j} - z^{LL}_{3ij}\overline{d^{C}_{L}}^{~i}S_{3}^{-2/3*}d^{j}_{L}\\
 &+ (V^{*}_{\text{CKM}}z_{3}^{LL}V_{\text{CKM}}^{\dagger})_{ij}\overline{u_{L}^{C}}^{~i}S_{3}^{4/3*}u_{L}^{j} - \frac{(V_{\text{CKM}}^{*}z_{3}^{LL})_{ij}}{\sqrt{2}}\overline{u_{L}^{C}}^{~i}S_{3}^{1/3*}d_{L}^{j}+\mathrm{h.c.},\\
\mathcal{L}_{S_1-F} = &+y^{LL}_{1ij} \overline{Q^{C}}^{~i,a}S_1 \epsilon_{ab} L^{j,b} + y^{RR}_{1 ij} \overline{u^{C}_R}^{i}S_1 e_R^j +z^{LL}_{1 ij} \overline{Q^{C}}^{i,a}S^*_1 \epsilon_{ab}Q^{j,b} \\
&+z^{RR}_{1 ij} \overline{u^{C}_R}^{~i}S^*_1 d_R^j +\mathrm{h.c.}\\
= &-y_{1ij}^{LL}\overline{d_{L}^{C}}^{~i}S_{1}\nu^{j} + (V_{\text{CKM}}^{*}y_{1}^{LL})_{ij}\overline{u_{L}^{C}}^{~i}S_{1}e_{L}^{j}+y_{1 ij}^{RR} \overline{u_{R}^{C}}^{~i}S_{1}e_{R}^{j} \\
&- (2z_{1}^{LL}V_{\text{CKM}}^{\dagger})_{ij}\overline{d_{L}^{C}}^{~i}S_{1}^{*}u_{L}^{j} +z^{RR}_{1 ij} \overline{u^{C}_R}^{~i}S^*_1 d_R^j + \mathrm{h.c.} 
\end{split}
\label{full lagrangian fermion couplings}
\end{equation}}%
There are no Yukawa interactions allowed by the SM gauge symmetries between the scalar $\phi_3$ and SM fermions.
The $\tau^k$, $k=1,2,3$, are the Pauli matrices; $i,j = 1, 2, 3$ are generation indices; $a,b = 1,2$ are $SU(2)$ flavour indices; $\epsilon^{ab} = (i\tau^2)^{ab}$; and $S_3^k$ are components of $S_3$ in $SU(2)$ space. The Levi-Civita tensor is needed in order to conserve charge for $SU(2)$ triplets and doublets. Superscript $C$ stands for the charge conjugation operation. For a fermion field $\psi$: $\psi_{R,L} = P_{R,L}\psi$, $\overline{\psi} = \psi^{\dagger}\gamma^{0}$, and $\psi^{C}= C\overline{\psi}^{T}$, where $P_{R,L} = (1 \pm \gamma^{5})/2$ and $C= i\gamma^{2}\gamma^{0}$. The diquark coupling to $S_{1}$, $z_{1}^{LL}$ is symmetric due to a combination of the antisymmetry of $SU(2)$ structure and of the colour structure of the fermion bilinear, while the diquark coupling to $S_{3}$, $z_{3}^{LL}$, must be antisymmetric for similar reasons. In the expansion of the $SU(2)$ structure of the Lagrangian, we have also rotated into the mass eigenbasis of the quarks, using the convention that $u_{L}^{i} \rightarrow (V_{\text{CKM}}^{\dagger})_{ij} u_{L}^{j}$ and $d_{L}^{i} \rightarrow d_{L}^{j}$.

\subsection{The scalar sector}
The scalar part of the interaction Lagrangian will contain quartic interactions with dimensionless couplings, and cubic and quadratic interactions with dimensionful couplings. The full scalar Lagrangian, without baryon number conservation imposed, is
{\setlength{\abovedisplayskip}{0pt}
\setlength{\belowdisplayskip}{0pt}
\begin{equation} \begin{split}
\mathcal{L}_{S-S^{(')}}  +\mathcal{L}_{S-H}  = \mathcal{L}_{4SB}+\mathcal{L}_{3SB} + \mathcal{L}_{2SB},
\label{eqn: scalar full lagrangian}
\end{split} \end{equation}}%
with
{\begin{align}
\begin{split}
\mathcal{L}_{4SB} = &\phantom{+} \lambda_{S_3 H}[S_3^\dagger  S_3]_1 [H^\dagger H]_1+\lambda_{S_3 H}'[S_3^\dagger  S_3]_3 [H^\dagger H]_3 + \lambda_{S_1 H}S_1^\dagger  S_1 [H^\dagger H]_1 \\
&+ \lambda_{\phi_3 H}[\phi_3^\dagger  \phi_3]_1[ H^\dagger H]_1+ \lambda_{\phi_3 H}'[\phi_3^\dagger  \phi_3]_3[ H^\dagger H]_3+ \lambda_{S_1 \phi_3}S_1^\dagger  S_1[ \phi_3^\dagger \phi_3]_1 \\
&+ \lambda_{S_3 \phi_3}[S_3^\dagger  S_3]_1 [\phi_3^\dagger \phi_3]_1+\lambda_{S_3 \phi_3}'[S_3^\dagger  S_3]_3 [\phi_3^\dagger \phi_3]_3+\lambda_{S_3 \phi_3}''[S_3^\dagger  S_3]_5 [\phi_3^\dagger \phi_3]_5 \\
&+\lambda_{S_3} [S_3^\dagger  S_3]_1[S_3^\dagger  S_3]_1+\lambda_{S_3}' [S_3^\dagger  S_3]_3[S_3^\dagger  S_3]_3+ \lambda_{S_1 S_3}[S_3^\dagger  S_3]_1S_1^\dagger S_1\\
&+ \lambda'_{S_1 S_3}[S_3^\dagger  S_3^{\dagger}]_1S_1 S_1 + \lambda_{S_1 \phi_3 S_{3}}[\phi_3^\dagger  \phi_3]_3S_{3} S_1^{\dagger} \\
& + \lambda_{S_1}(S_1^\dagger  S_1)^2 +\lambda_{\phi_3}[\phi_3^\dagger  \phi_3]_1[\phi_3^\dagger  \phi_3]_1+\lambda_{\phi_3}'[\phi_3^\dagger  \phi_3]_3[\phi_3^\dagger  \phi_3]_1\\
&+ \lambda_{S_1 S_3 H}S_1S_3^\dagger   [H^\dagger H]_3 + \lambda_{\phi_3 S_3 H}[\phi_3^\dagger  S_3^\dagger]_3 [H H]_3\\
&+ \lambda_{S_1 \phi_3 H}S_1^\dagger \phi_3^\dagger   [H H]_3  +\mathrm{h.c.}, \\
\mathcal{L}_{3SB} = &\phantom{+}m_{S_3 S_1 \phi_3}[S_3^\dagger  \phi_3 ]_{1}S_1^\dagger+\mathrm{h.c.}, \\
\mathcal{L}_{2SB} =&-\mu^2_{S_3}S_3^\dagger S_3 -\mu^2_{S_1}S_1^\dagger S_1 - \mu^2_{\phi_3}\phi_3^\dagger \phi_3.
\end{split}
\end{align}}%
The square brackets $[...]_{i}$ indicate that the scalars enclosed couple to form an $SU(2)$ singlet for $i=1$ or triplet for $i=3$ etc.
\section{Calculation of $\mathcal{I}_{kl}$}
\label{sec:Calculation of integral}
The two-loop integral in Equation \ref{eqn: two-loop integral for O47}  is actually a sum of integrals 
\begin{equation} \begin{split}
\mathcal{I}_{kl} = -\frac{8i \mathrm{cos~\theta}\, \mathrm{sin~\theta}}{(2\pi)^{8}}(\mathcal{I}_{kl31}^{k_{1} \cdot k_{2}} \mathrm{cos^{2}~\theta} - \mathcal{I}_{kl32}^{k_{1} \cdot k_{2}} \mathrm{cos^{2}~\theta} +\mathcal{I}_{kl41}^{k_{1} \cdot k_{2}} \mathrm{sin^{2}~\theta}-\mathcal{I}_{kl42}^{k_{1} \cdot k_{2}} \mathrm{sin^{2}~\theta}),
\end{split} \end{equation} 
with 
\begin{equation} \begin{split}
\mathcal{I}_{kl\alpha \beta} ^{k_{1}\cdot k_{2}}= \int d^{4}k_{1}\int d^{4}k_{2} \frac{k_{1}\cdot k_{2}}{(k_{1}^{2}- s_{k} )(k_{2}^{2}-s^{'}_{l})(k_{1}^{2}-t_{\alpha})(k_{2}^{2}-t_{\beta})([k_{1}-k_{2}]^{2}-1)}.
\end{split} 
\label{eq: The Integral}
\end{equation} 
Using Appendix C of \cite{Sierra:2014rxa} we find that integral $\mathcal{I}_{kl\alpha \beta}$ evaluates to
\begin{equation} \begin{split}
\mathcal{I}_{kl\alpha \beta}^{k_{1}\cdot k_{2}} = &-\frac{\pi^{4}}{2}  \Big[\hat{B}_{0}^{'}(0,s_{k},t_{\alpha})\hat{B}_{0}^{'}(0,s_{l}^{'},t_{\beta})+\\
& \frac{-\hat{g}(t_{\alpha},t_{\beta})(1+t_{\alpha}+t_{\beta})+ \hat{g}(t_{\alpha},s^{'}_{l})(1+t_{\alpha}+s^{'}_{l}) + \hat{g}(s_{k}, t_{\beta})(1+s_{k} +t_{\beta} ) - \hat{g}(s_{k},s^{'}_{l})(1+s_{k}+s^{'}_{l})}{(t_{\alpha}-s_{k})(t_{\beta}-s_{l}^{'})}\Big].
\end{split} \end{equation} 
The finite parts of $\hat{g}(s,t)$ and $\hat{B}_{0}^{'}(0,s,t)$ are defined as
\begin{equation} \begin{split}
\hat{B}_{0}^{'}(0,s,t) = i \left(\frac{s~\mathrm{ln}~s - t~\mathrm{ln}~t }{s-t} \right)
\end{split} \end{equation} 
and
\begin{equation} \begin{split}
\hat{g}(s,t) &= \frac{s}{2}\mathrm{ln}~s~\mathrm{ln}~t + \sum_{\pm} \pm \frac{s(1-s)+ 3st + 2(1-t)x_{\pm}}{2w}\\
& \times \left[\mathrm{Li}_{2}\left(\frac{x_{\pm}}{x_{\pm}-s} \right) - \mathrm{Li}_{2}\left(\frac{x_{\pm}-s}{x_{\pm}}\right)  +\mathrm{Li}_{2}\left(\frac{t-1}{x_{\pm}} \right)-\mathrm{Li}_{2}\left(\frac{t-1}{x_{\pm}-s} \right)\right],
\label{eqn: defn of g}
\end{split} \end{equation} 
where 
\begin{equation} \begin{split}
\mathrm{Li}_{2}(x) = - \int^{x}_{0}\frac{\ln(1-y)}{y}dy,
\end{split} \end{equation} 
and 
\begin{equation} \begin{split}
x_{\pm}= \frac{1}{2}(-1+s+t\pm w)\phantom{blank~spaces}w=\sqrt{1+s^{2} t^{2} - 2(s+t+st)}.
\label{eqn: defn of subparts of g}
\end{split} \end{equation} 
We need not worry about the non-finite parts of the integrals as we are guaranteed that they will cancel. \\\\
Finally, we give the evaluation of $\mathcal{I}_{kl}$ in the limiting case where the ratios of SM fermion squared masses on $m_{S_{1}}^{2}$ go to zero, specifically $s_{k},~ s_{l}^{'} \rightarrow 0$. In this limit 
\begin{equation} \begin{split}
\lim_{s_{k}\rightarrow 0, s^{'}_{l}\rightarrow 0}(\mathcal{I}_{kl\alpha \beta})=\mathcal{I}_{00\alpha \beta} =  \frac{\pi^{4}}{2}\left(\mathrm{ln}~t_{\alpha}~\mathrm{ln}~t_{\beta} + \frac{\hat{g}(t_{\alpha},t_{\beta})(1+t_{\alpha}+t_{\beta}) - \hat{g}(t_{\alpha},0)(1+t_{\alpha})}{t_{\alpha}t_{\beta}} \right),
\end{split} \end{equation} 
where taking the limit $\hat{g}(s,t)\xrightarrow{t\rightarrow 0}\hat{g}(s,0)$ given the definitions in Equations \ref{eqn: defn of g} and \ref{eqn: defn of subparts of g} gives 
{\begin{equation} \begin{split}
\hat{g}(s,0) = -s\frac{\pi^{2}}{6} - (1-s)\mathrm{ln}~s~\mathrm{ln}(1-s) - (1-s)\mathrm{Li_{2}}(s)).
\end{split} 
\label{eqn: defn of g0}
\end{equation} }
\newpage
\section{Explicit calculation of vanishing neutrino mass in Model 2}
\label{App: vanishing model}
\begin{figure}[t!]
\begin{subfigure}[t]{0.45\textwidth}
\centering
\begin{tikzpicture}[thick,scale=1]
	\coordinate[] (a1) at (-1,1.5) {};
	
	\coordinate[] (a2) at (-2,0.5) {};
		\node[vtx, label=180:$L$] at (-2,0.5) {};
	\coordinate[] (a3) at (-2,2.5) {};
		\node[vtx, label=180:$Q$] at (-2,2.5) {};
	\coordinate[] (mid1) at (0.5,1.5) {};
	
	\coordinate[] (mid2) at (0.5,2.5) {};
	\coordinate[] (c1) at (-0.5,3.5) {};
	\node[vtx, label=180:$\overline{Q}$] at (-0.5,3.5) {};
	\coordinate[] (c2) at (1.5,3.5) {};
	\node[vtx, label=0:$Q^{C}$] at (1.5,3.5) {};
	
	\coordinate[] (b1) at (2,1.5) {};
	\coordinate[] (b2) at (3,0.5) {};
		\node[vtx, label=0:$L$] at (3,0.5) {};
	\coordinate[] (b3) at (3,2.5) {};
		\node[vtx, label=0:$Q$] at (3,2.5) {};
	\coordinate[] (midH) at (1.25,1.5) {};
	\coordinate[] (H1) at (0.9,-0.2) {};
		\node[vtx, label=180:$\braket{H}$] at (0.9,-0.2) {};
	\coordinate[] (H2) at (1.6,-0.2) {};
		\node[vtx, label=0:$\braket{H}$] at (1.6,-0.2) {};
		\node[vtx, label=0:$S_{3}$] at (0.5,2) {};
		\node[vtx, label=0:$S_{1}$] at (-0.5,1.2) {};
		\node[vtx, label=0:$S_{1}$] at (1.5,1.2) {};
		\node[vtx, label=0:$\phi_{3}$] at (0.5,1.2) {};
		
		\graph[use existing nodes]{
		a1--[antifermion]a2;
		a1--[antifermion]a3;
		a1--[scalar]mid1;
		mid1--[scalar]mid2;
		mid1--[scalar]b1;
		b1--[antifermion]b2;
		b1--[antifermion]b3;
		mid2--[fermion]c1;
		mid2--[fermion]c2;
		midH--[scalar]H1;
		midH--[scalar]H2;
	};

\end{tikzpicture}
%
%
%
%
\caption{The UV-completion of operator $O_{47}$ with the introduction of three exotic scalars; $S_{1}$, $S_{3}$, $\phi_{3}$.}
\label{fig: neutrino mass diagram for our model 2 a}
\end{subfigure}
\hfill
\begin{subfigure}[t]{0.5\textwidth}
\centering
\begin{tikzpicture}[thick,scale=1.1]
	\coordinate[] (a1) at (-1,1.5) {};
		\node[vtx, label=0:$L$] at (-0.5,1.2) {};
	\coordinate[] (a2) at (-0.5,3) {};
		\node[vtx, label=0:$Q$] at (0,2.5) {};
	\coordinate[] (b) at (0.5,1.5) {};
		
	\coordinate[] (mid) at (2,1.5) {};
	\coordinate[] (mid2) at (2,3) {};
	\coordinate[] (d) at (3.5,1.5) {};
		\node[vtx, label=180:$S_{3}$] at (2.65,2.3) {};
	\coordinate[] (e1) at (5,1.5) {};
		\node[vtx, label=180:$L$] at (4.7,1.2) {};
	\coordinate[] (e2) at (4.5,3) {};
		\node[vtx, label=180:$Q$] at (4,2.5) {};
	\coordinate[vtx, label=180:$\braket{H}$] (H) at (2.45,0) {};
	\coordinate[vtx, label=0:$\braket{H}$] (H2) at (3.05,0) {};
	\coordinate[](quarter) at (2.75,1.5){};
		\node[vtx, label=180:$S_{1}$] at (1.6,1.2) {};
		\node[vtx, label=180:$\phi_{3}$] at (2.7,1.2) {};
		\node[vtx, label=180:$S_{1}$] at (3.5,1.2) {};

	\graph[use existing nodes]{
		a1--[fermion]b;
		b--[scalar]mid;
		mid--[scalar]mid2;
		mid--[scalar]d;
		H--[scalar]quarter;
		H2--[scalar]quarter;
		d--[antifermion]e1;
	};
	
    \draw[postaction={decorate},thick,decoration = {markings,
    mark=at position 0.5 with {\arrow {latex}}}] (2,3) arc (90:0:1.5);    
   \draw[postaction={decorate},thick,decoration = {markings,
    mark=at position 0.5 with {\arrow {latex}}}] (2,3) arc (90:180:1.5);
\end{tikzpicture}
\caption{ Closing the loops by joining the quarks leads to a neutrino mass diagram.}
\label{fig: neutrino mass diagram for our model 2 b}
\end{subfigure}
\begin{subfigure}[b]{0.99\textwidth}
\centering
\begin{tikzpicture}[thick,scale=1.1]
	\coordinate[] (a1) at (-1,1.5) {};
		\node[vtx, label=0:$\nu_{i}$] at (-0.5,1.2) {};
	\coordinate[] (a2) at (-0.5,3) {};
		\node[vtx, label=0:$d_{L}$] at (0,2.5) {};
	\coordinate[] (b) at (0.5,1.5) {};
		
	\coordinate[] (mid) at (2,1.5) {};
	\coordinate[] (mid2) at (2,3) {};
		\node[vtx, label=180:$S_{1}$] at (2.65,2.3) {};
	\coordinate[] (d) at (3.5,1.5) {};
		\node[vtx, label=180:$r_{\alpha}$] at (1.6,1.2) {};
		\node[vtx, label=180:$r_{\beta}$] at (3.2,1.2) {};

	\coordinate[] (e1) at (5,1.5) {};
		\node[vtx, label=180:$\nu_{j}$] at (4.7,1.2) {};
	\coordinate[] (e2) at (4.5,3) {};
		\node[vtx, label=180:$d_{L}$] at (4,2.5) {};
		
	\coordinate[](quarter) at (2.75,1.5){};
	\graph[use existing nodes]{
		a1--[fermion]b;
		b--[scalar]mid;
		mid--[scalar]mid2;
		mid--[scalar]d;
		d--[antifermion]e1;
	};
	
    \draw[postaction={decorate},thick,decoration = {markings,
    mark=at position 0.5 with {\arrow {latex}}}] (2,3) arc (90:0:1.5);    
   \draw[postaction={decorate},thick,decoration = {markings,
    mark=at position 0.5 with {\arrow {latex}}}] (2,3) arc (90:180:1.5);
\end{tikzpicture}
\caption{The neutrino self-energy, after electroweak symmetry breaking and rotating into the mass basis of the exotic scalars.}
\label{fig: neutrino mass diagram for our model 2 c}
\end{subfigure}
\caption{The UV--completion of operator $O_{47}$ by the exotic scalars $S_{3}$ (diquark), $S_{1}$ (leptoquark) and $\phi_{3}$ and its closure forming the neutrino self-energy Feynman diagrams for our model. The leptoquark mass states obtained after mixing are $r_{\alpha}$, for $\alpha = 1, 2.$}
\label{fig: neutrino mass diagram for our model 2}
\end{figure}
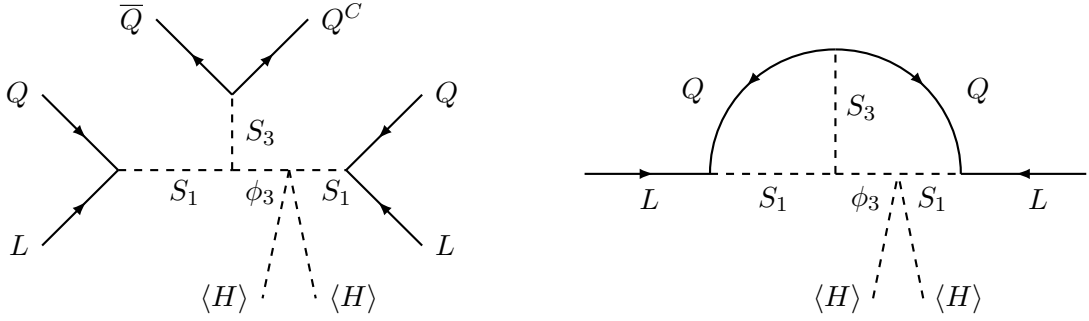
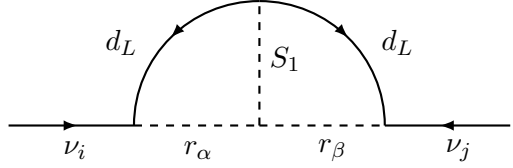
Here, we briefly present an explicit calculation of the vanishing neutrino mass model we refer to as Model 2. This neutrino mass model arises from the UV-completion of operator $O_{47}$, with $S_3$, coupling as a diquark, $S_1$ coupling as a leptoquark and $\phi_3$ only coupling to other scalars and gauge bosons. The part of the Lagrangian coupling fermions to the exotic scalars can be derived from Equation \ref{full lagrangian fermion couplings} by imposing baryon number conservation and fixing the baryon numbers to be $B(S_3)=2/3$, $B(S_1) = -1/3$ and $B(\phi_3)= 1/3$. We can see from Figure \ref{fig: neutrino mass diagram for our model 2} involves mixing between $S_1^*$ and $\phi_3^{-1/3}$. In a similar fashion to that detailed for Model 1 in Section \ref{sec: neutrino mass generation} neutrino mass is then obtained from the flavour sum of the self-energy diagrams with the freedom to set external momentum to zero:
\begin{equation} \begin{split}
i\Sigma_{ij} \propto m'_{S_{1}S_{3}\phi_{3}}\Big((y_{1}^{LL})^{T}_{ik}z_{3kl}^{LL}\mathcal{I}_{k l \alpha \beta}R_{\alpha \beta}y_{1lj}^{LL} \Big) + (i \leftrightarrow j),
\end{split} 
\label{eq: self energy diagrams model 2}
\end{equation} 
where $y^{LL}_1$ is the leptoquark coupling matrix, $z_3^{LL}$ is the diquark coupling matrix and $m'{S_1S_3\phi_3}$ represents the cubic exotic scalar coupling. The integral $I_{kl\alpha \beta}$ is defined as in Equation \ref{eq: The Integral}, with the $k_1 \cdot k_2$ being suppressed to reduce clutter. The indices $k,~l$ run over the three possible quark flavours and are summed over, while $\alpha,~\beta$ run over the two possible leptoquark mass states obtained from the mixing of $S_1^*$ and $\phi_3^{-1/3}$. We also define
\begin{equation} \begin{split}
R_{\alpha \beta}= \begin{pmatrix}
-\cos{\theta'}&\cos{\theta'}-\sin{\theta'}\\
\cos{\theta'}-\sin{\theta'}& \sin{\theta'}
\end{pmatrix},
\end{split} 
\label{eq: R matrix vanishing model}
\end{equation} 
where $\theta'$ is the mixing angle between $S_1^*$ and $\phi_3^{-1/3}$.\\\\
The argument showing that Equation \ref{eq: self energy diagrams model 2} leads to vanishing neutrino mass goes as follows :
\begin{enumerate}
\item The diquark coupling to $S_3$, $z_3^{LL}$ is antisymmetric in flavour space, i.e. $z_{3ij}^{LL}=-z_{3ji}^{LL}$. 
\item The integral $\mathcal{I}_{kl\alpha \beta}$ is symmetric under simultaneous relabelling of $k \leftrightarrow l$ and $\alpha \leftrightarrow \beta$, and the matrix $R_{\alpha \beta}$ is symmetric under $\alpha \leftrightarrow \beta.$
\item The neutrino mass then evaluates to
\begin{align} 
\begin{aligned}
i\Sigma_{ij} &\propto m'_{S_{1}S_{3}\phi_{3}}\Bigg((y_{1}^{LL})^{T}_{ik}z_{3kl}^{LL}\mathcal{I}_{k l \alpha \beta}R_{\alpha \beta}y_{1lj}^{LL}  + (y_{1}^{LL})^{T}_{jk}z_{3kl}^{LL}\mathcal{I}_{k l \alpha \beta}R_{\alpha \beta}y_{1li}^{LL} \Bigg)\\
&= m'_{S_{1}S_{3}\phi_{3}}\Bigg((y_{1}^{LL})^{T}_{ik}z_{3kl}^{LL}\mathcal{I}_{k l \alpha \beta}R_{\alpha \beta}y_{1lj}^{LL}  + (y_{1}^{LL})^{T}_{il}z_{3kl}^{LL}\mathcal{I}_{k l \alpha \beta}R_{\alpha \beta}y_{1jk}^{LL} \Bigg)&\text{reorder couplings}\\
&= m'_{S_{1}S_{3}\phi_{3}}\Bigg((y_{1}^{LL})^{T}_{ik}z_{3kl}^{LL}\mathcal{I}_{k l \alpha \beta}R_{\alpha \beta}y_{1lj}^{LL}  + (y_{1}^{LL})^{T}_{ik}z_{3lk}^{LL}\mathcal{I}_{lk \alpha \beta}R_{\alpha \beta}y_{1lj}^{LL} \Bigg)&\text{relabel $k\leftrightarrow l$}\\
 &=m'_{S_{1}S_{3}\phi_{3}}(y_{1}^{LL})^{T}_{ik}\Big[z_{3kl}^{LL}\mathcal{I}_{k l \alpha \beta}R_{\alpha \beta} 
 + (z_{3}^{LL})^{\dagger}_{kl}\mathcal{I}_{l k \alpha \beta}R_{\alpha \beta} \Big]y_{1lj}^{LL}  &\text{collect like terms} \\
 &=m'_{S_{1}S_{3}\phi_{3}}(y_{1}^{LL})^{T}_{ik}\Big[z_{3kl}^{LL}\mathcal{I}_{k l \alpha \beta}R_{\alpha \beta} 
 - z_{3kl}^{LL}\mathcal{I}_{l k  \beta \alpha}R_{\beta \alpha } \Big]y_{1lj}^{LL}&\text{reorder indices}\\
 &= 0,
\end{aligned}
\label{eq: self energy diagrams model 2 full}
\end{align}
where in the last line we used points 1. and 2., specifically that $z_{3kl}^{LL}=-z_{3lk}^{LL}$ and $\mathcal{I}_{k l\alpha \beta}= \mathcal{I}_{lk \beta \alpha}$. The neutrino mass arising from the UV-completion of $O_{47}$ with $S_3$ coupling as a diquark, $S_1$ coupling as a leptoquark and $\phi_3$ not coupling to any SM fermions vanishes. 
\end{enumerate}
\bibliographystyle{JHEP}

\bibliography{dim11}
%
%
%



\end{document}